\documentclass[aps,pra,twocolumn,superscriptaddress,footinbib]{revtex4-2}
\usepackage{amsmath,amssymb,bm}
\usepackage{graphicx}
\usepackage{epstopdf}
\usepackage{latexsym}
\usepackage[normalem]{ulem} 
\usepackage[usenames,dvipsnames]{color}
\usepackage{hyperref}
\usepackage{natbib}
\usepackage{braket}
\usepackage{comment}
\begin{document}

\newcommand{\Rev}[1]{{\color{blue}{#1}\normalcolor}} 
\newcommand{\Com}[1]{{\color{red}{#1}\normalcolor}} 

\title{Tailored generation of quantum states in an entangled spinor interferometer to overcome detection noise}
\date{\today}

\author{Q. Guan}
\affiliation{Homer L. Dodge Department of Physics and Astronomy, The University of Oklahoma, Norman, Oklahoma 73019, USA}
\affiliation{Center for Quantum Research and Technology, The University of Oklahoma, Norman, Oklahoma 73019, USA}

\author{G.~W. Biedermann}
\affiliation{Homer L. Dodge Department of Physics and Astronomy, The University of Oklahoma, Norman, Oklahoma 73019, USA}
\affiliation{Center for Quantum Research and Technology, The University of Oklahoma, Norman, Oklahoma 73019, USA}

\author{A. Schwettmann}
\affiliation{Homer L. Dodge Department of Physics and Astronomy, The University of Oklahoma, Norman, Oklahoma 73019, USA}
\affiliation{Center for Quantum Research and Technology, The University of Oklahoma, Norman, Oklahoma 73019, USA}

\author{R.~J. Lewis-Swan}
\affiliation{Homer L. Dodge Department of Physics and Astronomy, The University of Oklahoma, Norman, Oklahoma 73019, USA}
\affiliation{Center for Quantum Research and Technology, The University of Oklahoma, Norman, Oklahoma 73019, USA}

\begin{abstract}
We theoretically investigate how entangled atomic states generated via spin-changing collisions in a spinor Bose-Einstein condensate can be designed and controllably prepared for atom interferometry that is robust against common technical issues, such as limited detector resolution. We use analytic and numerical treatments of the spin-changing collision process to demonstrate that triggering the entangling collisions with a small classical seed rather than vacuum fluctuations leads to a more robust and superior sensitivity when technical noise is accounted for, despite the generated atomic state ideally featuring less metrologically useful entanglement. Our results are relevant for understanding how entangled atomic states are best designed and generated for use in quantum-enhanced matter-wave interferometry.
\end{abstract}

\maketitle  

\section{Introduction}
Entanglement, correlations and coherence have the potential to enable a quantum advantage in many tasks, including information processing, communications and metrology \cite{Cappellaro_2017}. However, due to the inherent fragility of such quantum phenomena to decoherence and technical imperfections, real-world examples of quantum-enhanced devices that outperform their state-of-the-art classical counterparts in meaningful applications remain rare \cite{AdvancedLigo2016}.

The use of cold atoms for quantum-enhanced sensors are a prominent example \cite{Pezze_2018}, as they have long been identified as a potential quantum platform with promising applications such as gravimetry \cite{Haine_Gravimeter_2020,bidel_2018,Peters1999}, time-keeping \cite{LudlowReview_2015}, navigation \cite{Bouyer_2018,GB_APL_2012} and resource exploration \cite{bongs_2019}. While there has been extensive progress in the generation of metrologically useful atomic entangled states \cite{EGW08,Gross2011a,Hamley2012a,Hosten_2016b,Bohnet2014,Strobel2014,Bohnet2014a,Schmiedmayer_2021}, including conceptual demonstrations of quantum-enhanced interferometry \cite{Lucke_2011,GZN10,Hosten_2016,Linnemann_2016,Vuletic_2020}, a myriad of technical challenges remain to be overcome to realize a quantum-enhanced device that is competitive with practical state-of-the-art sensors using separable atomic ensembles \cite{Haine_2021,GB_PRX_2020}. 

One relatively ubiquitous challenge is detection resolution, i.e., the ability to accurately resolve or count single atoms in large ensembles. Fundamentally, this limits the degree to which quantum states can be distinguished and thus inherently places bounds on how well small perturbations to a system can be inferred \cite{Braunstein1994}. To compound matters, one almost invariably finds that the demands on detection resolution increase in step with the degree of metrological enhancement that a quantum state can provide. An excellent example are macroscopic superposition states such as GHZ or NOON states, which in principle enable improvements in precision by a factor of $1/\sqrt{N}$ relative to current classical devices using $N$ probes, but typically require the ability to count single particles to enable measurements of parity or distribution functions \cite{bollinger1996optimal}. As such a capability is technically demanding, even in state-of-the-art experiments, and difficult to scale with particle number \cite{Evrard_2020} there have been efforts to overcome this limitation by developing novel methods such as interaction-based readout (IBR) \cite{Sekatski2015,Davis_2016,Nolan_2017,Fabian_2018}. Despite notable demonstrations \cite{Hosten_2016,colombo_2021,gilmore_2021}, IBR methods require a level of coherent control over the dynamics that can be demanding or impractical for many experimental platforms. Consequently, it is important to assess the metrological utility of quantum states with a practical viewpoint, striking a balance between idealized metrological enhancement and robustness to technical noise.

In this context, this manuscript presents a systematic investigation of the robustness and realistic metrological potential of a class of atomic entangled states for SU(2) atom interferometry with spinor BECs. Our study is targeted towards applications where entangled matter waves are spatially split and recombined to measure, e.g., gravitational acceleration, and are thus ill-suited to the widely studied IBR methods that underpin related SU$(1,1)$ atom interferometry \cite{Linnemann_2016,Jianwen_2019,Zhang_2019,Wrubel_2018}. We investigate the use of spin-changing collisions in a spinor BEC to generate atomic squeezing and entanglement, and show that entangled states generated by triggering the collisions with a small classical seed rather than vacuum fluctuations are more robust to realistic detection noise when using simple measurement observables. This is in contrast to the ideal scenario without technical noise, where seeding the entangling dynamics \emph{always} leads to a degradation of the metrological performance per particle. These results complement other favourable properties of seeding, such as an accelerated rate of entanglement generation due to bosonic stimulation and a broader dynamic range of sub-SQL sensitivity. Our findings are illustrated through an approximate analytic model of the spin-changing collisions, which enables us to derive insightful expressions for the sensitivity achievable with a range of experimentally relevant measurement signals. Moreover, the analytic predictions elucidate the dependence on initial state properties, such as the size and phase coherence of the seed. We also use numerical calculations of the exact quantum dynamics to verify our predictions and their relevance for experimentally realistic parameter regimes. Our results are pertinent for future demonstrations of quantum-enhanced atom interferometry using spin-changing collisions \cite{klempt2020spinor}, and demonstrate that judicious choices for initial state preparation can have important consequences for prospective quantum-enhancement \cite{davis2017advantages,Niezgoda_2018}. 

The manuscript is organized as follows. Section~\ref{sec:Model} introduces the physical model of spin-changing collisions in spinor BEC and briefly recaps the framework of an SU(2) atom interferometer. In Secs.~\ref{sec:Analytics} and \ref{sec:OptimalSensitivity} we use a simplified analytic model of the spin-changing collisions to obtain expressions for the ideal metrological performance and achievable sensitivity as a function of the quantum state and choice of measurement signal. We then expand our analysis to include the effect of deleterious technical noise and imperfect state preparation in Sec.~\ref{sec:DetectionNoise}. These predictions are compared to numerical calculations of the exact quantum dynamics for an experimentally relevant scenario in Sec.~\ref{sec:RealCalc}, before summarizing our results in Sec.~\ref{sec:Outlook}.

\section{Model System \label{sec:Model}}

\subsection{Few-mode Hamiltonian}
We consider the dynamics of a microwave dressed spin-$1$ Bose-Einstein condensate, such as that recently reported in Ref.~\cite{Arne2020}. The condensate is assumed to be confined in a deep trapping potential, which enables a simplified treatment where the spatial dynamics are frozen out and only the internal (spin) degrees of freedom need be considered. In this limit, known as the single-mode approximation (SMA) \cite{Law_1998,kawaguchi_2012, Jie2020}, the spinor dynamics are well described by the Hamiltonian $\hat{H} = \hat{H}_{\mathrm{inel}} + \hat{H}_{\mathrm{el}} + \hat{H}_{\mathrm{Z}}$ \cite{lewis2013epr} with:
\begin{equation}
\begin{gathered}
    \hat{H}_{\mathrm{inel}} = \hbar g\left( \hat{a}_0\hat{a}_0 \hat{a}^{\dagger}_1\hat{a}^{\dagger}_{-1} + h.c. \right),  \\
    \hat{H}_{\mathrm{el}} = \hbar g \hat{n}_0 \left( \hat{n}_1 + \hat{n}_{-1} \right) + \frac{\hbar g}{2}\left( \hat{n}_1 - \hat{n}_{-1} \right)^2 , \\
    \hat{H}_{\mathrm{Z}} = -\hbar p(\hat{n}_{1} - \hat{n}_{-1}) -\hbar q(\hat{n}_1 + \hat{n}_{-1}) . \label{eqn:H}
\end{gathered}
\end{equation}
Here, $n_{i} = \hat{a}^{\dagger}_i \hat{a}_i$ is the particle number operator for the $i=m_F=0,\pm1$ Zeeman sublevels. The Hamiltonian $\hat{H}$ in this form splits the interactions into two contributions: spin-changing ($\hat{H}_{\mathrm{inel}}$) and spin-preserving ($\hat{H}_{\mathrm{el}}$) collisions. The former describes a process where two $m_F = 0$ atoms scatter and generate a pair of atoms in $m_F = \pm1$, or vice versa, while the latter describes elastic scattering that preserves the occupation of each $m_F$ mode. The additional term, $\hat{H}_{\mathrm{Z}}$, arises due to an external magnetic field of magnitude $B$ and is decomposed into contributions from the linear and quadratic Zeeman shifts characterized by $p = g\mu_B B/\hbar$ and $q = p^2/\omega_{\mathrm{hf}}$, respectively, where $g$ is the Land\'{e} hyperfine g-factor, $\mu_B$ the Bohr magneton and $\omega_{\mathrm{hf}}$ the hyperfine frequency splitting. The quadratic term shifts both $m_F = \pm1$ states symmetrically with respect to $m_F = 0$, and can also be manipulated via complementary microwave dressing of the $m_F = 0$ state \cite{Arne2020,Linnemann_2016}, enabling the relative strengths of $p$ and $q$ to be tuned independently. Lastly, the Hamiltonian conserves the population difference $\hat{n}_1 - \hat{n}_{-1}$. As a consequence, in the analytic calculations presented in Sec.~\ref{sec:Analytics} and \ref{sec:DetectionNoise} we ignore the elastic scattering $\propto (\hat{n}_1 - \hat{n}_{-1})^2$ as a small irrelevant contribution (we validate this assumption by explicitly including it in Sec.~\ref{sec:RealCalc} and present a qualitative justification in Appendix \ref{app:Hamiltonian}). Moreover, we absorb the linear Zeeman shift by working in a frame rotating with it such that it falls out of our calculations (although we comment on the practical consequences of this where appropriate). Throughout the remainder of the manuscript we will set $\hbar = 1$.

\subsection{Initial state}
Typical experiments studying pair production dynamics in a spinor BEC focus on initial conditions where the majority of the condensate populates the $m_F = 0$ state and acts as a source for correlated pairs in $m_F = \pm1$. Here, we consider initial states where a BEC of $N$ atoms is prepared in the $m_F = 0$ mode, before a small number of atoms, $n_s$, are coherently transferred to either of the $m_F = \pm1$ modes by, e.g., resonant microwaves \cite{Lucke_2011}, to act as a coherent seed that stimulates the spin-changing collisions \cite{Jianwen_2019,Zhang_2019,Evrard_2021b}.

We distinguish two possible initial conditions stemming from such a preparation protocol. The first only considers a seed in the $m_F = 1$ mode, leading to the initial state,
\begin{equation}
    \vert \psi^s_0 \rangle = \vert \psi^s_{0,-1}, \psi^s_{0,0}, \psi^s_{0,1}\rangle = \vert 0, \sqrt{N-n_s},  \sqrt{n_s}e^{i\theta_s}\rangle . \label{eqn:Psi0}
\end{equation}
Here, we have assumed the $m_F = 0,1$ modes are described as coherent states with occupation $\langle \hat{n}_{1}(0) \rangle = \langle \psi_0 \vert \hat{n}_1 \vert \psi_0 \rangle = n_s$ and $\langle \hat{n}_0(0) \rangle = N-n_s$ \cite{Jianwen_2019}, respectively, while the $m_F = -1$ mode is prepared in the vacuum state with $\langle \hat{n}_{-1} \rangle = 0$. Without loss of generality we have taken the $m_F = 0$ coherent state to have a real amplitude, such that any relevant phase relationship between the $m_F = 0,\pm1$ modes is encoded in $\theta_s$. 

Secondly, we consider seeding both $m_F = \pm1$ modes (similar to a previous study, Ref.~\cite{Nolan_2016}). This initial condition offers a wide range of tunability in terms of, e.g., relative particle number and phase between the $m_F = \pm 1$ modes \cite{Jianwen_2019}, but we will choose to focus on the specific initial configuration described by,
\begin{equation}
    \vert \psi^d_0 \rangle = \vert \sqrt{n_s/2}e^{-i\theta_s}, \sqrt{N-n_s},  \sqrt{n_s/2}e^{i\theta_s}\rangle . \label{eqn:Psi0_double}
\end{equation}
Here, the $m_F = \pm1$ states are taken to be coherent states with identical occupation $\langle \hat{n}_{\pm 1}(0) \rangle = n_s/2$ and phase $\pm \theta_s$. The latter is chosen so that the phase coherence between the $m_F = \pm1$ modes ($\langle \psi^d_0 \vert \hat{a}^{\dagger}_1\hat{a}_{-1} \vert \psi^d_0 \rangle \sim e^{-2i\theta_s}$) can be tuned without impacting the phase coherence with respect to $m_F = 0$ (defined by the correlation $\langle \psi^d_0 \vert \hat{a}^{\dagger}_0\hat{a}^{\dagger}_0\hat{a}_1\hat{a}_{-1} \vert \psi^d_0 \rangle$ and related to the spinor phase \cite{Zhang_2019,Linnemann_2016}). The dynamics of both states will prove to be qualitatively similar so, for simplicity we will frequently focus our discussion on the case of a single seeded mode. 

The dynamics generated by $\hat{H}$ for the initial states (\ref{eqn:Psi0}) and (\ref{eqn:Psi0_double}) can be understood very simply in the limit where the quadratic Zeeman shift, $-q(\hat{n}_1 + \hat{n}_{-1})$, is tuned to approximately cancel the initial mean-field energy shift of the $m_F = \pm 1$ generated by the large $m_F = 0$ population, $\hat{n}_0 \left( \hat{n}_1 + \hat{n}_{-1} \right) \approx \hbar g(N-n_s)\left( \hat{n}_1 + \hat{n}_{-1} \right)$. Setting $q = g(N-n_s)$ eliminates the Zeeman and elastic contributions from the Hamiltonian to a first approximation and $\hat{H} \approx \hat{H}_{\mathrm{inel}}$. The initial dynamics is thus dominated by the resonant conversion of atoms from $m_F = 0$ to $m_F = \pm1$ pairs, in a process analogous to four-wave mixing or parametric down-conversion in quantum optics. These spin-changing collisions generate strong correlations and entanglement between the $m_F = \pm1$ modes, including squeezed fluctuations of the relative population difference, $\langle (\Delta \hat{N}_-)^2 \rangle < \langle \hat{N}_+ \rangle$ for $\hat{N}_{\pm} = \hat{n}_1 \pm \hat{n}_{-1}$ \cite{lewis2013epr,Linnemann_2016}, which, as we discuss in Sec.~\ref{sec:Analytics}, can be exploited for quantum-enhanced metrology.

\begin{figure}[!]
    \centering
    \includegraphics[width=8.2cm]{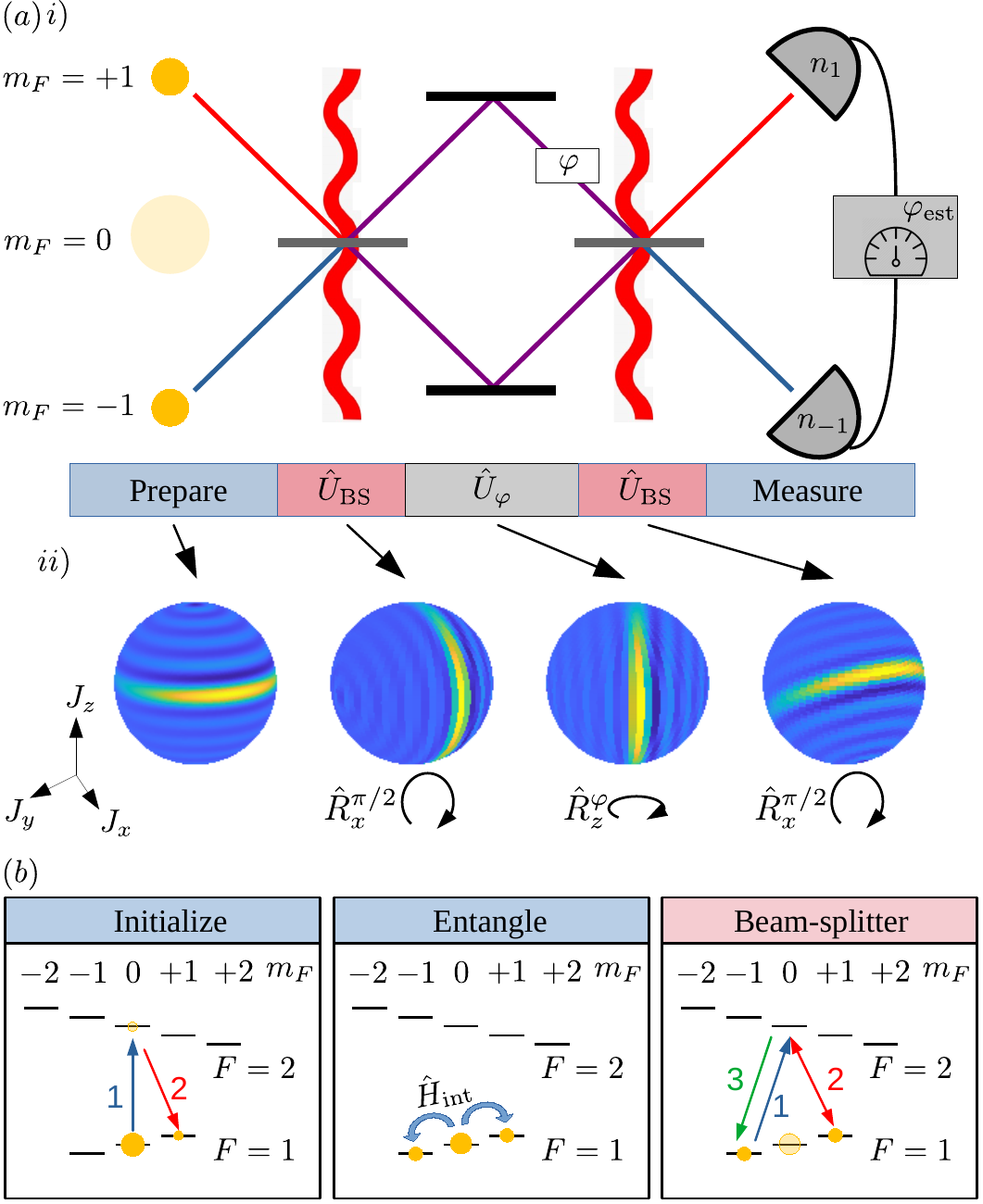}
    \caption{(a) Interferometric sequence. i) Illustration of a Mach-Zehnder scheme, where entangled pairs of $m_F = \pm1$ atoms are the input state of the upper and lower paths. Beam-splitters are realized by coherently mixing the $m_F=\pm1$ modes [see (b)] and a relative phase-shift $\varphi$ is imprinted. Accruing a phase-shift due to gravity requires combining state-dependent momentum kicks imparted by laser pulses (red curves) with the internal-state beam-splitters. An estimate of $\varphi$ is obtained by measuring populations $\hat{n}_{\pm1}$ at the outputs. ii) In an equivalent Ramsey sequence the beam-splitter and phase shift operations correspond to rotations (axes indicated) of the quantum state on a collective Bloch sphere, illustrated using the Wigner phase-space distribution of an example squeezed state (see Sec.~\ref{sec:Analytics} for details). (b) Example internal state dynamics for $F=1$ sodium spinor BEC. (Initialize) Coherent seeding is implemented via microwave pulses resonantly tuned to an ancillary $F=2$ manifold that transfer a fraction of the total population to the $m_F = \pm1$ modes. Pulse 1 transfers $n_s$ atoms from $(F,m_F) = (1,0) \to (2,0)$ and pulse 2 completes the transfer to $(1,1)$ [split equally to $(1,-1)$ for dual seeding]. (Entangle) Subsequent spin-changing collisions between $m_F = 0$ atoms produce entangled pairs in $m_F = \pm1$ to realize $\vert \psi^{s,d}_t \rangle$. (Beam-splitter) Coherent mixing of the $m_F = \pm1$ modes is also implemented via resonant microwaves. Pulse 1 transfers the entire population from $(1,-1) \to (2,0)$ before pulse 2 implements coherent $50-50$ mixing of $(2,0)\leftrightarrow (1,1)$. Finally, pulse 3 returns the remaining population from $(2,0)\to(1,-1)$.
    }
    \label{fig:fig1}
\end{figure}

\subsection{SU(2) atom interferometer}
In this manuscript we focus our investigation on the performance of states dynamically generated by spin-changing collisions in an SU(2) atom interferometer. A paradigmatic scheme is the atomic Mach-Zehnder (MZ) interferometer, an example of which is illustrated in Fig.~\ref{fig:fig1}. The MZ interferometer only actively involves the $m_F = \pm1$ modes and the $m_F = 0$ mode serves only to mediate the preparation of the input state $\vert \psi_t \rangle = e^{-i\hat{H}t} \vert \psi^{s,d}_0 \rangle$ for the interferometer. 

The atomic MZ sequence is composed of three key steps: i) an internal state beam-splitter that coherently mixes the $m_F = \pm1$ modes, ii) the accrual of a relative phase $\varphi$ between the $m_F = \pm1$ modes, and iii) a second beam-splitter to mix the $m_F = \pm1$ modes. The entire sequence can be equivalently described by the unitary $\hat{U}_{\mathrm{MZ}} \equiv \hat{U}_{\mathrm{BS}}(\pi/2) \hat{U}_{\varphi} \hat{U}_{\mathrm{BS}}(\pi/2)$ such that $\vert \psi_f \rangle = \hat{U}_{\mathrm{MZ}} \vert \psi_t \rangle$ is the final state at the output of the MZ interferometer. The beam-splitter operation between the $m_F=\pm1$ modes, $\hat{U}_{\mathrm{BS}}(\phi) = e^{i\phi(\hat{a}_1^{\dagger}\hat{a}_{-1} + \hat{a}_{-1}^{\dagger}\hat{a}_{1})/2}$ with $\phi = \pi/2$ corresponding to a balanced $50-50$ beam-splitter, can be realized by a series of resonant microwave pulses that couple internal states in different $F$ manifolds \cite{Lucke_2011}. Such a sequence can similarly be used for initial preparation of $\vert \psi^{s,d}_0 \rangle$ (see Fig.~\ref{fig:fig1}). The relative phase shift, $\hat{U}_{\varphi} = e^{-i\varphi(\hat{n}_{1} - \hat{n}_{-1})/2}$, can be generated by a number of different sources including the linear Zeeman shift or gravitational acceleration. The latter case requires additional state-selective momentum kicks after (before) step i) [iii)] of the MZ sequence to map the entanglement and correlations between internal (spin) to external (motional) degrees of freedom \cite{klempt2020spinor}. A subsequent free propagation time $T$ between the beam-splitters leads to the accrual of a phase shift $\varphi \sim \mathbf{g}\cdot\mathbf{k} T^2$ \cite{kasevich_1992}, where $\mathbf{g}$ characterizes the local gravitational acceleration and $\mathbf{k}$ the momentum kick.  

Alternatively, the MZ interferometer can be understood as analogous to a Ramsey interferometer for collective spin states [Fig.~\ref{fig:fig1}(a)]. Considering only the $m_F = \pm1$ modes that participate in the interferometer, one can map the bosonic problem to an equivalent collective spin picture using a Schwinger boson mapping,
\begin{eqnarray}
    \hat{J}_x = \frac{1}{2}(\hat{a}^{\dagger}_1\hat{a}_{-1} + \hat{a}^{\dagger}_{1}\hat{a}_{-1}) , \label{eqn:JxDefn} \\
    \hat{J}_y = \frac{1}{2i}(\hat{a}^{\dagger}_1\hat{a}_{-1} - \hat{a}^{\dagger}_{-1}\hat{a}_{1}) , \label{eqn:JyDefn} \\
    \hat{J}_z = \frac{1}{2}(\hat{a}^{\dagger}_1\hat{a}_1 - \hat{a}^{\dagger}_{-1}\hat{a}_{-1}) . \label{eqn:JzDefn} 
\end{eqnarray}
with accompanying raising and lowering operators $\hat{J}_{\pm} = \hat{J}_x \pm i\hat{J}_y$. In this picture the beam-splitters of the MZ interferometer correspond to the pair of $\pi/2$ rotations about $\hat{J}_x$ employed in a Ramsey sequence for an ensemble of spin-$1/2$ particles, e.g., $\hat{U}_{\mathrm{BS}} = e^{-i\pi\hat{J}_x/2}$  (equally, the rotations can be about $J_y$ depending on the phase convention chosen for the original beam-splitter operation), while the phase shift corresponds to a rotation about $\hat{J}_z$, $\hat{U}_{\varphi} \equiv e^{-i\varphi\hat{J}_z}$. This picture proves particularly useful as the quantum noise of the two-mode ($m_F = \pm1$) bosonic system can be readily visualized by plotting the Husimi or Wigner SU(2) phase-space distributions on a collective Bloch sphere, which enables a simple understanding of metrological performance of $\vert \psi_t \rangle$ for a Ramsey sequence in terms of, e.g., spin-squeezing. We defer a full discussion of this until Sec.~\ref{sec:Analytics}.

At the end of the interferometer the phase-shift $\varphi$ is estimated by measuring some signal $\hat{M}(\hat{n}_1,\hat{n}_{-1})$ that is a function of the populations $\hat{n}_{\pm1}$ of the internal states~\footnote{This also encompasses the measurement of e.g., coherences between the different $m_F$ levels by combining population measurements with linear operations that couple the internal states.}. The sensitivity to the phase-shift is characterized by the uncertainty $\Delta\varphi$ in the estimate of $\varphi$ due to quantum projection noise and can be computed by,
\begin{equation}
    (\Delta\varphi)^2 = \frac{\langle (\Delta\hat{M})^2 \rangle}{\vert \partial_{\varphi} \langle \hat{M} \rangle \vert^2} . \label{eqn:MSens}
\end{equation}
This sensitivity is minimized by choosing an optimal signal $\hat{M}$ and is fundamentally limited by the quantum Cramer-Rao bound, $(\Delta\varphi)^2 \geq 1/F_Q$ where $F_Q$ is the quantum Fisher information (QFI). In this manuscript we restrict our discussion to pure states, for which the QFI can be computed as the variance of the generator of the phase-shift \cite{Braunstein1994},
\begin{equation}
        F_Q = \langle (\Delta \hat{N}_-)^2 \rangle_{\mathrm{BS}} , \label{eqn:FQ_defn}
\end{equation}
where the subscript $\langle ... \rangle_{\mathrm{BS}}$ indicates that the expectation value is computed with respect to the quantum state \emph{after} the application of the first beam-splitter in the full MZ sequence. For an uncorrelated input state of $N_+ = \langle \hat{n}_1 \rangle + \langle \hat{n}_{-1} \rangle$ total atoms in the $m_F = \pm 1$ modes, the QCRB collapses to the standard quantum limit (SQL) $(\Delta\varphi)^2 \geq 1/N_+$ whereas we will show in the following discussion that when correlations and entanglement are allowed between the modes the QCRB leads instead to the Heisenberg limit (HL) of $(\Delta\varphi)^2 \geq 1/[N_+(N_+ + 2)]$ \cite{Pezze_2015}. Note that here we define the SQL and HL with respect to the occupation of only the $m_F = \pm1$ modes and not the complete system including the (passive) $m_F = 0$ mode. We will discuss comparisons to the SQL with respect to total atom number, $(\Delta\varphi)^2 \geq 1/N$, in Sec.~\ref{sec:RealCalc}.

\section{Dynamics and quantum Fisher information in the undepleted pump regime \label{sec:Analytics}}
In this section we discuss an analytic treatment of the entangling dynamics that is valid in the limit of large total particle number and suitably short interaction times. The tractability of the system in this limit enables us to derive insightful analytic expressions for the quantum Fisher information, and we are also able to use the SU(2) representation of the generated two-mode quantum state to better understand the metrological performance as a function of initial condition. 

We begin by making the simplifying assumption that the quadratic Zeeman shift is tuned to cancel the initial energy shift provided by $\hat{H}_{\mathrm{el}}$, i.e., $q = g(N - n_s)$. At short times the dynamics of the system is then dominated by resonant spin-changing collisions, e.g., $\hat{H} \approx \hat{H}_{\mathrm{inel}} = g(\hat{a}^{\dagger}_0\hat{a}^{\dagger}_0\hat{a}_1\hat{a}_{-1} + h.c.)$. If the $m_F = 0$ mode is macroscopically occupied, $N \gg n_s, 1$, we can invoke an undepleted pump approximation wherein we replace $\hat{a}_0, \hat{a}^{\dagger}_0 \to \sqrt{N}$ throughout $\hat{H}$. Together, these assumptions yield a quadratic effective Hamiltonian \cite{lewis2013epr}, 
\begin{equation}
    \hat{H}_{\mathrm{UP}} = gN\left( \hat{a}_1\hat{a}_{-1} + \hat{a}^{\dagger}_1\hat{a}^{\dagger}_{-1} \right) . \label{eqn:HUP}
\end{equation}
The undepleted pump approximation (and also the assumption that the Zeeman shift precisely cancels contributions from elastic interactions) is typically valid in the limit where the total number of atoms scattered into the $m_F = \pm1$ modes does not exceed $\sim 10$\% of the initial population of the $m_F = 0$ mode. Beyond this regime, the full form of $\hat{H}$ should be considered as the quantum nature of the $m_F = 0$ mode and processes described by $\hat{H}_{\mathrm{el}}$ and $\hat{H}_Z$ become relevant. 

The dynamics according to the simplified Hamiltonian (\ref{eqn:HUP}) can be exactly solved in either the Schr\"{o}dinger \cite{Caves_SqueezedCoherent_1991} or Heisenberg \cite{Nolan_2016,lewis2013epr} pictures using standard methods. We leave the details of such calculations to Appendix \ref{app:UPA} and simply present the key results here. First, for both seeded initial states the spin-changing collisions generate an exponential growth of the population of the $m_F = \pm1$ modes, which is identically given by
\begin{equation}
\begin{gathered}
    N_+ \equiv \langle \hat{n}_1(\tau) + \hat{n}_{-1}(\tau) \rangle = n_s + \bar{n}, \\
    \bar{n} = 2(n_s+1)\mathrm{sinh}^2(\tau) , \label{eqn:Np}
\end{gathered}
\end{equation}
where $\tau = gNt$ is the rescaled duration of spin-changing collisions. In our expression for $N_+$ we have adopted notation to emphasize the distinction between: i) the uncorrelated or classical population $n_s$ initially transferred to the $m_F = 1$ state to act as a coherent seed, and ii) the $\bar{n}$ atoms scattered into the $m_F = \pm 1$ modes by spin-changing collisions. Despite this separation the latter implicitly depends on $n_s$ as the coherent seed accelerates the pair production process through bosonic stimulation. For completeness, the population difference $N_- \equiv \langle \hat{n}_{1} \rangle - \langle \hat{n}_{-1} \rangle = n_s$ is trivially conserved during the collision dynamics. 

The fluctuations in both the population difference, $\langle (\Delta\hat{N}_{-})^2 \rangle$, and population sum, $\langle (\Delta\hat{N}_{+})^2 \rangle$, depend crucially on the introduction of the seed. For both initial states the dynamics identically preserves the initial fluctuations in the difference, 
\begin{equation}
    \langle (\Delta\hat{N}_{-})^2 \rangle = n_s , \label{eqn:NmFluct}
\end{equation}
which can be said to be suppressed, $\langle (\Delta\hat{N}_{-})^2 \rangle \ll N_+$, when $\bar{n} \gg n_s$. On the other hand, the fluctuations in the total population rapidly grow and for both initial states we find,
\begin{equation}
    \langle (\Delta\hat{N}_+)^2 \rangle  = n_s +  \frac{\bar{n}(1+2n_s)(\bar{n}+2+2n_s)}{(1+n_s)^2}. \label{eqn:NpFluct}
\end{equation}
Initially, or for $\bar{n} \ll n_s$, these fluctuations are Poissonian, $\langle (\Delta\hat{N}_+)^2 \rangle \sim n_s$, reflecting the initial seed population. As more atoms are scattered, such that $\bar{n} \gg n_s$, the right hand term of Eq.~(\ref{eqn:NpFluct}) dominates and the fluctuations become super-Poissonian, $\langle (\Delta\hat{N}_+)^2 \rangle \sim \bar{n}^2$.

Beyond these correlations, an illustrative understanding of the metrological utility of $\vert \psi^{s,d}_t \rangle$ as the input to the MZ interferometer can be provided by the collective spin basis. For the case of a single seed, the solution of the time-evolved bosonic state in the Schr\"{o}dinger picture can be expressed as \cite{Caves_SqueezedCoherent_1991}, 
\begin{equation}
    \vert \psi^s_{\tau} \rangle_{\mathrm{UP}} = \sum_{J = 0}^{\infty} \sum_{m_z = -J}^{J} c^s_{J,m_z}(\tau) \vert J, m_z \rangle , \label{eqn:psit_SU2}
\end{equation}
with expansion coefficients,
\begin{multline}
    c^s_{J,m_z}(\tau) = n_s^{m_z} e^{-n_s/2}e^{2i[(J-m_z)\frac{\pi}{4} + m_z\theta_s]} \sqrt{\frac{(J-m_z)!}{(J+m_z)!}}  \\
    \times \frac{\mathrm{sech}^{1+2m_z}(\tau) \left[ - \mathrm{tanh}(\tau) \right]^{J-m_z}}{(2m_z)!} . \label{eqn:cJmz}
\end{multline}
for $m_z \geq 0$ and $c^s_{J,m_z}(\tau) = 0$ otherwise. The state is written in the collective spin basis defined by $\hat{J}_z \vert J, m_z \rangle = m_z \vert J, m_z \rangle$ and $\hat{J}^2\vert J, m_z \rangle = J(J+1)\vert J, m_z \rangle$ with $\hat{J}^2 = \sum_{n = x,y,z} \hat{J}_n^2$. We point out that in this form the quantum label for total collective spin relates to the total occupation, $J \leftrightarrow (n_1 + n_{-1})/2$, while the spin-projection corresponds to the occupation difference, $m_z \leftrightarrow (n_1 - n_{-1})/2$. The results and discussion in the following are qualitatively analogous for the case of dual seeding.

\begin{figure}[!]
    \centering
    \includegraphics[width=8cm]{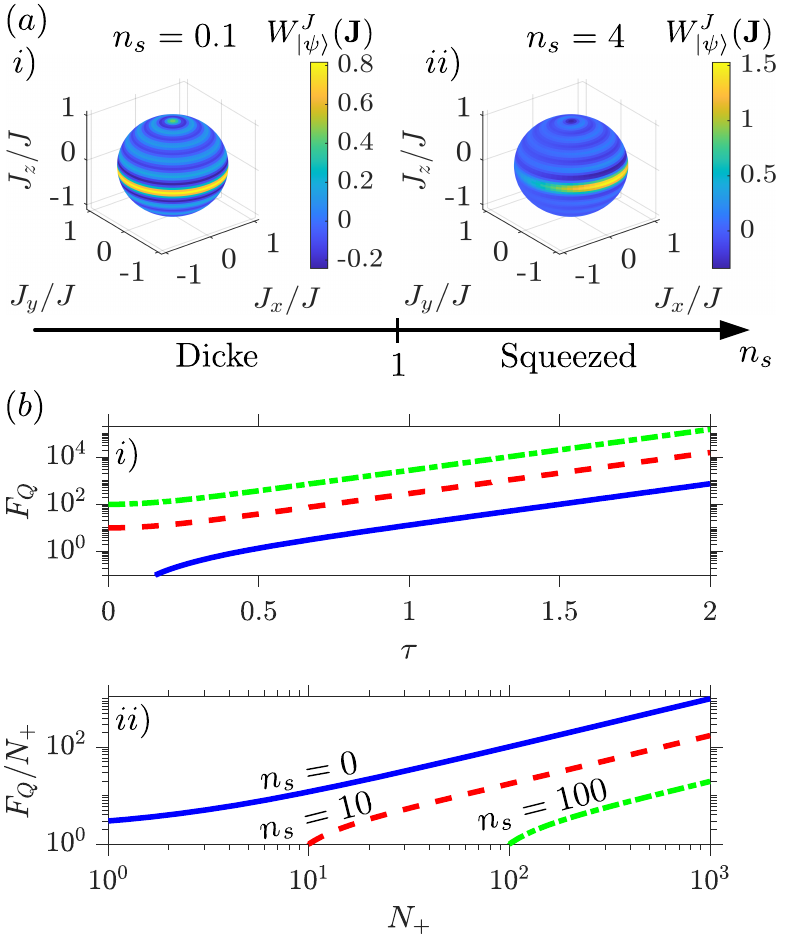}
    \caption{(a) SU($2$) Wigner distributions $W^{J}_{\vert \psi \rangle}(\mathbf{J})$ for $\vert \psi^s_t \rangle$ projected into the $J=10$ sector for: i) $n_s = 0.1$ and ii) $n_s = 4$ and fixed $N_+ = 10^2$. The distributions are renormalized to account for the projection onto the $J=10$ sector. (b) QFI, $F_Q$, as a function of: i) collision duration $\tau$ and ii) total population $N_+$ in $m_F = \pm 1$ modes. In panel ii) we plot the normalized QFI per particle, where $F_Q/N_+ > 1$ indicates sub-SQL sensitivity when only the $m_F = \pm1$ populations are considered. Predictions are from the undepleted pump approximation, Eq.~(\ref{eqn:QFI}), with initial seed ocupation: $n_s = 0$ (blue solid lines), $n_s = 10$ (red dashed lines) and $n_s = 100$ (green dot-dashed lines).
    }
    \label{fig:fig2}
\end{figure}

The state $\vert \psi^s_{\tau} \rangle_{\mathrm{UP}}$ is visualized by plotting the corresponding SU(2) Wigner quasiprobability distribution, $W_{\vert \psi \rangle}(\mathbf{J})$, on a collective Bloch sphere \cite{Dowling_1994,Koczor_2020}. We illustrate a pair of examples in Fig.~\ref{fig:fig2}(a) for $n_s = 0.1$ and $n_s = 4$ with $\bar{n}$ chosen such that $N_+ = \bar{n} + n_s = 100$ is fixed. For clarity, we only plot the Wigner function $W^{J}_{\vert \psi \rangle}(\mathbf{J})$ corresponding to the projection of the state $\vert \psi_t \rangle_{\mathrm{UP}}$ into a fixed $J$ subspace. This is because, in general, the large fluctuations of the total $m_F = \pm1$ population [Eq.~(\ref{eqn:NpFluct})] dictate that the quantum state spans a range of $J$ sectors and thus cannot be illustrated with a single sphere of fixed radius $J$. The Wigner distributions of the two examples showcase how the generated state $\vert \psi^s_t \rangle_{\mathrm{UP}}$ can be split into a pair of dominant cases depending the nature of the quantum fluctuations in the initial condition: i) a Dicke regime \cite{Holland_1993} for $n_s \lesssim 1$ and ii) a spin-squeezed state \cite{Kitagawa1993} regime for $n_s \gtrsim 1$.

The Dicke regime is understood by taking the extreme limit of $n_s = 0$, which corresponds to the previously studied case of two-mode squeezed vacuum \cite{Lucke_2011,Lucke2014}. The lack of fluctuations in $\hat{J}_z \propto \hat{N}_-$ means that the state $\vert \psi_t \rangle_{\mathrm{UP}}$ corresponds to a superposition of $m_z = 0$ Dicke states spanning multiple total spin sectors with integer $J = 0, 1, 2, ...$, and the Wigner function for a typical $J$ is dominated by a narrow ring of width $\Delta J_z \sim 1$ about the equator [see panel (i) of Fig.~\ref{fig:fig2}]. The radial symmetry of the distribution reflects that no well defined phase coherence is established between the $m_F = \pm1$ modes by the spin-changing collisions or the initial vacuum noise that triggers them, e.g., $\langle \hat{J}_+ \rangle \equiv \langle \hat{a}^{\dagger}_1\hat{a}_{-1} \rangle = 0$.

For $n_s \gtrsim 1$ the generated state changes qualitatively to a spin squeezed state \cite{Kitagawa1993,Nolan_2016}. Rigorously, a spin squeezed state satisfies $\xi^2 = N_+\langle (\Delta \hat{J}_z)^2 \rangle/\vert \langle \hat{J}_+ \rangle\vert^2 < 1$ where $\xi^2$ is the Wineland squeezing parameter \cite{Wineland1994a}. For $\vert \psi^s_t \rangle_{\mathrm{UP}}$ it is straightforward to compute, 
\begin{equation}
    \xi^2 = \frac{(1+n_s)^2}{n_s\bar{n}(2+2n_s + \bar{n})}
\end{equation}
which is less than one for $\bar{n},n_s \gg 1$. We understand the squeezing, in contrast to the Dicke regime, by noting that introducing a coherent seed generates a well-defined phase coherence between the $m_F = \pm1$ modes,
\begin{equation}
    \langle \hat{J}_+ \rangle = -i\frac{e^{-2i\theta_s}n_s}{2+2n_s}\sqrt{\bar{n}(2+2n_s + \bar{n})} .
\end{equation}
This means that the Wigner distribution is polarized along a specific direction in the $J_x-J_y$ plane [see panel (ii) of Fig.~\ref{fig:fig2}] but can still remain relatively narrow, $\Delta J_z \sim \sqrt{n_s}$, such that $\Delta J_z/\vert \langle \hat{J}_+ \rangle \vert \lesssim 1/\sqrt{N_+}$ and the state is squeezed.

In both cases, the Wigner distributions plotted in Fig.~\ref{fig:fig2} indicate that the quantum states feature reduced projection noise in the amplitude quadrature ($J_z$) and are thus suitable for distinguishing rotations (phase shifts) in a Ramsey (MZ) sequence. We can make this statement precise by computing the quantum Fisher information \ref{eqn:FQ_defn}, $F_Q = \langle (\Delta \hat{N}_-)^2 \rangle_{\mathrm{BS}} \equiv 4\langle (\Delta\hat{J}_z)^2 \rangle_{\mathrm{BS}}$,
\begin{eqnarray}
    F_Q & =& \frac{1+2n_s}{2}\mathrm{cosh}(4\tau) - \frac{1}{2} , \notag \\
    & = & n_s + \frac{\bar{n}(1+2n_s)(\bar{n}+2+2n_s)}{(1+n_s)^2} . \label{eqn:QFI}
\end{eqnarray}
An identical expression for the QFI is obtained for the case of dual seeds. The independence of Eq.~(\ref{eqn:QFI}) with respect to $\theta_s$ might be surprising given the simplistic collective spin interpretation we have presented so far. For the Dicke regime, $n_s \lesssim 1$, it is trivial that the QFI does not depend on $\theta_s$ as the Wigner distribution becomes an increasingly symmetric ring about the equator as $n_s \to 0$ [Fig.~\ref{fig:fig2}(a)i)]. On the other hand, for the squeezed regime [Fig.~\ref{fig:fig2}(a)ii)], $n_s \gtrsim 1$, one could expect that $\theta_s$ must be chosen to align the orientation of the collective spin (defined by $\langle \hat{J}_+ \rangle$) with the axis of rotation corresponding to the first beam-splitter, such that the squeezed projection noise optimally matches the subsequent rotation about $\hat{J}_z$. However, this intuition is incorrect as it neglects that the state $\vert \psi^s_{\tau} \rangle_{\mathrm{UP}}$ spans multiple $J$ sectors and thus \emph{does not} live on a single Bloch sphere. 

Some further important remarks should be made about the two equivalent formulations of the QFI presented in Eq.~(\ref{eqn:QFI}). First, the second line indicates the QFI always predicts sub-SQL sensitivity, e.g., $1/F_Q \leq 1/N_+$, for any $\bar{n} > 0$. Similarly, the HL is only explicitly saturated when a vacuum state is used as the initial condition, $n_s = 0$, leading to $1/F_Q = 1/[\bar{n}(\bar{n}+2)]$. The latter condition demonstrates that in principle the introduction of any arbitrarily small coherent seed degrades the ideal sensitivity. 

Second, and despite the former observations, we point out that one must carefully offset any apparent loss in relative sensitivity against the accelerated rate at which pairs are produced due to the bosonic stimulation provided by a coherent seed [see Eq.~(\ref{eqn:Np})]. In Fig.~\ref{fig:fig2}(a), we illustrate that within the undepleted pump regime the QFI with a coherent seed ($n_s \neq 0$) is \emph{always} superior to the unseeded ($n_s = 0$) case as a function of $\tau$. In fact, inspection of the first line of Eq.~(\ref{eqn:QFI}) demonstrates that $F_Q \equiv F_Q\vert_{n_s = 0} + n_s\mathrm{cosh}(4\tau)$. Nevertheless, it is similarly important to recognize that the undepleted pump regime specifically ignores that in real experimental systems there is always a finite total number of particles available from the initial $m_F = 0$ BEC that can be converted into pairs, e.g., $N_+ \leq N$. Thus, the relative QFI per particle, $F_Q/N_+$ shown in Fig.~\ref{fig:fig2}(b), is also an important metric. 

While Dicke and spin-squeezed states are relatively well understood in terms of their broad metrological utility, the analysis of the QFI already makes clear that in the spinor BEC system we must carefully understand how the initial seed $n_s$ tunes us between these regimes. This is not only true in terms of the achievable metrological sensitivity given, e.g., a fixed particle resource $N_+$ or time $\tau$, but also in terms of robustness to sources of technical noise. In the following sections we investigate this more systematically by considering a range of measurement strategies for the MZ interferometer and the impact of technical noise.

\section{Optimal measurements and attainable sensitivity \label{sec:OptimalSensitivity}}
The understanding of $\vert \psi^s_{\tau} \rangle_{\mathrm{UP}}$ provided by the collective spin picture also enables us to readily identify measurements that should allow for an optimal estimate of $\varphi$. Specifically, in the squeezed regime a rotation can be inferred by simply monitoring the change in the spin projection $J_z$ \cite{Nolan_2016}, while in the Dicke regime one needs to track $J_z^2$ due to the symmetry of the Wigner distribution about the Bloch sphere \cite{Lucke_2011}. These observables are readily accessible in a spinor BEC experiment from measurements of the $m_F = \pm1$ occupations.

The ideal sensitivity attainable with either measurement of $\hat{J}_z$ or $\hat{J}_z^2$ is straightforward, albeit sometimes cumbersome, within the undepleted pump approximation. For brevity and simplicity, we only present analytic expressions for the former measurement but show example calculations and analysis for both in Fig.~\ref{fig:fig3}. Further details of the calculations and more extensive expressions can be found in Appendix \ref{app:UPA}.

The mean and variance of $\hat{J}_z$ at the end of the MZ sequence can be computed for either single or dual initial seeds and expressed entirely in terms of the phase-shift $\varphi$, initial seed $n_s$ and scattered population $\bar{n}$. For a single seed we obtain,
\begin{widetext}
\begin{equation}
    \begin{gathered}
        \langle \hat{J}_z(\varphi) \rangle_s =  -\frac{n_s}{2}\cos (\varphi ) + \frac{n_s\cos(2\theta_s)  \sqrt{\bar{n} \left(\bar{n}+2 n_s+2\right)}}{2(n_s+1)}\sin (\varphi ) , \label{eqn:JzJz2Phi} \\
        \langle [\Delta \hat{J}_z(\varphi)]^2 \rangle_s = \frac{(2n_s+1)\bar{n}}{2(n_s + 1)}\sin^2(\varphi) + \frac{(2 n_s+1) \bar{n}^2 }{4(n_s+1)^2}\sin
        ^2(\varphi ) + \frac{n_s \left[1+n_s-\cos (2 \theta_s ) \sin (2 \varphi ) \sqrt{\bar{n} \left(\bar{n}+2 n_s+2\right)}\right]}{4 (n_s+1)} , 
    \end{gathered}
\end{equation}
and for dual seeds (see also Ref.~\cite{Nolan_2016}), 
\begin{equation}
    \begin{gathered}
        \langle \hat{J}_z(\varphi) \rangle_d = -\frac{n_s(1+n_s + \bar{n})\mathrm{sin}(2\theta_s)}{2(1+n_s)} \mathrm{sin}(\varphi) , \label{eqn:JzJz2Phi_d} \\
        \langle [\Delta \hat{J}_z(\varphi)]^2 \rangle_d = \frac{ \bar{n}(1+2n_s)(2+2n_s + \bar{n}) }{4(1+n_s)^2}\sin^2(\varphi) + \frac{n_s\left[ 1 + n_s - \cos(2\theta_s)\sin(2\varphi)\sqrt{\bar{n}\left(\bar{n}+2 n_s+2\right)} \right]}{4(1+n_s)} .
    \end{gathered}
\end{equation}
\end{widetext}
We use the subscript $\langle ... \rangle_{s,d}$ to differentiate expectation values computed for the initial conditions $\vert \psi^{s,d}_0 \rangle$. 

Unlike the prior result for the QFI, the form of $\langle \hat{J}_z(\varphi) \rangle_{s,d}$ confirms the important role played by the seed phase $\theta_s$. For the case of a single seed we observe that for $\theta_s = 0,\pi/2,\pi,...$ the interferometric signal $\langle \hat{J}_z(\varphi) \rangle_s$ is maximally boosted by the spin-changing collisions -- the contrast scales as $\bar{n}$ for $\bar{n} \gg n_s,1$) -- whereas for $\theta_s = \pi/4,3\pi/4,5\pi/4...$ the signal depends only on $n_s$. Similar analysis is true for the case of dual seeds, albeit with the corresponding values of $\theta_s$ interchanged.

\begin{figure}[!]
    \centering
    \includegraphics[width=8cm]{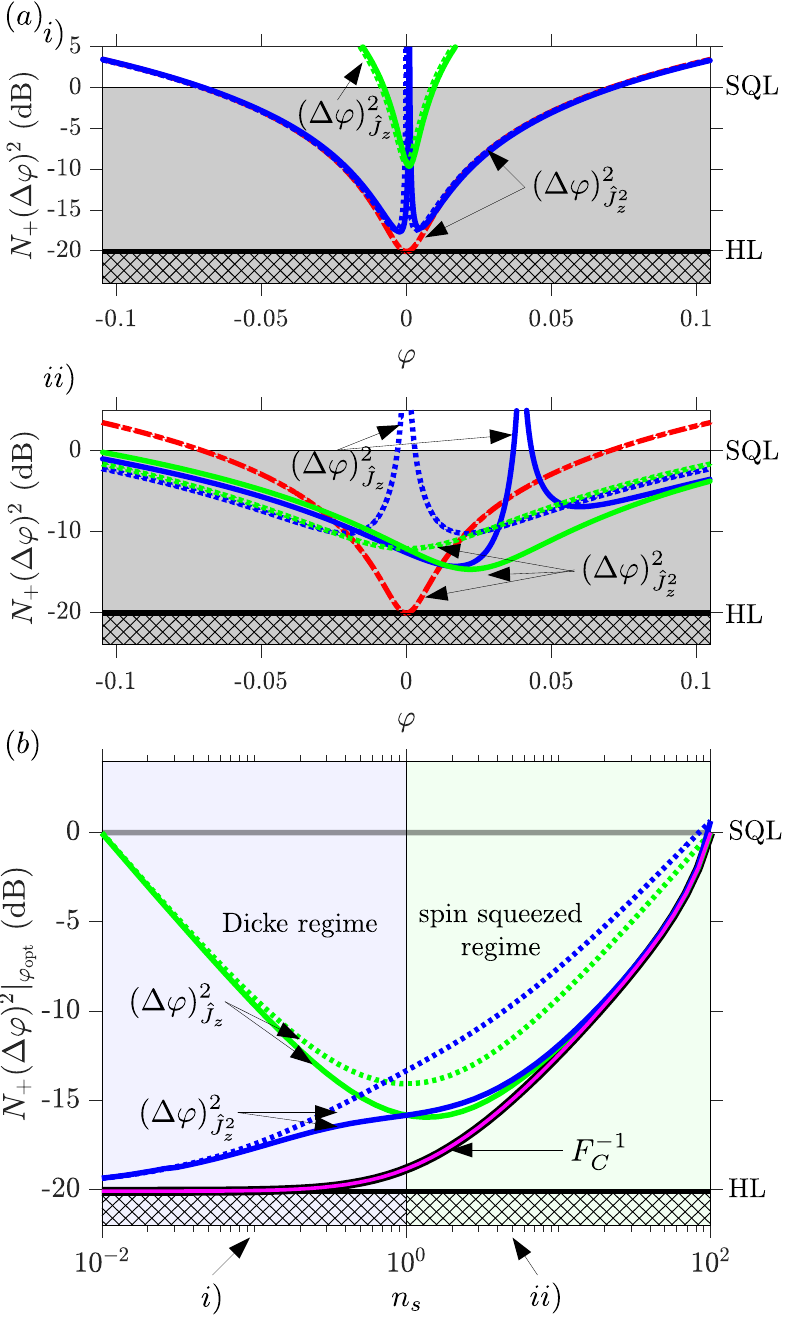}
    \caption{(a) Example interferometric sensitivities as a function of phase-shift $\varphi$ for different initial conditions: i) $n_s = 0.1$ and ii) $n_s = 4$. The sensitivities are computed from measurements of $\hat{J}_z$ (green lines) and $\hat{J}^2_z$ (blue lines) as labelled. Line style indicates single (solid) and dual seed (dotted) initial conditions. We also plot the case of $n_s = 0$ with a $\hat{J}^2_z$ measurement (red dot-dashed line), which saturates the HL, as a reference in both panels. For all data $\theta_s$ is optimally chosen (see main text) and $N_+ = 10^2$ is fixed. (b) Optimal sensitivity as a function of initial seed occupation $n_s$ and $N_+ = 10^2$. We indicate with arrows the results of measurements of $\hat{J}_z$ (green lines) and $\hat{J}_z^2$ (blue lines) against the CFI $F_C^{-1}$ [Eq.~(\ref{eqn:CFI}), narrow magenta line] that also saturates the QCRB [Eq.~(\ref{eqn:QFI}), thick black line underlying CFI]. The line styles indicate single (solid) and dual seed (dotted) initial conditions. For clarity, we also indicate the Heisenberg (HL) and standard quantum limits (SQL) for $N_+= 10^2$ in all panels.
    }
    \label{fig:fig3}
\end{figure}

The attainable sensitivity from a measurement of $\hat{J}_z$ is calculated by directly substituting Eqs.~(\ref{eqn:JzJz2Phi}) and (\ref{eqn:JzJz2Phi_d}) into Eq.~(\ref{eqn:MSens}). We focus on the optimal cases of: i) $\theta_s = 0$ (single seed) and ii) $\theta_s = \pi/4$ (dual seed), corresponding to the phases which maximize the contrast of $\langle \hat{J}_z(\varphi) \rangle_{s,d}$, while results for arbitrary $\theta_s$ can be found in Appendix \ref{app:UPA}. For a single seed we obtain,
\begin{widetext}
\begin{equation}
    (\Delta \varphi)_{\hat{J}_z,s}^2 = \frac{n_s (n_s+1) \left[1 + n_s - \sin (2 \varphi) \sqrt{\bar{n} \left(\bar{n}+2 n_s+2\right)}\right]+(2 n_s+1) \bar{n}^2 \sin
   ^2(\varphi )+2 (n_s+1) (2 n_s+1) \bar{n} \sin ^2(\varphi )}{n_s^2 \left[\cos (\varphi ) \sqrt{\bar{n} \left(\bar{n}+2
   n_s+2\right)}+(n_s+1) \sin (\varphi )\right]^2} , \label{eqn:SensJz_s}
\end{equation}
and for dual seeds we obtain \cite{Nolan_2016},
\begin{equation}
    (\Delta \varphi)_{\hat{J}_z,d}^2 = \frac{(2 n_s+1) \bar{n} \tan ^2(\varphi ) \left(\bar{n}+2 n_s+2\right)+n_s (n_s+1)^2 \sec ^2(\varphi )}{n_s^2 \left(\bar{n}+n_s+1\right)^2} . \label{eqn:SensJz_d}
\end{equation}
\end{widetext}

Although the expressions are lengthy, a few simple statements can be made. First, for $n_s \to 0$ both sensitivities (\ref{eqn:SensJz_s}) and (\ref{eqn:SensJz_d}) diverge for arbitrary $\varphi$. This is consistent with the fact that the amplitude of the interferometric signal rapidly vanishes with decreasing $n_s$, $\langle \hat{J}_z(\varphi) \rangle_{s,d} \propto n_s \to 0$. Conversely, when the coherent seed dominates, $n_s \gg \bar{n}$, both Eqs.~(\ref{eqn:SensJz_s}) and (\ref{eqn:SensJz_d}) limit to the archetypal example of a coherent spin state of $n_s$ atoms (e.g., squeezing parameter $\xi^2 = 1$) input to a Ramsey interferometer: $(\Delta\varphi)^2_{\hat{J}_z,s} \simeq [\sin^2(\varphi)n_s]^{-1}$ and $(\Delta\varphi)^2_{\hat{J}_z,d} \simeq [\cos^2(\varphi)n_s]^{-1}$. These expressions have a minimum at the optimal points $\varphi_{\mathrm{opt}} = \pi/2$ and $\varphi_{\mathrm{opt}} = 0$, respectively, where the sensitivity reaches the associated SQL, $(\Delta\varphi)^2_{\hat{J}_z,s,d} \simeq 1/n_s$.

The sensitivity for intermediate values of $\bar{n}$ and $n_s$ is more complex, but we show representative examples in Fig.~\ref{fig:fig3}(a) for $n_s = 0.1$ [panel i)] and $n_s = 4$ [panel ii)] with $N_+ = n_s + \bar{n} = 100$. For dual seeding we always observe a minimum sensitivity at $\varphi = 0$, while in the case of a single seed the location of the minimum shifts with $n_s$ but remains close to $\varphi = 0$. Moreover, the achievable sensitivity appears to be superior for a single seed in the squeezed regime, $n_s = 4$. 

We make these representative observations more rigorous by computing the optimal (minimum) sensitivity in the limit $\bar{n} \gg n_s, 1$. In this case the expressions (\ref{eqn:SensJz_s}) and (\ref{eqn:SensJz_d}) can be expanded in powers of $\varphi$ about the point $\varphi = 0$ (Appendix \ref{app:UPA}), and we obtain:  
\begin{equation}
    (\Delta \varphi)_{\hat{J}_z,s}^2\big\vert_{\varphi_{\mathrm{opt}}} = \frac{(1+n_s)^3}{n_s(1+2n_s)} \frac{1}{\bar{n}^2} , \label{eqn:JzSensOpt}
\end{equation}
and
\begin{equation}
    (\Delta \varphi)_{\hat{J}_z,d}^2\big\vert_{\varphi_{\mathrm{opt}}} = \frac{(1+n_s)^2}{n_s(1+n_s +\bar{n})^2} , \label{eqn:JzSensOpt_d}
\end{equation}
which occur at $\varphi_{\mathrm{opt}} = [n_s(1+n_s)]/[\bar{n}(1+2n_s)]$ and $\varphi_{\mathrm{opt}} = 0$, respectively. 

The results (\ref{eqn:JzSensOpt}) and (\ref{eqn:JzSensOpt_d}) indicate optimal choices of initial seed population, $n^{\mathrm{opt}}_s \simeq (1+\sqrt{3})/2$ for a single seed and $n^{\mathrm{opt}}_s \simeq 1$ for dual seeds (for $\bar{n} \gg n_s, 1$), that minimize the sensitivity: 
\begin{equation}
    \begin{gathered}
    \mathrm{min}_{n_s}\left[ (\Delta \varphi)_{\hat{J}_z,s}^2\big\vert_{\varphi_{\mathrm{opt}}} \right] \simeq \frac{3\sqrt{3}}{2\bar{n}^2} , \label{eqn:IdealSensJzOpt} \\
    \mathrm{min}_{n_s}\left[ (\Delta \varphi)_{\hat{J}_z,d}^2\big\vert_{\varphi_{\mathrm{opt}}} \right] \simeq \frac{4}{\bar{n}^2}.
    \end{gathered}
\end{equation}
As previously observed in Ref.~\cite{Nolan_2016} for the latter case, this demonstrates that a measurement of $\hat{J}_z$ can in principle lead to a sensitivity that is within a $\mathcal{O}(1)$ prefactor of the HL, $\sim 1/\bar{n}^2$, for a fixed $n_s$ without any fine tuning as the total population $N_+$ is varied. 

The optimal value of $n_s \sim 1$ is equivalent to the definitional separation between the regime of squeezed states, identifiable by the Wineland squeezing parameter $\xi^2 < 1$, as opposed to Dicke-like oversqueezed states where the Wigner distribution begins to wrap around the Bloch sphere and $\langle \hat{J}_z(\varphi) \rangle \to 0$ due to symmetry. Moreover, the insensitivity of the optimal $n_s$ to total particle number $N_+ = \bar{n} + n_s$ arises because the state $\vert \psi^s_t \rangle_{\mathrm{UP}}$ can be crudely approximated by considering a representative Wigner distribution on a single Bloch sphere of radius $\bar{J} = \sqrt{\langle \hat{J}^2 \rangle} \sim N_+/2$. For a minimum uncertainty state we will have that $\Delta J_{\perp} \Delta J_z \sim N_+$ where $\Delta J_{\perp}$ is the rms width of the state in the $J_x-J_y$ equatorial plane. Substituting $\Delta J_z \sim \sqrt{n_s}$ we rearrange to obtain $n_s \sim N^2_+/(\Delta J_{\perp})^2$, for which a maximally squeezed state, $\Delta J_{\perp} \sim \bar{J}$, yields $n_s \sim 1$ independent of $N_+$.  

The sensitivity obtained with a measurement of $\hat{J}_z^2$ can also be obtained for arbitrary $n_s$. The resulting expressions are lengthy and not insightful so we refer the interested reader to Appendix \ref{app:UPA}. However, it is useful to reproduce the well understood limiting case of $n_s = 0$ as a reference \cite{Lucke_2011,Zhang_2013}, 
\begin{equation}
    (\Delta\varphi)^2_{\hat{J}_z^2} = \frac{1 + [2\bar{n}(\bar{n}+2) + 1]\mathrm{tan}^2(\varphi)}{\bar{n}(\bar{n} + 2)} , \label{eqn:SensJz2}
\end{equation}
which has the optimal sensitivity 
\begin{equation}
    (\Delta \varphi)_{\hat{J}^2_z}^2\big\vert_{\varphi_{\mathrm{opt}}} = \frac{1}{\bar{n}(\bar{n}+2)}, 
\end{equation}
at $\varphi_{\mathrm{opt}} = 0$.

In Fig.~\ref{fig:fig3}(a) we plot representative examples of $(\Delta\varphi)^2_{\hat{J}^2_z}$ to compare against the prior expressions for $(\Delta\varphi)^2_{\hat{J}_z}$. As previously discussed, we always choose $\theta_s = 0$ (single-sided seed) and $\theta_s = \pi/4$ (dual seed) to optimize the achievable sensitivity. For the case of $n_s = 0.1$ in the Dicke regime, the sensitivity achievable with $\hat{J}_z^2$ is similar for either seed configuration and is predictably superior to that attainable via measurement of only $\hat{J}_z$. We note that a divergence develops that is located around the idealized $n_s = 0$ working point of $\varphi = 0$ (understood by the fact that for $n_s = 0$ it corresponds to a limit where both $\partial_{\varphi}\langle \hat{J}_z^2(\varphi) \rangle$ and $\langle (\Delta \hat{J}_z^2)^2\rangle$ vanish). In the squeezed regime, $n_s = 4$, we observe that the sensitivity achievable with $\hat{J}_z$ and $\hat{J}_z^2$ is similar and the shift of the optimal working point is approximately the same.

In Fig.~\ref{fig:fig3} we validate Eqs.~(\ref{eqn:JzSensOpt}) and (\ref{eqn:JzSensOpt_d}) by exactly computing the minimum sensitivity, optimized over $\varphi$, for measurements of both $\hat{J}_z$ and $\hat{J}_z^2$. We show results as a function of $n_s$ with fixed $N_+ = 100$, although the behaviour of $(\Delta \varphi)_{\hat{J}_z}^2$ and $(\Delta \varphi)^2_{\hat{J}_z^2}$ are qualitatively unchanged as $N_+$ is increased. The predicted minima of $(\Delta \varphi)^2_{\hat{J}_z}$ for either initial seed configuration are clearly observable near $n_s \sim 1$, in agreement with Eq.~(\ref{eqn:IdealSensJzOpt}). In the squeezed regime, $n_s \gtrsim 1$, we observe that while $(\Delta \varphi)_{\hat{J}_z,s}^2$ and $(\Delta \varphi)^2_{\hat{J}_z^2,s}$ collapse together for the single seed and approach the QCRB, $(\Delta \varphi)_{\hat{J}_z,d}^2$ and $(\Delta \varphi)^2_{\hat{J}_z^2,d}$ are visibly worse (until $n_s$ approaches $N_+$). On the other hand, in the Dicke regime, $n_s \lesssim 1$, the sensitivity attained with $\hat{J}_z^2$ quickly becomes optimal, and the difference between the choices of initial state vanish. Of note is that for trace amounts of seeding $n_s \ll 1$ the sensitivity achievable with $\hat{J}_z^2$ does not saturate the QCRB (in fact it only strictly saturates the bound for $n_s = 0$).

For completeness, we also compute the \emph{classical} Fisher information (CFI) $F_C$, which bounds the attainable sensitivity in the case that one has access to the complete distribution function $P_{n_1,n_{-1}}(\varphi)$ of the populations at the end of the MZ sequence (equivalently, access to all possible moments of $\hat{N}_-$ and $\hat{N}_+$). The CFI is defined as
\begin{equation}
    F_C(\varphi) = \sum_{n_1,n_{-1}} \frac{1}{P_{n_1,n_{-1}}(\varphi)} \left[ \frac{\partial P_{n_1,n_{-1}}(\varphi)}{\partial \varphi} \right]^2 \label{eqn:CFI}
\end{equation}
and is related to the sensitivity through $(\Delta\varphi)^2 = 1/F_C \geq 1/F_Q$. The CFI can be analytically computed in the limit of $n_s=0$ \cite{Lucke_2011,Rowe_2001}, but for generic $n_s \neq 0$ we numerically evaluate $F_C$ by efficiently simulating the full MZ sequence with $\vert \psi^s_{\mathrm{UP}}(\tau) \rangle$ as the input state (see Appendix \ref{app:CFI} for details). The CFI is independent of $\varphi$ and so we only include it in Fig.~\ref{fig:fig3}(b) as a function of $n_s$. We find it saturates the QCRB for all $n_s$. While the CFI is a demanding quantity to extract in an experiment, it serves here to confirm that for $n_s \lesssim 1$ measurements of the $m_F = \pm1$ populations remain an optimal signal, although the phase-shift is encoded in higher-order moments than we consider (e.g., $\hat{J}_z$ and $\hat{J}_z^2$).

\section{Robustness to experimental imperfections \label{sec:DetectionNoise}}
The results of the previous sections indicate that triggering the spin-changing collisions with vacuum noise will generically lead to the optimal generation of metrologically useful entanglement. Adding a coherent seed alters the nature of the state but nevertheless always fundamentally leads to a degradation of the achievable sensitivity per particle. However, even in current state-of-the-art experimental systems this perspective is too simplistic as it discounts a myriad of technical imperfections and limitations. In particular, it is accepted that Dicke states are typically more susceptible to, e.g., detection noise, than spin-squeezed states \cite{davis2017advantages}. In the following discussion we demonstrate that when detection noise is incorporated it becomes favourable to use a coherent seed to generate squeezed states that offer less \emph{ideal} metrological potential but are nevertheless more robust and thus provide a meaningful practical advantage in metrological performance. We give estimates for the optimal seed $n_s$ in this scenario, and also discuss other favourable features of seeded initial states such as an increased dynamic range.  

\subsection{Detection noise}
In ultracold atomic gases imperfect detection limits the ability to precisely count atoms and thus measure, e.g., moments of $\hat{J}_z$. We assume this can be modeled as random noise on population measurements in each shot, e.g., $J_z \to J_z +\zeta_{\mathrm{dn}}$ where $\zeta_{\mathrm{dn}}$ is Gaussian noise with variance $\sigma$ and zero mean. For our theoretical calculations this is equivalent to making the substitution, 
\begin{equation}
    \begin{gathered}
        \langle \hat{J}_z \rangle_{\sigma} = \langle \hat{J}_z \rangle_{\sigma=0} , \\
        \langle (\Delta \hat{J}_z)^2 \rangle_{\sigma} = \langle (\Delta \hat{J}_z)^2 \rangle_{\sigma=0} + \sigma^2 , \\
        \langle (\Delta \hat{J}^2_z)^2 \rangle_{\sigma} = \langle (\Delta \hat{J}^2_z)^2 \rangle_{\sigma=0} + 4\sigma^2\langle \hat{J}_z^2 \rangle_{\sigma=0} + 2\sigma^4 , \label{eqn:sigma_relations}
    \end{gathered}
\end{equation}
where the subscript $\langle .... \rangle_{\sigma}$ indicates the expectation value includes averaging over the detection noise characterized by $\sigma$. 

It is most illuminating to first examine the case where there is no seed, for which a measurement of $\hat{J}_z^2$ is minimally required and a useful analytic expression can be given. The ideal ($\sigma = 0$) working point $\varphi_{\mathrm{opt}} = 0$ corresponds to a case where both the variance $\langle (\Delta \hat{J}^2_z)^2 \rangle$ and slope of the signal $\partial_{\varphi} \langle \hat{J}_z^2(\varphi) \rangle$ vanish. Thus, the introduction of detection noise leads to a divergent sensitivity at $\varphi = 0$ and we instead compute the shifted optimal working point to be $\varphi_{\mathrm{opt},\sigma} \simeq 2\sigma^2/\bar{n}^2$, for which the sensitivity is
\begin{equation}
    (\Delta \varphi)_{\hat{J}^2_z,\sigma}^2\big\vert_{\varphi_{\mathrm{opt}}} \simeq \frac{1 + 12\sigma^2}{\bar{n}(\bar{n}+2)} . \label{eqn:Jz2SensSigma_TMSV}
\end{equation}
The top line of Eq.~(\ref{eqn:Jz2SensSigma_TMSV}) clearly illustrates that to achieve the true HL one must satisfy the highly restrictive requirement $\sigma \ll 1/\sqrt{12}$ or, in simpler language, possess the ability to precisely count the number of atoms in the ensemble at the single particle level. While there has been notable progress in this direction for spinor BECs \cite{Evrard_2020}, this is so far limited to ensembles equivalent to $N_+ \lesssim 10^3$ atoms. 

For the case of seeded initial states, $n_s \neq 0$, it is relatively straightforward to obtain analytic expressions for the sensitivity attainable from both $\hat{J}_z$ and $\hat{J}_z^2$ measurements. However, we again find that only the former expressions have an insightful form. For weak detection noise, $\sigma \ll \sqrt{N_+}$, the working point $\varphi_{\mathrm{opt}}$ is approximately unmoved from the ideal ($\sigma = 0$) scenario regardless of the initial seed configuration, and we obtain the optimal sensitivities,
\begin{equation}
    (\Delta \varphi)^2_{\hat{J}_z,s,\sigma} \vert_{\varphi_{\mathrm{opt}}} \simeq \frac{(1+n_s)^3}{n_s(1+2n_s)} \frac{1}{\bar{n}^2} + \frac{4\sigma^2}{n_s^2}\frac{(1+n_s)^2}{\bar{n}^2} , \label{eqn:SigmaSensJz}
\end{equation}
and 
\begin{equation}
    (\Delta \varphi)^2_{\hat{J}_z,d,\sigma} \vert_{\varphi_{\mathrm{opt}}} \simeq \frac{(1+n_s)^2}{n_s(1+n_s+\bar{n})^2}  + 
    \frac{4\sigma^2}{n_s^2}\frac{(1+n_s)^2}{(1+n_s+\bar{n})^2} , \label{eqn:SigmaSensJz_d}
\end{equation}
for $\bar{n} \gg 1$. Both equations, particularly the latter Eq.~(\ref{eqn:SigmaSensJz_d}), are of a form that suggests choosing a suitable seed, e.g., $n_s \sim \sigma$, might suppress the effects of modest detection noise. 

\begin{figure}[!]
    \centering
    \includegraphics[width=8cm]{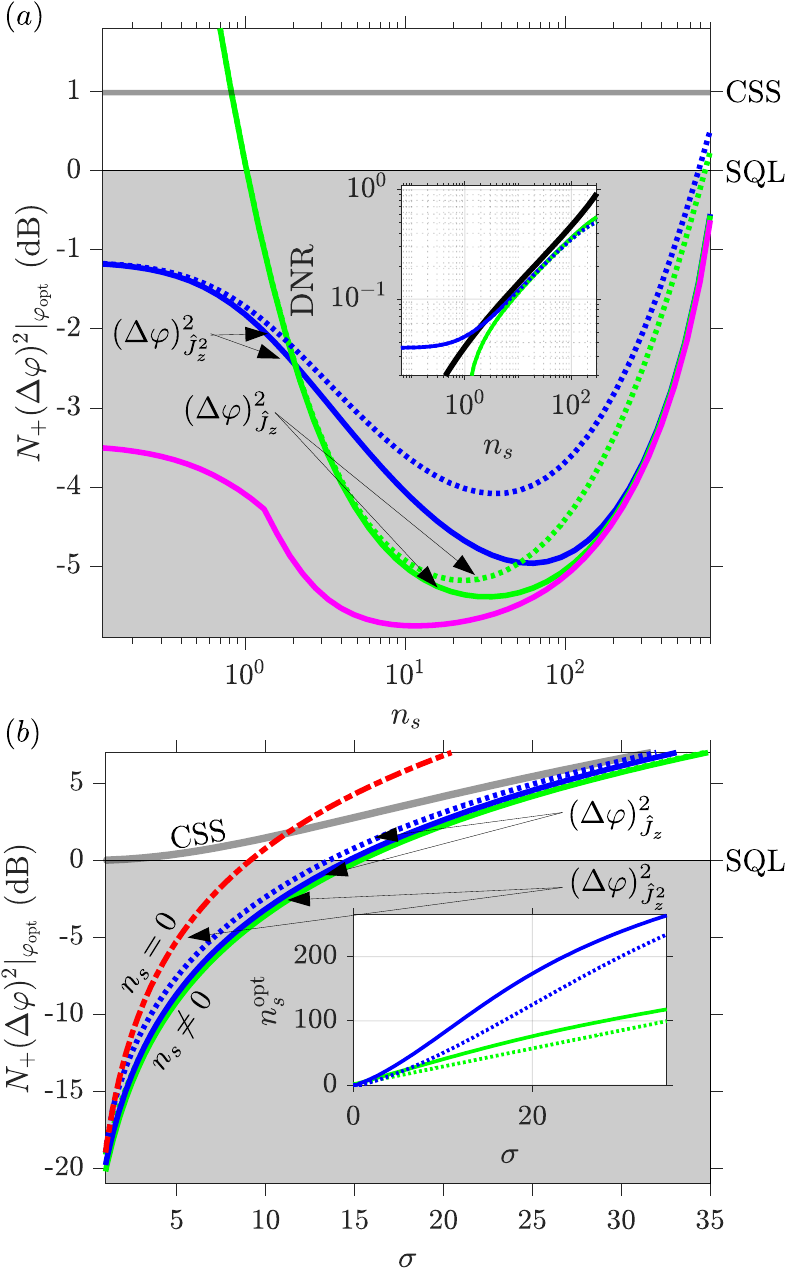}
    \caption{(a) Optimal sensitivity as a function of seed occupation $n_s$ with fixed detection noise $\sigma = 8$. We label results for measurements of $\hat{J}_z$ (green lines) (green solid and dashed lines for single and dual seeds), $\hat{J}^2_z$ (blue lines) and the CFI $F_C^{-1}$ [Eq.~(\ref{eqn:CFI}), magenta line]. Line style indicates single (solid line) and dual (dashed line) seed initial conditions. For reference, we also indicate the sensitivity achievable with a coherent spin state (CSS) including identical detection noise (faded gray line). Inset: Corresponding dynamic range (DNR) depending on seed occupation and measurement signal (same line styles as main panel). Results are indistinguishable for single or dual seed initial conditions. We compare to the ideal ($\sigma = 0$) result (\ref{eqn:JzDNR}) (black line). (b) Best sensitivity (optimized over both $\varphi$ and $n_s$) as a function of detection noise $\sigma$. Line styles are the same as (a) with the additional comparison to the reference case of $n_s = 0$ (red dot-dashed line) and measurement of $\hat{J}_z^2$ (other signals are labelled in plot). Inset: Optimal seed occupation $n^{\mathrm{opt}}_s$ as a function of detection noise. The shaded grey region in (a) and (b) indicates sensitivity below the SQL, $(\Delta\varphi)^2_{\mathrm{SQL}} = 1/N_+$.
    }
    \label{fig:fig4}
\end{figure}

We explore this prediction by plotting the optimal sensitivity as a function of seed occupation $n_s$ for fixed $N_+ = 1000$ and realistic $\sigma = 8$ in Fig.~\ref{fig:fig4}(a). We compare the sensitivity attainable with both $\hat{J}_z$ and $\hat{J}_z^2$ where the results are obtained by numerical optimization of the exact analytic expressions for $(\Delta\varphi)_{\hat{J}_z,\sigma}$ and $(\Delta\varphi)_{\hat{J}^2_z,\sigma}$ in the undepleted pump regime with no approximations. For this case, our calculations clearly demonstrate introducing a coherent seed provides a marked advantage over the unseeded case. In fact, for this value of $\sigma$ we highlight that the sensitivity attainable without seeding is limited to a negligible $\sim 1$~dB below the SQL, whereas for a broad regime around $n_s \approx 4\sigma^2$ we comparatively observe $\sim 5-6$~dB below the SQL. 

We also compute the CFI as a function of $n_s$ for the same parameters, to better probe the distinction between seeded and unseeded states in the presence of detection noise. Detection noise can be included by convolving the true distribution function $P_{n_1,n_{-1}}(\varphi)$ with a Gaussian function of width $\sim \sigma$ (see Appendix \ref{app:CFI} for details). Also note that, unlike our previous calculations for $F_C$, when detection noise is included the CFI depends on the phase-shift $\varphi$ and so we plot the minimum value of $1/F_C(\varphi)$ in panel (a). The optimal sensitivity obtained with the CFI follows the same trend as simpler measurement signals, with a clear improvement in sensitivity when a small seed is included albeit at a slightly smaller value of $n_s \approx \sigma$. However, the improvement between the seeded and unseeded states is comparatively reduced to only $\sim 2$~dB. This latter observation suggests that the improvement we observe in $(\Delta\varphi)_{\hat{J}_z,\sigma}$ and $(\Delta\varphi)_{\hat{J}^2_z,\sigma}$ when a seed is included should not entirely be attributed to an enhanced robustness of the state for arbitrary measurements of populations. Rather, the benefit of seeding is driven by the fact that squeezed states encode the phase rotation $\varphi$ in simple (e.g., low-order) moments of the populations in a very robust way.

In panel (b) we compare the best attainable sensitivity without a seed (using a $\hat{J}_z^2$ measurement) and with coherent seeding (either $\hat{J}_z$ or $\hat{J}_z^2$ is measured) as a function of detection noise $\sigma$. In the latter case we optimize the chosen value of $n_s$ to provide the largest gain (corresponding values are plotted inset). Again, we observe that seeding provides a robust enhancement to detection noise. Comparing to the SQL we find that seeded states can tolerate up to $50$\% larger detection noise ($\sigma \approx 9$ for $n_s = 0$ compared to $\sigma \approx 15$ for $n_s \neq 0$) for $N_+ = 10^3$ while still retaining a pure quantum advantage. A fairer comparison is to define a practical classical limit in terms of the sensitivity achievable with a coherent spin state (which is typically used to define the SQL) and subject to equivalent detection noise. With respect to this standard we observe that unseeded initial states lead to an entanglement-enhanced sensitivity up to $\sigma \approx 11$, whereas seeding retains a quantum advantage up to $\sigma \approx 35$~\footnote{In fact, seeded states are technically always superior according to this metric for $n_s < N_+$. However, this improvement is vanishingly small beyond $\sigma \approx 35$.}, or a $>300$\% improvement in acceptable detection noise. 

The inset of panel (b) shows the optimal seed population $n_s$. For the $\hat{J}_z$ measurement we approximately find that $n_s \sim 4\sigma$ (single seed) and $n_s \sim 3\sigma$ (dual seed) are optimal for weak $\sigma$, consistent with the cursory inspection of Eqs.~(\ref{eqn:SigmaSensJz}) and (\ref{eqn:SigmaSensJz_d}). As the particular example of panel (a) illustrates, this choice is not fine tuned and in fact as $\sigma$ increases we find the minimum of $(\Delta\varphi)_{\hat{J}_z,\sigma}$ becomes increasingly broad and less sensitive to the precise value of $n_s$. In contrast, it is interesting to observe that the optimal $n_s$ for the $\hat{J}_z^2$ measurement appears to be quite different and favours larger seeds, even though the optimal sensitivity is almost indistinguishable. Insight into this difference is unfortunately constrained by the complexity of the analytic expression for $(\Delta\varphi)_{\hat{J}_z^2,\sigma}$.

Finally, it is worth commenting on the apparent paradox of suggesting that detection noise can be offset by using a small seed $n_s \sim \sigma$ that itself could be barely resolved by standard imaging due to detection noise. This is in fact not a contradiction, for two key reasons. Firstly, the microwave pulse sequence used to transfer atoms out of the condensate into the $m_F = \pm1$ modes [Fig.~\ref{fig:fig1}] can be well calibrated at large values of $n_s$ before extrapolating to low power such that small $n_s$ can be reliably prepared. Secondly, the small seed population can be inferred indirectly by analysis of the population dynamics as a result of spin-changing collisions rather than by direct imaging after the initial state is prepared. Specifically, the sum population $N_+$ has a well defined and strong dependence on $n_s$, as established in Eq.~(\ref{eqn:Np}), that can be used to estimate the seed population to a degree much better than the direct imaging may allow, i.e., $n_s \ll \sigma$ is resolvable. 

\subsection{Dynamic range}
Another important but often overlooked consideration for quantum sensing is the dynamical range (DNR), i.e., the range of $\varphi$ over which each state provides a quantum advantage compared to the SQL, $(\Delta\varphi) < 1/N_+$. Again, this quantity can be analytically computed in the limit of $n_s = 0$ for $\hat{J}_z^2$ and generic $n_s \gtrsim 1$ for $\hat{J}_z$, both in the absence of detection noise. For the former case we use Eq.~(\ref{eqn:SensJz2}) to directly obtain 
\begin{equation}
    \mathrm{DNR}_{\hat{J}_z^2} = \sqrt{\frac{2}{\bar{n}}} ,
\end{equation}
for $\bar{n} \gg 1$, while in the case of $n_s \gtrsim 1$ manipulation of Eqs.~(\ref{eqn:SensJz_s}) and (\ref{eqn:SensJz_d}), 
\begin{equation}
    \mathrm{DNR}_{\hat{J}_z} = \sqrt{\frac{2n_s^2}{1+2n_s}} \sqrt{\frac{2}{\bar{n}}} , \label{eqn:JzDNR}
\end{equation}
independent of the choice of initial single or dual seeds but we have assumed $\bar{n} \gg n_s, 1$. Thus, in principle the dynamic range increases by a factor $\sim \sqrt{n_s}$ when an appreciable seed $n_s \gtrsim 1$ is introduced. While this is not necessarily a meaningful advantage in the absence of detection noise, as the optimal sensitivity is for $n_s \sim 1$, it suggests that for $\sigma \neq 0$ then we might predict an enhanced DNR by a factor $\sim \sqrt{\sigma}$ (given the optimal seed is $n_s \sim \sigma$). 

This speculation is validated by explicit computation of the DNR in the inset of panel (a) including detection noise of $\sigma = 8$ (other parameters are identical to the main plot). We observe that the DNR grows with $n_s$ regardless of measurement choice (discounting the redundant region of divergent sensitivity for $n_s \lesssim 1$ using $\hat{J}_z$) and for this case the optimal sensitivity at $n_s \sim 30$ is accompanied by an approximately tenfold improvement in the DNR. This indicates that not only do we generically expect improved optimal sensitivity with seeded states but also a broader range of $\varphi$ for which the sensitivity is sub-SQL.

\subsection{Errors in state preparation}
Following the spirit of the previous section we present a brief analysis demonstrating that our conclusions are not sensitive to small errors in preparation of the initial seed. We separately consider the impact of number fluctuations and phase fluctuations and, for simplicity, focus only on results for the sensitivity $(\Delta\varphi)^2_{\hat{J}_z}$ obtained for single-sided seeding.

Spurious fluctuations of the seed phase $\theta_s$ can be caused by, e.g., imprecise characterization of Zeeman shifts in either the dynamics or state preparation and noise in the phase of applied microwaves that realize the internal-state beam-splitter operations. To understand the former, recall that the results of the previous sections are calculated in a frame rotating with the linear Zeeman shift, which in the original frame of $\hat{H}$ [Eq.~(\ref{eqn:H}) manifests as a shift $\theta_s \to \theta_s + pt$. This is easily accounted for and removed by experimental calibration, but unwanted fluctuations of the magnetic field or imprecise experimental timing could lead to shot-to-shot variations in $\theta_s$. We qualitatively account for these effects in our calculations by a simple model wherein $\theta_s$ is taken to be a Gaussian random variable with mean $\theta_{s,0}$ and variance $\delta\theta_s^2$. It is straightforward to substitute this definition of $\theta_s$ into the previously derived results for $\langle \hat{J}_z(\varphi) \rangle$ and $\langle \hat{J}_z^2(\varphi) \rangle$ [see Eq.~(\ref{eqn:JzJz2Phi})] and analytically compute the average over $\theta_s$. Considering small fluctuations $\delta\theta_s \ll 1$ and choosing $\theta_{0,s} = 0$, the best attainable sensitivity is found to be
\begin{equation}
    (\Delta\varphi)^2_{\hat{J}_z}\vert_{\varphi_{\mathrm{opt}}} \simeq \frac{(1+n_s)^3}{n_s(1+2n_s)}\frac{1}{\bar{n}^2}\left( 1 + \frac{4+8n_s}{4+4n_s} \delta\theta_s^2 \right) , \label{eqn:JzSens_thetasfluct}
\end{equation}
at 
\begin{equation}
    \varphi_{\mathrm{opt}} = \frac{n_s(1+n_s)}{\bar{n}(1+2n_s)}(1-2\delta\theta_s^2).
\end{equation}
Both expressions are only perturbed weakly by phase fluctuations at second-order, with an $\mathcal{O}(1)$ prefactor depending on $n_s$. This indicates that our findings are robust to spurious variations of $\theta_s$.  

Fluctuations in the number of seed atoms can also arise due to, e.g., shot-to-shot variations in the applied microwave power or duration during the transfer of atoms from $m_F=0$ to $m_F = \pm1$ initially. The effect of this can be investigated with a crude model where $n_s$ fluctuates shot-to-shot as a Gaussian variable with variance $\delta n_s^2 \ll n^2_s$. This is sufficient to demonstrate the robustness of our results, although in practice a more quantitative treatment could be designed based on the precise technical source of the fluctuations (i.e., one should formally model the source of the fluctuations, such as the microwave power, instead of the output atom number $n_s$). Moreover, the condition on $\delta n_s$ (enforced to ensure contributions from unphysical $n_s < 0$ do not skew the result) is not overly restrictive when realistic values of $\sigma$ are taken into account. Substituting this model into the results for $\langle \hat{J}_z(\varphi) \rangle$ and $\langle \hat{J}_z^2(\varphi) \rangle$ [see Eq.~(\ref{eqn:JzJz2Phi})] the same as previous leads to an optimal attainable sensitivity, 
\begin{equation}
    (\Delta\varphi)^2_{\hat{J}_z}\vert_{\varphi_{\mathrm{opt}}} \simeq \frac{(1+n_s)^3}{n_s(1+2n_s)}\frac{1}{\bar{n}^2} \left( 1 + \frac{1+n_s}{1+2n_s} \frac{\delta n_s^2}{n_s}\right) , \label{eqn:JzSens_nsfluct}
\end{equation}
at 
\begin{equation}
    \varphi_{\mathrm{opt}} \simeq \frac{n_s(1+n_s)}{\bar{n}(1+2n_s)}\left( 1 + \frac{1+n_s}{1+2n_s}\delta n_s^2 \right) ,
\end{equation}
for $\delta n_s \ll n_s$. The former result for $(\Delta\varphi)^2_{\hat{J}_z}\vert_{\varphi_{\mathrm{opt}}}$ indicates that one should have $\delta n_s^2 \lesssim n_s$, e.g., small noise compared to the Poissonian quantum fluctuations of the initial coherent seed, to retain good sensitivity.

\section{Numerical analysis of realistic system \label{sec:RealCalc}}
The results and analysis of the previous section is insightful but it is ultimately limited by the validity of the undepleted pump approximation. Here, we extend our investigation by simulating the full quantum dynamics of large, experimentally relevant systems and including the effects of depletion on the pair production process. 

We numerically integrate the quantum dynamics of a system of $N\sim 10^4$ particles that evolves according to the full Hamiltonian $\hat{H} = \hat{H}_{\mathrm{inel}} + \hat{H}_{\mathrm{el}} + \hat{H}_{\mathrm{Z}}$ as given in Eq.~(\ref{eqn:H}). While this still assumes the spatial dynamics are frozen, our treatment now properly treats depletion of the $m_F = 0$ mode and the interplay of the quadratic Zeeman shift with the elastic collisions. Moreover, we explicitly include the term $\propto g(\hat{n}_1 - \hat{n}_{-1})^2$, although we only find it is not quantitatively relevant (see also Appendix \ref{app:Hamiltonian}). Our calculations solve for the time-evolved state $\vert \psi^s_t \rangle = e^{-i\hat{H}t} \vert \psi^s_0 \rangle$ expanded in the Fock basis, based on an efficient Chebyshev scheme \cite{Kosloff_JCP1984} (see Appendix \ref{app:CFI} for further details).

We assume a condensate of $N = 10^4$ atoms is prepared in the $m_F = 0$ mode before a small number of atoms is coherently transferred to seed the $m_F = 1$ mode. We model this transfer process formally such that the total number of atoms $N$ is fixed in our calculations and we do not truncate our Fock basis \cite{Jianwen_2019}. For this large particle number our initial state very well approximates $\vert \psi^s_0 \rangle = \vert 0, \sqrt{N-n_s},  \sqrt{n_s}e^{i\theta_s}\rangle$ considered in previous sections. Spin-changing collisions are abruptly commenced by quenching the quadratic Zeeman shift to resonance, $q=g(N-n_s)$. After a time $t$ the generated state $\vert \psi^s_t\rangle$ is then input to a MZ sequence before the relevant expectation values are computed. 

\begin{figure}[!]
    \centering
    \includegraphics[width=8cm]{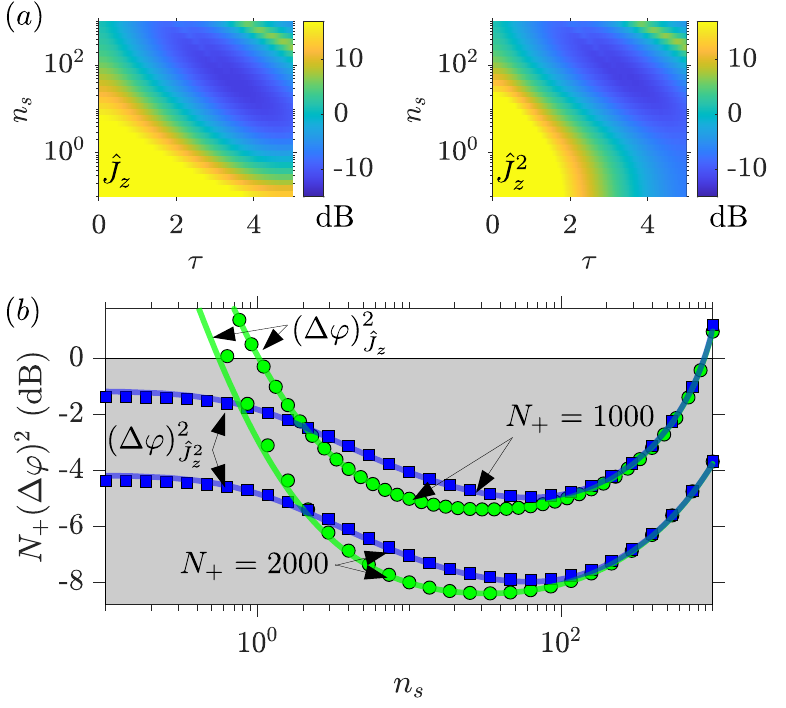}
    \caption{(a) Normalized sensitivity $N_+(\Delta\phi)^2$ obtained from full quantum dynamics [governed by $\hat{H}$, Eq.~(\ref{eqn:H})] of an initial BEC of $N = 10^4$ atoms as a function of initial seed occupation $n_s$ and interaction time $\tau$. Measurement signals are indicated on panels. Note that the plotted data saturates the colorscale in lower left corner of panels (i.e., $N_+(\Delta\phi)^2 > 10$~dB). (b) Optimal sensitivity as a function of seed $n_s$, with $\tau$ chosen such that $N_+ = 1000$ or $2000$ (indicated by arrows). Markers indicate results of full numerical simulations for a measurement of $\hat{J}_z$ (circles) or $\hat{J}_z^2$ (squares), while the tracking solid lines are equivalent analytic predictions from the undepleted pump approximation [e.g., Eq.~(\ref{eqn:SensJz_s})] with $\bar{n}$ and $n_s$ chosen to match numerical results. In all the panels of (a) and (b) we include fixed detection noise $\sigma = 8$.
    }
    \label{fig:fig5}
\end{figure}

We compute the optimal sensitivities $(\Delta\varphi)^2_{\hat{J}_z}\vert_{\varphi_{\mathrm{opt}}}$ and $(\Delta\varphi)^2_{\hat{J}^2_z}\vert_{\varphi_{\mathrm{opt}}}$ as a function of both seed size $n_s$ and interaction time $\tau = gNt$ and plot the results in Fig.~\ref{fig:fig5}(a). For best comparison to prior results (e.g., Fig.~\ref{fig:fig3}) the calculated sensitivities include detection noise of $\sigma = 8$. Both $(\Delta\varphi)^2_{\hat{J}_z}$ and $(\Delta\varphi)^2_{\hat{J}^2_z}$ show similar behaviour, including a pronounced minimum in the sensitivity as a function of time that approximately corresponds to the point where the maximum occupation of the $m_F = \pm1$ modes is first reached (before the collision process dynamically reverses and atoms re-populate the $m_F = 0$ mode). This minimum occurs faster as $n_s$ is increased, reflecting the bosonic stimulation of the scattering process provided by an initial coherent seed. Moreover, the minimum is clearly enhanced for $n_s \sim 10^2$, consistent with our prior analysis of Fig.~\ref{fig:fig3}. 

Figure \ref{fig:fig5}(b) shows the best attainable sensitivity as a function of $n_s$ with $\tau$ chosen for each $n_s$ such that the total population of the $m_F = \pm1$ modes is fixed to $N_+ = 1000$ or $2000$, corresponding to $10$\% and $20$\% depletion, respectively. In both cases we compare the results of the full numerical calculations to the analytic expressions Eqs.~(\ref{eqn:SensJz_s}) and (\ref{eqn:SensJz_d}) (we use the exact analytic expressions and optimize numerically over $\varphi$) and find superb agreement despite the large depletion. We comment that comparing the full numerical results with these analytic expressions of course does not actually assess their validity with respect to the simplest effects of \emph{pump depletion}, such as the slowdown in scattering to the $m_F = \pm 1$ modes [as clearly we are neglecting the analytic prediction linking $\bar{n}$ and $\tau$, Eq.~(\ref{eqn:Np})]. Rather, it simply makes clear that the connection established by the analytic model between various correlation functions and the number of scattered atoms $\bar{n}$ can still give tremendous insight into the structure of correlations and associated robustness of the generated states. 

\begin{figure}[!]
    \centering
    \includegraphics[width=8cm]{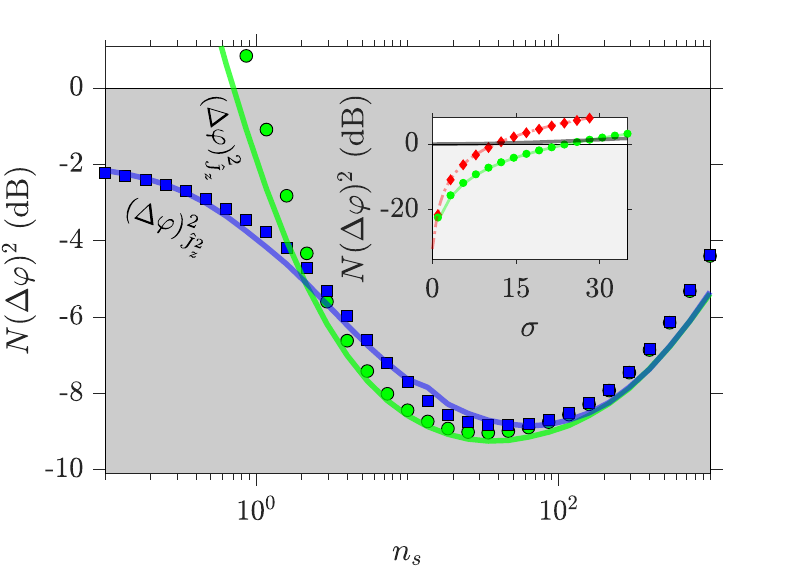}
    \caption{Best achievable sensitivity as a function of seed $n_s$ after optimization over interaction time $\tau$. The sensitivity is rescaled relative to the absolute SQL $(\Delta\varphi)^2_{\mathrm{SQL}} = 1/N$. We compare results of full numerical simulations for a measurement of $\hat{J}_z$ (green circles) or $\hat{J}_z^2$ (blue squares) with analytic expressions (corresponding blue and green lines overlaying the markers) derived from the undepleted pump approximation [e.g., optimization of Eq.~(\ref{eqn:SensJz_s})] with $\bar{n}$ matched to numerical simulations. We include fixed detection noise $\sigma = 8$. Inset: Sensitivity as a function of $\sigma$ obtained from numerical calculations with optimally chosen initial seed $n_s$ (green squares) compared to $n_s = 0$ prediction (red data). Lines (red dashed for $n_s = 0$ and green solid for $n_s \neq 0$) are analytic predictions from undepleted pump approximation with matching $N_+$.
    }
    \label{fig:fig6}
\end{figure}

In Fig.~\ref{fig:fig6} we go even further beyond our prior analytic analysis and compute the best attainable sensitivity as a function of $n_s$ after optimising over interaction time $\tau$. For simplicity, we restrict the optimization to times $\tau \leq \tau_{\mathrm{max}}$ where $\tau_{\mathrm{max}}$ corresponds to the time at which the first maximum in the population of the $m_F = \pm1$ modes is reached. This is reasonable as it is typically challenging for experiments to reliably probe correlations beyond this timescale without including quantitative corrections due to the spatial dynamics, although there has been notable recent progress in this direction \cite{Evrard_2021a,Evrard_2021b}. We find excellent qualitative agreement with our previous conclusions: In the presence of detection noise the addition of a coherent seed improves the achievable sensitivity. Moreover, while our results are clearly beyond the validity of the undepleted pump regime [typical depletion of $m_F = 0$ for the data in panel (b) is $\sim 50$\%] we nevertheless find substantial quantitative agreement by substitution of $n_s$ and $\bar{n}$ obtained from the numerical calculations directly into Eq.~(\ref{eqn:SensJz_s}) and optimizing over $\varphi$. This demonstrates that the analytic insight provided by the undepleted pump regime can remain relevant for realistic experimental conditions. 

Quantitatively, we highlight that the calculation of the full dynamics predicts a $\sim 7$~dB improvement in sensitivity by adding a suitable seed, to be compared to $\gtrsim 4.5$~dB in Fig.~\ref{fig:fig3} based on undepleted pump calculations. This improvement can be attributed to another favourable feature of seeded initial states -- they tend to support a larger total depletion of the $m_F = 0$ mode and thus better utilize the total available particle resource for metrology.

\section{Discussion and outlook \label{sec:Outlook}}
Our results can be contrasted with recent studies of seeded spin-exchange dynamics both in the limit of the undepleted pump approximation and longer timescales \cite{Jianwen_2019,Zhang_2019}. These works focused on the metrological potential of the generated states for SU$(1,1)$ interferometry, which is a well-known example of interaction based readout \cite{Linnemann_2016,Zhou_2020}. In this case, the main benefit of seeding is taken to be the acceleration of the dynamics and thus naive protection against sources of technical noise and decoherence such as particle loss. On the other hand,  other work has shown that the interaction-based readout intrinsically provides robustness to detection noise $\sigma \sim \mathcal{O}(\sqrt{N})$ \cite{Szigeti_2017,niezgoda2019optimal}, so introducing a seed simply degrades the metrologically useful entanglement (per particle) within the undepleted pump regime.

In summary, we have investigated how the initial quantum fluctuations that trigger spin-changing collisions in a spinor BEC can influence their practical utility for generating metrologically useful states for matter-wave interferometry. Particularly, introducing a coherent seed allows one to tune controllably between regimes of Dicke-like states and spin-squeezed states \cite{Nolan_2016}. The former states are inherently superior in an idealized setting, as they posses an optimal distribution of quantum projection noise that enables saturation of the Heisenberg limit in an SU(2) atom interferometer \cite{Holland_1993}, but the latter are more robust to ever present detection noise \cite{davis2017advantages}. A careful analysis demonstrates that while introducing a weak seed does inevitably reduce the ideal per-particle interferometric performance of the state generated by spin-changing collisions, superior practical performance with a coherent seed is robustly obtained for any reasonable value of $\sigma$. 

Our results can be directly relevant for current efforts to realize entanglement-enhanced interferometry using the dynamics of spinor BECs \cite{klempt2020spinor,Guo_2021,Zhang_2019,Evrard_2020}, although the conclusions are broad. In particular, our work demonstrates the importance of tailoring the generation of entanglement to survive technical noise and imperfections in realistic quantum systems, rather than for idealized properties and promised metrological potential. 

\begin{acknowledgments}
\noindent{\textit{Acknowledgements:}} We thank S.~Szigeti and S.~Haine for helpful discussions. Numerical calculations for this project were performed at the OU Supercomputing Center for Education \& Research (OSCER) at the University of Oklahoma. A.~S., G.~W.~B and R.~J. L-S acknowledge support from AFOSR Grant No. FA9550-20-1-007. A.~S. gratefully acknowledges support from the National Science Foundation through Grant No. PHY-1846965.  R.~J.~L-S acknowledges support from the Research Council of the University of Oklahoma Norman Campus.
\end{acknowledgments}

\bibliography{library}

\begin{thebibliography}{70}%
\makeatletter
\providecommand \@ifxundefined [1]{%
 \@ifx{#1\undefined}
}%
\providecommand \@ifnum [1]{%
 \ifnum #1\expandafter \@firstoftwo
 \else \expandafter \@secondoftwo
 \fi
}%
\providecommand \@ifx [1]{%
 \ifx #1\expandafter \@firstoftwo
 \else \expandafter \@secondoftwo
 \fi
}%
\providecommand \natexlab [1]{#1}%
\providecommand \enquote  [1]{``#1''}%
\providecommand \bibnamefont  [1]{#1}%
\providecommand \bibfnamefont [1]{#1}%
\providecommand \citenamefont [1]{#1}%
\providecommand \href@noop [0]{\@secondoftwo}%
\providecommand \href [0]{\begingroup \@sanitize@url \@href}%
\providecommand \@href[1]{\@@startlink{#1}\@@href}%
\providecommand \@@href[1]{\endgroup#1\@@endlink}%
\providecommand \@sanitize@url [0]{\catcode `\\12\catcode `\$12\catcode
  `\&12\catcode `\#12\catcode `\^12\catcode `\_12\catcode `\%12\relax}%
\providecommand \@@startlink[1]{}%
\providecommand \@@endlink[0]{}%
\providecommand \url  [0]{\begingroup\@sanitize@url \@url }%
\providecommand \@url [1]{\endgroup\@href {#1}{\urlprefix }}%
\providecommand \urlprefix  [0]{URL }%
\providecommand \Eprint [0]{\href }%
\providecommand \doibase [0]{https://doi.org/}%
\providecommand \selectlanguage [0]{\@gobble}%
\providecommand \bibinfo  [0]{\@secondoftwo}%
\providecommand \bibfield  [0]{\@secondoftwo}%
\providecommand \translation [1]{[#1]}%
\providecommand \BibitemOpen [0]{}%
\providecommand \bibitemStop [0]{}%
\providecommand \bibitemNoStop [0]{.\EOS\space}%
\providecommand \EOS [0]{\spacefactor3000\relax}%
\providecommand \BibitemShut  [1]{\csname bibitem#1\endcsname}%
\let\auto@bib@innerbib\@empty
\bibitem [{\citenamefont {Degen}\ \emph {et~al.}(2017)\citenamefont {Degen},
  \citenamefont {Reinhard},\ and\ \citenamefont
  {Cappellaro}}]{Cappellaro_2017}%
  \BibitemOpen
  \bibfield  {author} {\bibinfo {author} {\bibfnamefont {C.~L.}\ \bibnamefont
  {Degen}}, \bibinfo {author} {\bibfnamefont {F.}~\bibnamefont {Reinhard}},\
  and\ \bibinfo {author} {\bibfnamefont {P.}~\bibnamefont {Cappellaro}},\
  }\bibfield  {title} {\bibinfo {title} {Quantum sensing},\ }\href
  {https://doi.org/10.1103/RevModPhys.89.035002} {\bibfield  {journal}
  {\bibinfo  {journal} {Rev. Mod. Phys.}\ }\textbf {\bibinfo {volume} {89}},\
  \bibinfo {pages} {035002} (\bibinfo {year} {2017})}\BibitemShut {NoStop}%
\bibitem [{\citenamefont {{Abbott, B. P., \emph{et.
  al.}}}(2016)}]{AdvancedLigo2016}%
  \BibitemOpen
  \bibfield  {author} {\bibinfo {author} {\bibnamefont {{Abbott, B. P.,
  \emph{et. al.}}}} (\bibinfo {collaboration} {LIGO Scientific Collaboration
  and Virgo Collaboration}),\ }\bibfield  {title} {\bibinfo {title} {Gw150914:
  The advanced {LIGO} detectors in the era of first discoveries},\ }\href
  {https://doi.org/10.1103/PhysRevLett.116.131103} {\bibfield  {journal}
  {\bibinfo  {journal} {Phys. Rev. Lett.}\ }\textbf {\bibinfo {volume} {116}},\
  \bibinfo {pages} {131103} (\bibinfo {year} {2016})}\BibitemShut {NoStop}%
\bibitem [{\citenamefont {Pezz\`e}\ \emph {et~al.}(2018)\citenamefont
  {Pezz\`e}, \citenamefont {Smerzi}, \citenamefont {Oberthaler}, \citenamefont
  {Schmied},\ and\ \citenamefont {Treutlein}}]{Pezze_2018}%
  \BibitemOpen
  \bibfield  {author} {\bibinfo {author} {\bibfnamefont {L.}~\bibnamefont
  {Pezz\`e}}, \bibinfo {author} {\bibfnamefont {A.}~\bibnamefont {Smerzi}},
  \bibinfo {author} {\bibfnamefont {M.~K.}\ \bibnamefont {Oberthaler}},
  \bibinfo {author} {\bibfnamefont {R.}~\bibnamefont {Schmied}},\ and\ \bibinfo
  {author} {\bibfnamefont {P.}~\bibnamefont {Treutlein}},\ }\bibfield  {title}
  {\bibinfo {title} {Quantum metrology with nonclassical states of atomic
  ensembles},\ }\href {https://doi.org/10.1103/RevModPhys.90.035005} {\bibfield
   {journal} {\bibinfo  {journal} {Rev. Mod. Phys.}\ }\textbf {\bibinfo
  {volume} {90}},\ \bibinfo {pages} {035005} (\bibinfo {year}
  {2018})}\BibitemShut {NoStop}%
\bibitem [{\citenamefont {Szigeti}\ \emph {et~al.}(2020)\citenamefont
  {Szigeti}, \citenamefont {Nolan}, \citenamefont {Close},\ and\ \citenamefont
  {Haine}}]{Haine_Gravimeter_2020}%
  \BibitemOpen
  \bibfield  {author} {\bibinfo {author} {\bibfnamefont {S.~S.}\ \bibnamefont
  {Szigeti}}, \bibinfo {author} {\bibfnamefont {S.~P.}\ \bibnamefont {Nolan}},
  \bibinfo {author} {\bibfnamefont {J.~D.}\ \bibnamefont {Close}},\ and\
  \bibinfo {author} {\bibfnamefont {S.~A.}\ \bibnamefont {Haine}},\ }\bibfield
  {title} {\bibinfo {title} {High-precision quantum-enhanced gravimetry with a
  {Bose-Einstein} condensate},\ }\href
  {https://doi.org/10.1103/PhysRevLett.125.100402} {\bibfield  {journal}
  {\bibinfo  {journal} {Phys. Rev. Lett.}\ }\textbf {\bibinfo {volume} {125}},\
  \bibinfo {pages} {100402} (\bibinfo {year} {2020})}\BibitemShut {NoStop}%
\bibitem [{\citenamefont {Bidel}\ \emph {et~al.}(2018)\citenamefont {Bidel},
  \citenamefont {Zahzam}, \citenamefont {Blanchard}, \citenamefont {Bonnin},
  \citenamefont {Cadoret}, \citenamefont {Bresson}, \citenamefont {Rouxel},\
  and\ \citenamefont {Lequentrec-Lalancette}}]{bidel_2018}%
  \BibitemOpen
  \bibfield  {author} {\bibinfo {author} {\bibfnamefont {Y.}~\bibnamefont
  {Bidel}}, \bibinfo {author} {\bibfnamefont {N.}~\bibnamefont {Zahzam}},
  \bibinfo {author} {\bibfnamefont {C.}~\bibnamefont {Blanchard}}, \bibinfo
  {author} {\bibfnamefont {A.}~\bibnamefont {Bonnin}}, \bibinfo {author}
  {\bibfnamefont {M.}~\bibnamefont {Cadoret}}, \bibinfo {author} {\bibfnamefont
  {A.}~\bibnamefont {Bresson}}, \bibinfo {author} {\bibfnamefont
  {D.}~\bibnamefont {Rouxel}},\ and\ \bibinfo {author} {\bibfnamefont
  {M.}~\bibnamefont {Lequentrec-Lalancette}},\ }\bibfield  {title} {\bibinfo
  {title} {Absolute marine gravimetry with matter-wave interferometry},\ }\href
  {https://doi.org/10.1038/s41467-018-03040-2} {\bibfield  {journal} {\bibinfo
  {journal} {Nature communications}\ }\textbf {\bibinfo {volume} {9}},\
  \bibinfo {pages} {1} (\bibinfo {year} {2018})}\BibitemShut {NoStop}%
\bibitem [{\citenamefont {Peters}\ \emph {et~al.}(1999)\citenamefont {Peters},
  \citenamefont {Chung},\ and\ \citenamefont {Chu}}]{Peters1999}%
  \BibitemOpen
  \bibfield  {author} {\bibinfo {author} {\bibfnamefont {A.}~\bibnamefont
  {Peters}}, \bibinfo {author} {\bibfnamefont {K.~Y.}\ \bibnamefont {Chung}},\
  and\ \bibinfo {author} {\bibfnamefont {S.}~\bibnamefont {Chu}},\ }\bibfield
  {title} {\bibinfo {title} {Measurement of gravitational acceleration by
  dropping atoms},\ }\href {https://doi.org/10.1038/23655} {\bibfield
  {journal} {\bibinfo  {journal} {Nature}\ }\textbf {\bibinfo {volume} {400}},\
  \bibinfo {pages} {849} (\bibinfo {year} {1999})}\BibitemShut {NoStop}%
\bibitem [{\citenamefont {Ludlow}\ \emph {et~al.}(2015)\citenamefont {Ludlow},
  \citenamefont {Boyd}, \citenamefont {Ye}, \citenamefont {Peik},\ and\
  \citenamefont {Schmidt}}]{LudlowReview_2015}%
  \BibitemOpen
  \bibfield  {author} {\bibinfo {author} {\bibfnamefont {A.~D.}\ \bibnamefont
  {Ludlow}}, \bibinfo {author} {\bibfnamefont {M.~M.}\ \bibnamefont {Boyd}},
  \bibinfo {author} {\bibfnamefont {J.}~\bibnamefont {Ye}}, \bibinfo {author}
  {\bibfnamefont {E.}~\bibnamefont {Peik}},\ and\ \bibinfo {author}
  {\bibfnamefont {P.~O.}\ \bibnamefont {Schmidt}},\ }\bibfield  {title}
  {\bibinfo {title} {Optical atomic clocks},\ }\href
  {https://doi.org/10.1103/RevModPhys.87.637} {\bibfield  {journal} {\bibinfo
  {journal} {Rev. Mod. Phys.}\ }\textbf {\bibinfo {volume} {87}},\ \bibinfo
  {pages} {637} (\bibinfo {year} {2015})}\BibitemShut {NoStop}%
\bibitem [{\citenamefont {Cheiney}\ \emph {et~al.}(2018)\citenamefont
  {Cheiney}, \citenamefont {Fouch\'e}, \citenamefont {Templier}, \citenamefont
  {Napolitano}, \citenamefont {Battelier}, \citenamefont {Bouyer},\ and\
  \citenamefont {Barrett}}]{Bouyer_2018}%
  \BibitemOpen
  \bibfield  {author} {\bibinfo {author} {\bibfnamefont {P.}~\bibnamefont
  {Cheiney}}, \bibinfo {author} {\bibfnamefont {L.}~\bibnamefont {Fouch\'e}},
  \bibinfo {author} {\bibfnamefont {S.}~\bibnamefont {Templier}}, \bibinfo
  {author} {\bibfnamefont {F.}~\bibnamefont {Napolitano}}, \bibinfo {author}
  {\bibfnamefont {B.}~\bibnamefont {Battelier}}, \bibinfo {author}
  {\bibfnamefont {P.}~\bibnamefont {Bouyer}},\ and\ \bibinfo {author}
  {\bibfnamefont {B.}~\bibnamefont {Barrett}},\ }\bibfield  {title} {\bibinfo
  {title} {Navigation-compatible hybrid quantum accelerometer using a {Kalman}
  filter},\ }\href {https://doi.org/10.1103/PhysRevApplied.10.034030}
  {\bibfield  {journal} {\bibinfo  {journal} {Phys. Rev. Applied}\ }\textbf
  {\bibinfo {volume} {10}},\ \bibinfo {pages} {034030} (\bibinfo {year}
  {2018})}\BibitemShut {NoStop}%
\bibitem [{\citenamefont {McGuinness}\ \emph {et~al.}(2012)\citenamefont
  {McGuinness}, \citenamefont {Rakholia},\ and\ \citenamefont
  {Biedermann}}]{GB_APL_2012}%
  \BibitemOpen
  \bibfield  {author} {\bibinfo {author} {\bibfnamefont {H.~J.}\ \bibnamefont
  {McGuinness}}, \bibinfo {author} {\bibfnamefont {A.~V.}\ \bibnamefont
  {Rakholia}},\ and\ \bibinfo {author} {\bibfnamefont {G.~W.}\ \bibnamefont
  {Biedermann}},\ }\bibfield  {title} {\bibinfo {title} {High data-rate atom
  interferometer for measuring acceleration},\ }\href
  {https://doi.org/10.1063/1.3673845} {\bibfield  {journal} {\bibinfo
  {journal} {Applied Physics Letters}\ }\textbf {\bibinfo {volume} {100}},\
  \bibinfo {pages} {011106} (\bibinfo {year} {2012})}\BibitemShut {NoStop}%
\bibitem [{\citenamefont {Bongs}\ \emph {et~al.}(2019)\citenamefont {Bongs},
  \citenamefont {Holynski}, \citenamefont {Vovrosh}, \citenamefont {Bouyer},
  \citenamefont {Condon}, \citenamefont {Rasel}, \citenamefont {Schubert},
  \citenamefont {Schleich},\ and\ \citenamefont {Roura}}]{bongs_2019}%
  \BibitemOpen
  \bibfield  {author} {\bibinfo {author} {\bibfnamefont {K.}~\bibnamefont
  {Bongs}}, \bibinfo {author} {\bibfnamefont {M.}~\bibnamefont {Holynski}},
  \bibinfo {author} {\bibfnamefont {J.}~\bibnamefont {Vovrosh}}, \bibinfo
  {author} {\bibfnamefont {P.}~\bibnamefont {Bouyer}}, \bibinfo {author}
  {\bibfnamefont {G.}~\bibnamefont {Condon}}, \bibinfo {author} {\bibfnamefont
  {E.}~\bibnamefont {Rasel}}, \bibinfo {author} {\bibfnamefont
  {C.}~\bibnamefont {Schubert}}, \bibinfo {author} {\bibfnamefont {W.~P.}\
  \bibnamefont {Schleich}},\ and\ \bibinfo {author} {\bibfnamefont
  {A.}~\bibnamefont {Roura}},\ }\bibfield  {title} {\bibinfo {title} {Taking
  atom interferometric quantum sensors from the laboratory to real-world
  applications},\ }\href@noop {} {\bibfield  {journal} {\bibinfo  {journal}
  {Nature Reviews Physics}\ }\textbf {\bibinfo {volume} {1}},\ \bibinfo {pages}
  {731} (\bibinfo {year} {2019})}\BibitemShut {NoStop}%
\bibitem [{\citenamefont {Esteve}\ \emph {et~al.}(2008)\citenamefont {Esteve},
  \citenamefont {Gross}, \citenamefont {Weller}, \citenamefont {Giovanazzi},\
  and\ \citenamefont {Oberthaler}}]{EGW08}%
  \BibitemOpen
  \bibfield  {author} {\bibinfo {author} {\bibfnamefont {J.}~\bibnamefont
  {Esteve}}, \bibinfo {author} {\bibfnamefont {C.}~\bibnamefont {Gross}},
  \bibinfo {author} {\bibfnamefont {A.}~\bibnamefont {Weller}}, \bibinfo
  {author} {\bibfnamefont {S.}~\bibnamefont {Giovanazzi}},\ and\ \bibinfo
  {author} {\bibfnamefont {M.~K.}\ \bibnamefont {Oberthaler}},\ }\bibfield
  {title} {\bibinfo {title} {{Squeezing and entanglement in a Bose-Einstein
  condensate}},\ }\href {https://doi.org/10.1038/nature07332} {\bibfield
  {journal} {\bibinfo  {journal} {Nature}\ }\textbf {\bibinfo {volume} {455}},\
  \bibinfo {pages} {1216} (\bibinfo {year} {2008})}\BibitemShut {NoStop}%
\bibitem [{\citenamefont {Gross}\ \emph {et~al.}(2011)\citenamefont {Gross},
  \citenamefont {Strobel}, \citenamefont {Nicklas}, \citenamefont {Zibold},
  \citenamefont {Bar-Gill}, \citenamefont {Kurizki},\ and\ \citenamefont
  {Oberthaler}}]{Gross2011a}%
  \BibitemOpen
  \bibfield  {author} {\bibinfo {author} {\bibfnamefont {C.}~\bibnamefont
  {Gross}}, \bibinfo {author} {\bibfnamefont {H.}~\bibnamefont {Strobel}},
  \bibinfo {author} {\bibfnamefont {E.}~\bibnamefont {Nicklas}}, \bibinfo
  {author} {\bibfnamefont {T.}~\bibnamefont {Zibold}}, \bibinfo {author}
  {\bibfnamefont {N.}~\bibnamefont {Bar-Gill}}, \bibinfo {author}
  {\bibfnamefont {G.}~\bibnamefont {Kurizki}},\ and\ \bibinfo {author}
  {\bibfnamefont {M.~K.}\ \bibnamefont {Oberthaler}},\ }\bibfield  {title}
  {\bibinfo {title} {{Atomic homodyne detection of continuous-variable
  entangled twin-atom states}},\ }\href {https://doi.org/10.1038/nature10654}
  {\bibfield  {journal} {\bibinfo  {journal} {Nature}\ }\textbf {\bibinfo
  {volume} {480}},\ \bibinfo {pages} {219} (\bibinfo {year}
  {2011})}\BibitemShut {NoStop}%
\bibitem [{\citenamefont {Hamley}\ \emph {et~al.}(2012)\citenamefont {Hamley},
  \citenamefont {Gerving}, \citenamefont {Hoang}, \citenamefont {Bookjans},\
  and\ \citenamefont {Chapman}}]{Hamley2012a}%
  \BibitemOpen
  \bibfield  {author} {\bibinfo {author} {\bibfnamefont {C.~D.}\ \bibnamefont
  {Hamley}}, \bibinfo {author} {\bibfnamefont {C.~S.}\ \bibnamefont {Gerving}},
  \bibinfo {author} {\bibfnamefont {T.~M.}\ \bibnamefont {Hoang}}, \bibinfo
  {author} {\bibfnamefont {E.~M.}\ \bibnamefont {Bookjans}},\ and\ \bibinfo
  {author} {\bibfnamefont {M.~S.}\ \bibnamefont {Chapman}},\ }\bibfield
  {title} {\bibinfo {title} {{Spin-nematic squeezed vacuum in a quantum gas}},\
  }\href {https://doi.org/10.1038/nphys2245} {\bibfield  {journal} {\bibinfo
  {journal} {Nat. Phys.}\ }\textbf {\bibinfo {volume} {8}},\ \bibinfo {pages}
  {305} (\bibinfo {year} {2012})}\BibitemShut {NoStop}%
\bibitem [{\citenamefont {Hosten}\ \emph
  {et~al.}(2016{\natexlab{a}})\citenamefont {Hosten}, \citenamefont {Engelsen},
  \citenamefont {Krishnakumar},\ and\ \citenamefont {Kasevich}}]{Hosten_2016b}%
  \BibitemOpen
  \bibfield  {author} {\bibinfo {author} {\bibfnamefont {O.}~\bibnamefont
  {Hosten}}, \bibinfo {author} {\bibfnamefont {N.~J.}\ \bibnamefont
  {Engelsen}}, \bibinfo {author} {\bibfnamefont {R.}~\bibnamefont
  {Krishnakumar}},\ and\ \bibinfo {author} {\bibfnamefont {M.~A.}\ \bibnamefont
  {Kasevich}},\ }\bibfield  {title} {\bibinfo {title} {Measurement noise 100
  times lower than the quantum-projection limit using entangled atoms},\ }\href
  {https://doi.org/10.1038/nature16176} {\bibfield  {journal} {\bibinfo
  {journal} {Nature}\ }\textbf {\bibinfo {volume} {529}},\ \bibinfo {pages}
  {505} (\bibinfo {year} {2016}{\natexlab{a}})}\BibitemShut {NoStop}%
\bibitem [{\citenamefont {Bohnet}\ \emph
  {et~al.}(2014{\natexlab{a}})\citenamefont {Bohnet}, \citenamefont {Cox},
  \citenamefont {Norcia}, \citenamefont {Weiner}, \citenamefont {Chen},\ and\
  \citenamefont {Thompson}}]{Bohnet2014}%
  \BibitemOpen
  \bibfield  {author} {\bibinfo {author} {\bibfnamefont {J.~G.}\ \bibnamefont
  {Bohnet}}, \bibinfo {author} {\bibfnamefont {K.~C.}\ \bibnamefont {Cox}},
  \bibinfo {author} {\bibfnamefont {M.~A.}\ \bibnamefont {Norcia}}, \bibinfo
  {author} {\bibfnamefont {J.~M.}\ \bibnamefont {Weiner}}, \bibinfo {author}
  {\bibfnamefont {Z.}~\bibnamefont {Chen}},\ and\ \bibinfo {author}
  {\bibfnamefont {J.~K.}\ \bibnamefont {Thompson}},\ }\bibfield  {title}
  {\bibinfo {title} {Reduced spin measurement back-action for a phase
  sensitivity ten times beyond the standard quantum limit},\ }\href
  {https://doi.org/10.1038/nphoton.2014.151} {\bibfield  {journal} {\bibinfo
  {journal} {Nature Photonics}\ }\textbf {\bibinfo {volume} {8}},\ \bibinfo
  {pages} {731} (\bibinfo {year} {2014}{\natexlab{a}})}\BibitemShut {NoStop}%
\bibitem [{\citenamefont {Strobel}\ \emph {et~al.}(2014)\citenamefont
  {Strobel}, \citenamefont {Muessel}, \citenamefont {Linnemann}, \citenamefont
  {Zibold}, \citenamefont {Hume}, \citenamefont {Pezze}, \citenamefont
  {Smerzi},\ and\ \citenamefont {Oberthaler}}]{Strobel2014}%
  \BibitemOpen
  \bibfield  {author} {\bibinfo {author} {\bibfnamefont {H.}~\bibnamefont
  {Strobel}}, \bibinfo {author} {\bibfnamefont {W.}~\bibnamefont {Muessel}},
  \bibinfo {author} {\bibfnamefont {D.}~\bibnamefont {Linnemann}}, \bibinfo
  {author} {\bibfnamefont {T.}~\bibnamefont {Zibold}}, \bibinfo {author}
  {\bibfnamefont {D.~B.}\ \bibnamefont {Hume}}, \bibinfo {author}
  {\bibfnamefont {L.}~\bibnamefont {Pezze}}, \bibinfo {author} {\bibfnamefont
  {A.}~\bibnamefont {Smerzi}},\ and\ \bibinfo {author} {\bibfnamefont {M.~K.}\
  \bibnamefont {Oberthaler}},\ }\bibfield  {title} {\bibinfo {title} {{Fisher
  information and entanglement of non-Gaussian spin states}},\ }\href
  {https://doi.org/10.1126/science.1250147} {\bibfield  {journal} {\bibinfo
  {journal} {Science}\ }\textbf {\bibinfo {volume} {345}},\ \bibinfo {pages}
  {424} (\bibinfo {year} {2014})}\BibitemShut {NoStop}%
\bibitem [{\citenamefont {Bohnet}\ \emph
  {et~al.}(2014{\natexlab{b}})\citenamefont {Bohnet}, \citenamefont {Cox},
  \citenamefont {Norcia}, \citenamefont {Weiner}, \citenamefont {Chen},\ and\
  \citenamefont {Thompson}}]{Bohnet2014a}%
  \BibitemOpen
  \bibfield  {author} {\bibinfo {author} {\bibfnamefont {J.~G.}\ \bibnamefont
  {Bohnet}}, \bibinfo {author} {\bibfnamefont {K.~C.}\ \bibnamefont {Cox}},
  \bibinfo {author} {\bibfnamefont {M.~A.}\ \bibnamefont {Norcia}}, \bibinfo
  {author} {\bibfnamefont {J.~M.}\ \bibnamefont {Weiner}}, \bibinfo {author}
  {\bibfnamefont {Z.}~\bibnamefont {Chen}},\ and\ \bibinfo {author}
  {\bibfnamefont {J.~K.}\ \bibnamefont {Thompson}},\ }\bibfield  {title}
  {\bibinfo {title} {{Reduced spin measurement back-action for a phase
  sensitivity ten times beyond the standard quantum limit}},\ }\href
  {https://doi.org/10.1038/nphoton.2014.151} {\bibfield  {journal} {\bibinfo
  {journal} {Nat. Photonics}\ }\textbf {\bibinfo {volume} {8}},\ \bibinfo
  {pages} {731} (\bibinfo {year} {2014}{\natexlab{b}})}\BibitemShut {NoStop}%
\bibitem [{\citenamefont {Borselli}\ \emph {et~al.}(2021)\citenamefont
  {Borselli}, \citenamefont {Maiw\"oger}, \citenamefont {Zhang}, \citenamefont
  {Haslinger}, \citenamefont {Mukherjee}, \citenamefont {Negretti},
  \citenamefont {Montangero}, \citenamefont {Calarco}, \citenamefont {Mazets},
  \citenamefont {Bonneau},\ and\ \citenamefont
  {Schmiedmayer}}]{Schmiedmayer_2021}%
  \BibitemOpen
  \bibfield  {author} {\bibinfo {author} {\bibfnamefont {F.}~\bibnamefont
  {Borselli}}, \bibinfo {author} {\bibfnamefont {M.}~\bibnamefont
  {Maiw\"oger}}, \bibinfo {author} {\bibfnamefont {T.}~\bibnamefont {Zhang}},
  \bibinfo {author} {\bibfnamefont {P.}~\bibnamefont {Haslinger}}, \bibinfo
  {author} {\bibfnamefont {V.}~\bibnamefont {Mukherjee}}, \bibinfo {author}
  {\bibfnamefont {A.}~\bibnamefont {Negretti}}, \bibinfo {author}
  {\bibfnamefont {S.}~\bibnamefont {Montangero}}, \bibinfo {author}
  {\bibfnamefont {T.}~\bibnamefont {Calarco}}, \bibinfo {author} {\bibfnamefont
  {I.}~\bibnamefont {Mazets}}, \bibinfo {author} {\bibfnamefont
  {M.}~\bibnamefont {Bonneau}},\ and\ \bibinfo {author} {\bibfnamefont
  {J.}~\bibnamefont {Schmiedmayer}},\ }\bibfield  {title} {\bibinfo {title}
  {Two-particle interference with double twin-atom beams},\ }\href
  {https://doi.org/10.1103/PhysRevLett.126.083603} {\bibfield  {journal}
  {\bibinfo  {journal} {Phys. Rev. Lett.}\ }\textbf {\bibinfo {volume} {126}},\
  \bibinfo {pages} {083603} (\bibinfo {year} {2021})}\BibitemShut {NoStop}%
\bibitem [{\citenamefont {L{\"u}cke}\ \emph {et~al.}(2011)\citenamefont
  {L{\"u}cke}, \citenamefont {Scherer}, \citenamefont {Kruse}, \citenamefont
  {Pezz{\'e}}, \citenamefont {Deuretzbacher}, \citenamefont {Hyllus},
  \citenamefont {Topic}, \citenamefont {Peise}, \citenamefont {Ertmer},
  \citenamefont {Arlt}, \citenamefont {Santos}, \citenamefont {Smerzi},\ and\
  \citenamefont {Klempt}}]{Lucke_2011}%
  \BibitemOpen
  \bibfield  {author} {\bibinfo {author} {\bibfnamefont {B.}~\bibnamefont
  {L{\"u}cke}}, \bibinfo {author} {\bibfnamefont {M.}~\bibnamefont {Scherer}},
  \bibinfo {author} {\bibfnamefont {J.}~\bibnamefont {Kruse}}, \bibinfo
  {author} {\bibfnamefont {L.}~\bibnamefont {Pezz{\'e}}}, \bibinfo {author}
  {\bibfnamefont {F.}~\bibnamefont {Deuretzbacher}}, \bibinfo {author}
  {\bibfnamefont {P.}~\bibnamefont {Hyllus}}, \bibinfo {author} {\bibfnamefont
  {O.}~\bibnamefont {Topic}}, \bibinfo {author} {\bibfnamefont
  {J.}~\bibnamefont {Peise}}, \bibinfo {author} {\bibfnamefont
  {W.}~\bibnamefont {Ertmer}}, \bibinfo {author} {\bibfnamefont
  {J.}~\bibnamefont {Arlt}}, \bibinfo {author} {\bibfnamefont {L.}~\bibnamefont
  {Santos}}, \bibinfo {author} {\bibfnamefont {A.}~\bibnamefont {Smerzi}},\
  and\ \bibinfo {author} {\bibfnamefont {C.}~\bibnamefont {Klempt}},\
  }\bibfield  {title} {\bibinfo {title} {Twin matter waves for interferometry
  beyond the classical limit},\ }\href
  {https://doi.org/10.1126/science.1208798} {\bibfield  {journal} {\bibinfo
  {journal} {Science}\ }\textbf {\bibinfo {volume} {334}},\ \bibinfo {pages}
  {773} (\bibinfo {year} {2011})}\BibitemShut {NoStop}%
\bibitem [{\citenamefont {Gross}\ \emph {et~al.}(2010)\citenamefont {Gross},
  \citenamefont {Zibold}, \citenamefont {Nicklas}, \citenamefont
  {Est{\`{e}}ve},\ and\ \citenamefont {Oberthaler}}]{GZN10}%
  \BibitemOpen
  \bibfield  {author} {\bibinfo {author} {\bibfnamefont {C.}~\bibnamefont
  {Gross}}, \bibinfo {author} {\bibfnamefont {T.}~\bibnamefont {Zibold}},
  \bibinfo {author} {\bibfnamefont {E.}~\bibnamefont {Nicklas}}, \bibinfo
  {author} {\bibfnamefont {J.}~\bibnamefont {Est{\`{e}}ve}},\ and\ \bibinfo
  {author} {\bibfnamefont {M.~K.}\ \bibnamefont {Oberthaler}},\ }\bibfield
  {title} {\bibinfo {title} {{Nonlinear atom interferometer surpasses classical
  precision limit}},\ }\href {https://doi.org/10.1038/nature08919} {\bibfield
  {journal} {\bibinfo  {journal} {Nature}\ }\textbf {\bibinfo {volume} {464}},\
  \bibinfo {pages} {1165} (\bibinfo {year} {2010})}\BibitemShut {NoStop}%
\bibitem [{\citenamefont {Hosten}\ \emph
  {et~al.}(2016{\natexlab{b}})\citenamefont {Hosten}, \citenamefont
  {Krishnakumar}, \citenamefont {Engelsen},\ and\ \citenamefont
  {Kasevich}}]{Hosten_2016}%
  \BibitemOpen
  \bibfield  {author} {\bibinfo {author} {\bibfnamefont {O.}~\bibnamefont
  {Hosten}}, \bibinfo {author} {\bibfnamefont {R.}~\bibnamefont
  {Krishnakumar}}, \bibinfo {author} {\bibfnamefont {N.~J.}\ \bibnamefont
  {Engelsen}},\ and\ \bibinfo {author} {\bibfnamefont {M.~A.}\ \bibnamefont
  {Kasevich}},\ }\bibfield  {title} {\bibinfo {title} {Quantum phase
  magnification},\ }\href {https://doi.org/10.1126/science.aaf3397} {\bibfield
  {journal} {\bibinfo  {journal} {Science}\ }\textbf {\bibinfo {volume}
  {352}},\ \bibinfo {pages} {1552} (\bibinfo {year}
  {2016}{\natexlab{b}})}\BibitemShut {NoStop}%
\bibitem [{\citenamefont {Linnemann}\ \emph {et~al.}(2016)\citenamefont
  {Linnemann}, \citenamefont {Strobel}, \citenamefont {Muessel}, \citenamefont
  {Schulz}, \citenamefont {Lewis-Swan}, \citenamefont {Kheruntsyan},\ and\
  \citenamefont {Oberthaler}}]{Linnemann_2016}%
  \BibitemOpen
  \bibfield  {author} {\bibinfo {author} {\bibfnamefont {D.}~\bibnamefont
  {Linnemann}}, \bibinfo {author} {\bibfnamefont {H.}~\bibnamefont {Strobel}},
  \bibinfo {author} {\bibfnamefont {W.}~\bibnamefont {Muessel}}, \bibinfo
  {author} {\bibfnamefont {J.}~\bibnamefont {Schulz}}, \bibinfo {author}
  {\bibfnamefont {R.~J.}\ \bibnamefont {Lewis-Swan}}, \bibinfo {author}
  {\bibfnamefont {K.~V.}\ \bibnamefont {Kheruntsyan}},\ and\ \bibinfo {author}
  {\bibfnamefont {M.~K.}\ \bibnamefont {Oberthaler}},\ }\bibfield  {title}
  {\bibinfo {title} {Quantum-enhanced sensing based on time reversal of
  nonlinear dynamics},\ }\href {https://doi.org/10.1103/PhysRevLett.117.013001}
  {\bibfield  {journal} {\bibinfo  {journal} {Phys. Rev. Lett.}\ }\textbf
  {\bibinfo {volume} {117}},\ \bibinfo {pages} {013001} (\bibinfo {year}
  {2016})}\BibitemShut {NoStop}%
\bibitem [{\citenamefont {Pedrozo-Pe{\~{n}}afiel}\ \emph
  {et~al.}(2020)\citenamefont {Pedrozo-Pe{\~{n}}afiel}, \citenamefont
  {Colombo}, \citenamefont {Shu}, \citenamefont {Adiyatullin}, \citenamefont
  {Li}, \citenamefont {Mendez}, \citenamefont {Braverman}, \citenamefont
  {Kawasaki}, \citenamefont {Akamatsu}, \citenamefont {Xiao},\ and\
  \citenamefont {Vuleti{\'{c}}}}]{Vuletic_2020}%
  \BibitemOpen
  \bibfield  {author} {\bibinfo {author} {\bibfnamefont {E.}~\bibnamefont
  {Pedrozo-Pe{\~{n}}afiel}}, \bibinfo {author} {\bibfnamefont {S.}~\bibnamefont
  {Colombo}}, \bibinfo {author} {\bibfnamefont {C.}~\bibnamefont {Shu}},
  \bibinfo {author} {\bibfnamefont {A.~F.}\ \bibnamefont {Adiyatullin}},
  \bibinfo {author} {\bibfnamefont {Z.}~\bibnamefont {Li}}, \bibinfo {author}
  {\bibfnamefont {E.}~\bibnamefont {Mendez}}, \bibinfo {author} {\bibfnamefont
  {B.}~\bibnamefont {Braverman}}, \bibinfo {author} {\bibfnamefont
  {A.}~\bibnamefont {Kawasaki}}, \bibinfo {author} {\bibfnamefont
  {D.}~\bibnamefont {Akamatsu}}, \bibinfo {author} {\bibfnamefont
  {Y.}~\bibnamefont {Xiao}},\ and\ \bibinfo {author} {\bibfnamefont
  {V.}~\bibnamefont {Vuleti{\'{c}}}},\ }\bibfield  {title} {\bibinfo {title}
  {Entanglement on an optical atomic-clock transition},\ }\href
  {https://doi.org/10.1038/s41586-020-3006-1} {\bibfield  {journal} {\bibinfo
  {journal} {Nature}\ }\textbf {\bibinfo {volume} {588}},\ \bibinfo {pages}
  {414} (\bibinfo {year} {2020})}\BibitemShut {NoStop}%
\bibitem [{\citenamefont {Szigeti}\ \emph {et~al.}(2021)\citenamefont
  {Szigeti}, \citenamefont {Hosten},\ and\ \citenamefont {Haine}}]{Haine_2021}%
  \BibitemOpen
  \bibfield  {author} {\bibinfo {author} {\bibfnamefont {S.~S.}\ \bibnamefont
  {Szigeti}}, \bibinfo {author} {\bibfnamefont {O.}~\bibnamefont {Hosten}},\
  and\ \bibinfo {author} {\bibfnamefont {S.~A.}\ \bibnamefont {Haine}},\
  }\bibfield  {title} {\bibinfo {title} {Improving cold-atom sensors with
  quantum entanglement: Prospects and challenges},\ }\href
  {https://doi.org/10.1063/5.0050235} {\bibfield  {journal} {\bibinfo
  {journal} {Applied Physics Letters}\ }\textbf {\bibinfo {volume} {118}},\
  \bibinfo {pages} {140501} (\bibinfo {year} {2021})}\BibitemShut {NoStop}%
\bibitem [{\citenamefont {Brif}\ \emph {et~al.}(2020)\citenamefont {Brif},
  \citenamefont {Ruzic},\ and\ \citenamefont {Biedermann}}]{GB_PRX_2020}%
  \BibitemOpen
  \bibfield  {author} {\bibinfo {author} {\bibfnamefont {C.}~\bibnamefont
  {Brif}}, \bibinfo {author} {\bibfnamefont {B.~P.}\ \bibnamefont {Ruzic}},\
  and\ \bibinfo {author} {\bibfnamefont {G.~W.}\ \bibnamefont {Biedermann}},\
  }\bibfield  {title} {\bibinfo {title} {Characterization of errors in
  interferometry with entangled atoms},\ }\href
  {https://doi.org/10.1103/PRXQuantum.1.010306} {\bibfield  {journal} {\bibinfo
   {journal} {PRX Quantum}\ }\textbf {\bibinfo {volume} {1}},\ \bibinfo {pages}
  {010306} (\bibinfo {year} {2020})}\BibitemShut {NoStop}%
\bibitem [{\citenamefont {Braunstein}\ and\ \citenamefont
  {Caves}(1994)}]{Braunstein1994}%
  \BibitemOpen
  \bibfield  {author} {\bibinfo {author} {\bibfnamefont {S.~L.}\ \bibnamefont
  {Braunstein}}\ and\ \bibinfo {author} {\bibfnamefont {C.~M.}\ \bibnamefont
  {Caves}},\ }\bibfield  {title} {\bibinfo {title} {Statistical distance and
  the geometry of quantum states},\ }\href
  {https://doi.org/10.1103/PhysRevLett.72.3439} {\bibfield  {journal} {\bibinfo
   {journal} {Phys. Rev. Lett.}\ }\textbf {\bibinfo {volume} {72}},\ \bibinfo
  {pages} {3439} (\bibinfo {year} {1994})}\BibitemShut {NoStop}%
\bibitem [{\citenamefont {Bollinger}\ \emph {et~al.}(1996)\citenamefont
  {Bollinger}, \citenamefont {Itano}, \citenamefont {Wineland},\ and\
  \citenamefont {Heinzen}}]{bollinger1996optimal}%
  \BibitemOpen
  \bibfield  {author} {\bibinfo {author} {\bibfnamefont {J.~J.}\ \bibnamefont
  {Bollinger}}, \bibinfo {author} {\bibfnamefont {W.~M.}\ \bibnamefont
  {Itano}}, \bibinfo {author} {\bibfnamefont {D.~J.}\ \bibnamefont
  {Wineland}},\ and\ \bibinfo {author} {\bibfnamefont {D.~J.}\ \bibnamefont
  {Heinzen}},\ }\bibfield  {title} {\bibinfo {title} {Optimal frequency
  measurements with maximally correlated states},\ }\href
  {https://doi.org/10.1103/PhysRevA.54.R4649} {\bibfield  {journal} {\bibinfo
  {journal} {Physical Review A}\ }\textbf {\bibinfo {volume} {54}},\ \bibinfo
  {pages} {R4649} (\bibinfo {year} {1996})}\BibitemShut {NoStop}%
\bibitem [{\citenamefont {Qu}\ \emph {et~al.}(2020)\citenamefont {Qu},
  \citenamefont {Evrard}, \citenamefont {Dalibard},\ and\ \citenamefont
  {Gerbier}}]{Evrard_2020}%
  \BibitemOpen
  \bibfield  {author} {\bibinfo {author} {\bibfnamefont {A.}~\bibnamefont
  {Qu}}, \bibinfo {author} {\bibfnamefont {B.}~\bibnamefont {Evrard}}, \bibinfo
  {author} {\bibfnamefont {J.}~\bibnamefont {Dalibard}},\ and\ \bibinfo
  {author} {\bibfnamefont {F.}~\bibnamefont {Gerbier}},\ }\bibfield  {title}
  {\bibinfo {title} {Probing spin correlations in a {Bose-Einstein} condensate
  near the single-atom level},\ }\href
  {https://doi.org/10.1103/PhysRevLett.125.033401} {\bibfield  {journal}
  {\bibinfo  {journal} {Phys. Rev. Lett.}\ }\textbf {\bibinfo {volume} {125}},\
  \bibinfo {pages} {033401} (\bibinfo {year} {2020})}\BibitemShut {NoStop}%
\bibitem [{\citenamefont {Fr\"owis}\ \emph {et~al.}(2016)\citenamefont
  {Fr\"owis}, \citenamefont {Sekatski},\ and\ \citenamefont
  {D\"ur}}]{Sekatski2015}%
  \BibitemOpen
  \bibfield  {author} {\bibinfo {author} {\bibfnamefont {F.}~\bibnamefont
  {Fr\"owis}}, \bibinfo {author} {\bibfnamefont {P.}~\bibnamefont {Sekatski}},\
  and\ \bibinfo {author} {\bibfnamefont {W.}~\bibnamefont {D\"ur}},\ }\bibfield
   {title} {\bibinfo {title} {Detecting large quantum fisher information with
  finite measurement precision},\ }\href
  {https://doi.org/10.1103/PhysRevLett.116.090801} {\bibfield  {journal}
  {\bibinfo  {journal} {Phys. Rev. Lett.}\ }\textbf {\bibinfo {volume} {116}},\
  \bibinfo {pages} {090801} (\bibinfo {year} {2016})}\BibitemShut {NoStop}%
\bibitem [{\citenamefont {Davis}\ \emph {et~al.}(2016)\citenamefont {Davis},
  \citenamefont {Bentsen},\ and\ \citenamefont {Schleier-Smith}}]{Davis_2016}%
  \BibitemOpen
  \bibfield  {author} {\bibinfo {author} {\bibfnamefont {E.}~\bibnamefont
  {Davis}}, \bibinfo {author} {\bibfnamefont {G.}~\bibnamefont {Bentsen}},\
  and\ \bibinfo {author} {\bibfnamefont {M.}~\bibnamefont {Schleier-Smith}},\
  }\bibfield  {title} {\bibinfo {title} {Approaching the heisenberg limit
  without single-particle detection},\ }\href
  {https://doi.org/10.1103/PhysRevLett.116.053601} {\bibfield  {journal}
  {\bibinfo  {journal} {Phys. Rev. Lett.}\ }\textbf {\bibinfo {volume} {116}},\
  \bibinfo {pages} {053601} (\bibinfo {year} {2016})}\BibitemShut {NoStop}%
\bibitem [{\citenamefont {Nolan}\ \emph {et~al.}(2017)\citenamefont {Nolan},
  \citenamefont {Szigeti},\ and\ \citenamefont {Haine}}]{Nolan_2017}%
  \BibitemOpen
  \bibfield  {author} {\bibinfo {author} {\bibfnamefont {S.~P.}\ \bibnamefont
  {Nolan}}, \bibinfo {author} {\bibfnamefont {S.~S.}\ \bibnamefont {Szigeti}},\
  and\ \bibinfo {author} {\bibfnamefont {S.~A.}\ \bibnamefont {Haine}},\
  }\bibfield  {title} {\bibinfo {title} {Optimal and robust quantum metrology
  using interaction-based readouts},\ }\href
  {https://doi.org/10.1103/PhysRevLett.119.193601} {\bibfield  {journal}
  {\bibinfo  {journal} {Phys. Rev. Lett.}\ }\textbf {\bibinfo {volume} {119}},\
  \bibinfo {pages} {193601} (\bibinfo {year} {2017})}\BibitemShut {NoStop}%
\bibitem [{\citenamefont {Anders}\ \emph {et~al.}(2018)\citenamefont {Anders},
  \citenamefont {Pezz\`e}, \citenamefont {Smerzi},\ and\ \citenamefont
  {Klempt}}]{Fabian_2018}%
  \BibitemOpen
  \bibfield  {author} {\bibinfo {author} {\bibfnamefont {F.}~\bibnamefont
  {Anders}}, \bibinfo {author} {\bibfnamefont {L.}~\bibnamefont {Pezz\`e}},
  \bibinfo {author} {\bibfnamefont {A.}~\bibnamefont {Smerzi}},\ and\ \bibinfo
  {author} {\bibfnamefont {C.}~\bibnamefont {Klempt}},\ }\bibfield  {title}
  {\bibinfo {title} {Phase magnification by two-axis countertwisting for
  detection-noise robust interferometry},\ }\href
  {https://doi.org/10.1103/PhysRevA.97.043813} {\bibfield  {journal} {\bibinfo
  {journal} {Phys. Rev. A}\ }\textbf {\bibinfo {volume} {97}},\ \bibinfo
  {pages} {043813} (\bibinfo {year} {2018})}\BibitemShut {NoStop}%
\bibitem [{\citenamefont {Colombo}\ \emph {et~al.}(2021)\citenamefont
  {Colombo}, \citenamefont {Pedrozo-Peñafiel}, \citenamefont {Adiyatullin},
  \citenamefont {Li}, \citenamefont {Mendez}, \citenamefont {Shu},\ and\
  \citenamefont {Vuletic}}]{colombo_2021}%
  \BibitemOpen
  \bibfield  {author} {\bibinfo {author} {\bibfnamefont {S.}~\bibnamefont
  {Colombo}}, \bibinfo {author} {\bibfnamefont {E.}~\bibnamefont
  {Pedrozo-Peñafiel}}, \bibinfo {author} {\bibfnamefont {A.~F.}\ \bibnamefont
  {Adiyatullin}}, \bibinfo {author} {\bibfnamefont {Z.}~\bibnamefont {Li}},
  \bibinfo {author} {\bibfnamefont {E.}~\bibnamefont {Mendez}}, \bibinfo
  {author} {\bibfnamefont {C.}~\bibnamefont {Shu}},\ and\ \bibinfo {author}
  {\bibfnamefont {V.}~\bibnamefont {Vuletic}},\ }\bibfield  {title} {\bibinfo
  {title} {Time-{Reversal}-{Based} {Quantum} {Metrology} with {Many}-{Body}
  {Entangled} {States}},\ }\href {http://arxiv.org/abs/2106.03754} {\bibfield
  {journal} {\bibinfo  {journal} {arXiv:2106.03754}\ } (\bibinfo {year}
  {2021})}\BibitemShut {NoStop}%
\bibitem [{\citenamefont {Gilmore}\ \emph {et~al.}(2021)\citenamefont
  {Gilmore}, \citenamefont {Affolter}, \citenamefont {Lewis-Swan},
  \citenamefont {Barberena}, \citenamefont {Jordan}, \citenamefont {Rey},\ and\
  \citenamefont {Bollinger}}]{gilmore_2021}%
  \BibitemOpen
  \bibfield  {author} {\bibinfo {author} {\bibfnamefont {K.~A.}\ \bibnamefont
  {Gilmore}}, \bibinfo {author} {\bibfnamefont {M.}~\bibnamefont {Affolter}},
  \bibinfo {author} {\bibfnamefont {R.~J.}\ \bibnamefont {Lewis-Swan}},
  \bibinfo {author} {\bibfnamefont {D.}~\bibnamefont {Barberena}}, \bibinfo
  {author} {\bibfnamefont {E.}~\bibnamefont {Jordan}}, \bibinfo {author}
  {\bibfnamefont {A.~M.}\ \bibnamefont {Rey}},\ and\ \bibinfo {author}
  {\bibfnamefont {J.~J.}\ \bibnamefont {Bollinger}},\ }\bibfield  {title}
  {\bibinfo {title} {Quantum-enhanced sensing of displacements and electric
  fields with large trapped-ion crystals},\ }\href
  {http://arxiv.org/abs/2103.08690} {\bibfield  {journal} {\bibinfo  {journal}
  {arXiv:2103.08690}\ } (\bibinfo {year} {2021})}\BibitemShut {NoStop}%
\bibitem [{\citenamefont {Jie}\ \emph {et~al.}(2019)\citenamefont {Jie},
  \citenamefont {Guan},\ and\ \citenamefont {Blume}}]{Jianwen_2019}%
  \BibitemOpen
  \bibfield  {author} {\bibinfo {author} {\bibfnamefont {J.}~\bibnamefont
  {Jie}}, \bibinfo {author} {\bibfnamefont {Q.}~\bibnamefont {Guan}},\ and\
  \bibinfo {author} {\bibfnamefont {D.}~\bibnamefont {Blume}},\ }\bibfield
  {title} {\bibinfo {title} {Spinor {Bose-Einstein} condensate interferometer
  within the undepleted pump approximation: Role of the initial state},\ }\href
  {https://doi.org/10.1103/PhysRevA.100.043606} {\bibfield  {journal} {\bibinfo
   {journal} {Phys. Rev. A}\ }\textbf {\bibinfo {volume} {100}},\ \bibinfo
  {pages} {043606} (\bibinfo {year} {2019})}\BibitemShut {NoStop}%
\bibitem [{\citenamefont {Zhang}\ and\ \citenamefont
  {Schwettmann}(2019)}]{Zhang_2019}%
  \BibitemOpen
  \bibfield  {author} {\bibinfo {author} {\bibfnamefont {Q.}~\bibnamefont
  {Zhang}}\ and\ \bibinfo {author} {\bibfnamefont {A.}~\bibnamefont
  {Schwettmann}},\ }\bibfield  {title} {\bibinfo {title} {Quantum
  interferometry with microwave-dressed $f=1$ spinor {Bose-Einstein}
  condensates: Role of initial states and long-time evolution},\ }\href
  {https://doi.org/10.1103/PhysRevA.100.063637} {\bibfield  {journal} {\bibinfo
   {journal} {Phys. Rev. A}\ }\textbf {\bibinfo {volume} {100}},\ \bibinfo
  {pages} {063637} (\bibinfo {year} {2019})}\BibitemShut {NoStop}%
\bibitem [{\citenamefont {Wrubel}\ \emph {et~al.}(2018)\citenamefont {Wrubel},
  \citenamefont {Schwettmann}, \citenamefont {Fahey}, \citenamefont {Glassman},
  \citenamefont {Pechkis}, \citenamefont {Griffin}, \citenamefont {Barnett},
  \citenamefont {Tiesinga},\ and\ \citenamefont {Lett}}]{Wrubel_2018}%
  \BibitemOpen
  \bibfield  {author} {\bibinfo {author} {\bibfnamefont {J.~P.}\ \bibnamefont
  {Wrubel}}, \bibinfo {author} {\bibfnamefont {A.}~\bibnamefont {Schwettmann}},
  \bibinfo {author} {\bibfnamefont {D.~P.}\ \bibnamefont {Fahey}}, \bibinfo
  {author} {\bibfnamefont {Z.}~\bibnamefont {Glassman}}, \bibinfo {author}
  {\bibfnamefont {H.~K.}\ \bibnamefont {Pechkis}}, \bibinfo {author}
  {\bibfnamefont {P.~F.}\ \bibnamefont {Griffin}}, \bibinfo {author}
  {\bibfnamefont {R.}~\bibnamefont {Barnett}}, \bibinfo {author} {\bibfnamefont
  {E.}~\bibnamefont {Tiesinga}},\ and\ \bibinfo {author} {\bibfnamefont
  {P.~D.}\ \bibnamefont {Lett}},\ }\bibfield  {title} {\bibinfo {title} {Spinor
  bose-einstein-condensate phase-sensitive amplifier for su(1,1)
  interferometry},\ }\href {https://doi.org/10.1103/PhysRevA.98.023620}
  {\bibfield  {journal} {\bibinfo  {journal} {Phys. Rev. A}\ }\textbf {\bibinfo
  {volume} {98}},\ \bibinfo {pages} {023620} (\bibinfo {year}
  {2018})}\BibitemShut {NoStop}%
\bibitem [{\citenamefont {Anders}\ \emph {et~al.}(2020)\citenamefont {Anders},
  \citenamefont {Idel}, \citenamefont {Feldmann}, \citenamefont {Bondarenko},
  \citenamefont {Loriani}, \citenamefont {Lange}, \citenamefont {Peise},
  \citenamefont {Gersemann}, \citenamefont {Meyer}, \citenamefont {Abend},
  \citenamefont {Schubert}, \citenamefont {Schlippert}, \citenamefont {Santos},
  \citenamefont {Rasel},\ and\ \citenamefont {Klempt}}]{klempt2020spinor}%
  \BibitemOpen
  \bibfield  {author} {\bibinfo {author} {\bibfnamefont {F.}~\bibnamefont
  {Anders}}, \bibinfo {author} {\bibfnamefont {A.}~\bibnamefont {Idel}},
  \bibinfo {author} {\bibfnamefont {P.}~\bibnamefont {Feldmann}}, \bibinfo
  {author} {\bibfnamefont {D.}~\bibnamefont {Bondarenko}}, \bibinfo {author}
  {\bibfnamefont {S.}~\bibnamefont {Loriani}}, \bibinfo {author} {\bibfnamefont
  {K.}~\bibnamefont {Lange}}, \bibinfo {author} {\bibfnamefont
  {J.}~\bibnamefont {Peise}}, \bibinfo {author} {\bibfnamefont
  {M.}~\bibnamefont {Gersemann}}, \bibinfo {author} {\bibfnamefont
  {B.}~\bibnamefont {Meyer}}, \bibinfo {author} {\bibfnamefont
  {S.}~\bibnamefont {Abend}}, \bibinfo {author} {\bibfnamefont
  {C.}~\bibnamefont {Schubert}}, \bibinfo {author} {\bibfnamefont
  {D.}~\bibnamefont {Schlippert}}, \bibinfo {author} {\bibfnamefont
  {L.}~\bibnamefont {Santos}}, \bibinfo {author} {\bibfnamefont
  {E.}~\bibnamefont {Rasel}},\ and\ \bibinfo {author} {\bibfnamefont
  {C.}~\bibnamefont {Klempt}},\ }\bibfield  {title} {\bibinfo {title} {Momentum
  entanglement for atom interferometry},\ }\href
  {https://arxiv.org/abs/2010.15796} {\bibfield  {journal} {\bibinfo  {journal}
  {arXiv:2010.15796}\ } (\bibinfo {year} {2020})}\BibitemShut {NoStop}%
\bibitem [{\citenamefont {Davis}\ \emph {et~al.}(2017)\citenamefont {Davis},
  \citenamefont {Bentsen}, \citenamefont {Li},\ and\ \citenamefont
  {Schleier-Smith}}]{davis2017advantages}%
  \BibitemOpen
  \bibfield  {author} {\bibinfo {author} {\bibfnamefont {E.}~\bibnamefont
  {Davis}}, \bibinfo {author} {\bibfnamefont {G.}~\bibnamefont {Bentsen}},
  \bibinfo {author} {\bibfnamefont {T.}~\bibnamefont {Li}},\ and\ \bibinfo
  {author} {\bibfnamefont {M.}~\bibnamefont {Schleier-Smith}},\ }\bibfield
  {title} {\bibinfo {title} {Advantages of interaction-based readout for
  quantum sensing},\ }in\ \href@noop {} {\emph {\bibinfo {booktitle} {Advances
  in Photonics of Quantum Computing, Memory, and Communication X}}},\ Vol.\
  \bibinfo {volume} {10118}\ (\bibinfo {organization} {International Society
  for Optics and Photonics},\ \bibinfo {year} {2017})\ p.\ \bibinfo {pages}
  {101180Z}\BibitemShut {NoStop}%
\bibitem [{\citenamefont {Niezgoda}\ \emph {et~al.}(2018)\citenamefont
  {Niezgoda}, \citenamefont {Kajtoch},\ and\ \citenamefont
  {Witkowska}}]{Niezgoda_2018}%
  \BibitemOpen
  \bibfield  {author} {\bibinfo {author} {\bibfnamefont {A.}~\bibnamefont
  {Niezgoda}}, \bibinfo {author} {\bibfnamefont {D.}~\bibnamefont {Kajtoch}},\
  and\ \bibinfo {author} {\bibfnamefont {E.}~\bibnamefont {Witkowska}},\
  }\bibfield  {title} {\bibinfo {title} {Efficient two-mode interferometers
  with spinor {Bose-Einstein} condensates},\ }\href
  {https://doi.org/10.1103/PhysRevA.98.013610} {\bibfield  {journal} {\bibinfo
  {journal} {Phys. Rev. A}\ }\textbf {\bibinfo {volume} {98}},\ \bibinfo
  {pages} {013610} (\bibinfo {year} {2018})}\BibitemShut {NoStop}%
\bibitem [{\citenamefont {Morgenstern}\ \emph {et~al.}(2020)\citenamefont
  {Morgenstern}, \citenamefont {Zhong}, \citenamefont {Zhang}, \citenamefont
  {Baker}, \citenamefont {Norris}, \citenamefont {Tran},\ and\ \citenamefont
  {Schwettmann}}]{Arne2020}%
  \BibitemOpen
  \bibfield  {author} {\bibinfo {author} {\bibfnamefont {I.}~\bibnamefont
  {Morgenstern}}, \bibinfo {author} {\bibfnamefont {S.}~\bibnamefont {Zhong}},
  \bibinfo {author} {\bibfnamefont {Q.}~\bibnamefont {Zhang}}, \bibinfo
  {author} {\bibfnamefont {L.}~\bibnamefont {Baker}}, \bibinfo {author}
  {\bibfnamefont {J.}~\bibnamefont {Norris}}, \bibinfo {author} {\bibfnamefont
  {B.}~\bibnamefont {Tran}},\ and\ \bibinfo {author} {\bibfnamefont
  {A.}~\bibnamefont {Schwettmann}},\ }\bibfield  {title} {\bibinfo {title} {A
  versatile microwave source for cold atom experiments controlled by a field
  programmable gate array},\ }\href {https://doi.org/10.1063/1.5127880}
  {\bibfield  {journal} {\bibinfo  {journal} {Review of Scientific
  Instruments}\ }\textbf {\bibinfo {volume} {91}},\ \bibinfo {pages} {023202}
  (\bibinfo {year} {2020})}\BibitemShut {NoStop}%
\bibitem [{\citenamefont {Law}\ \emph {et~al.}(1998)\citenamefont {Law},
  \citenamefont {Pu},\ and\ \citenamefont {Bigelow}}]{Law_1998}%
  \BibitemOpen
  \bibfield  {author} {\bibinfo {author} {\bibfnamefont {C.~K.}\ \bibnamefont
  {Law}}, \bibinfo {author} {\bibfnamefont {H.}~\bibnamefont {Pu}},\ and\
  \bibinfo {author} {\bibfnamefont {N.~P.}\ \bibnamefont {Bigelow}},\
  }\bibfield  {title} {\bibinfo {title} {Quantum {Spins} {Mixing} in {Spinor}
  {Bose}-{Einstein} {Condensates}},\ }\href
  {https://doi.org/10.1103/PhysRevLett.81.5257} {\bibfield  {journal} {\bibinfo
   {journal} {Phys. Rev. Lett.}\ }\textbf {\bibinfo {volume} {81}},\ \bibinfo
  {pages} {5257} (\bibinfo {year} {1998})}\BibitemShut {NoStop}%
\bibitem [{\citenamefont {Kawaguchi}\ and\ \citenamefont
  {Ueda}(2012)}]{kawaguchi_2012}%
  \BibitemOpen
  \bibfield  {author} {\bibinfo {author} {\bibfnamefont {Y.}~\bibnamefont
  {Kawaguchi}}\ and\ \bibinfo {author} {\bibfnamefont {M.}~\bibnamefont
  {Ueda}},\ }\bibfield  {title} {\bibinfo {title} {Spinor {Bose}–{Einstein}
  condensates},\ }\href
  {https://doi.org/https://doi.org/10.1016/j.physrep.2012.07.005} {\bibfield
  {journal} {\bibinfo  {journal} {Physics Reports}\ }\textbf {\bibinfo {volume}
  {520}},\ \bibinfo {pages} {253} (\bibinfo {year} {2012})}\BibitemShut
  {NoStop}%
\bibitem [{\citenamefont {Jie}\ \emph {et~al.}(2020)\citenamefont {Jie},
  \citenamefont {Guan}, \citenamefont {Zhong}, \citenamefont {Schwettmann},\
  and\ \citenamefont {Blume}}]{Jie2020}%
  \BibitemOpen
  \bibfield  {author} {\bibinfo {author} {\bibfnamefont {J.}~\bibnamefont
  {Jie}}, \bibinfo {author} {\bibfnamefont {Q.}~\bibnamefont {Guan}}, \bibinfo
  {author} {\bibfnamefont {S.}~\bibnamefont {Zhong}}, \bibinfo {author}
  {\bibfnamefont {A.}~\bibnamefont {Schwettmann}},\ and\ \bibinfo {author}
  {\bibfnamefont {D.}~\bibnamefont {Blume}},\ }\bibfield  {title} {\bibinfo
  {title} {Mean-field spin-oscillation dynamics beyond the single-mode
  approximation for a harmonically trapped spin-1 {Bose}-{Einstein}
  condensate},\ }\href {https://doi.org/10.1103/PhysRevA.102.023324} {\bibfield
   {journal} {\bibinfo  {journal} {Phys. Rev. A}\ }\textbf {\bibinfo {volume}
  {102}},\ \bibinfo {pages} {023324} (\bibinfo {year} {2020})}\BibitemShut
  {NoStop}%
\bibitem [{\citenamefont {Lewis-Swan}\ and\ \citenamefont
  {Kheruntsyan}(2013)}]{lewis2013epr}%
  \BibitemOpen
  \bibfield  {author} {\bibinfo {author} {\bibfnamefont {R.~J.}\ \bibnamefont
  {Lewis-Swan}}\ and\ \bibinfo {author} {\bibfnamefont {K.~V.}\ \bibnamefont
  {Kheruntsyan}},\ }\bibfield  {title} {\bibinfo {title} {Sensitivity to
  thermal noise of atomic einstein-podolsky-rosen entanglement},\ }\href
  {https://doi.org/10.1103/PhysRevA.87.063635} {\bibfield  {journal} {\bibinfo
  {journal} {Phys. Rev. A}\ }\textbf {\bibinfo {volume} {87}},\ \bibinfo
  {pages} {063635} (\bibinfo {year} {2013})}\BibitemShut {NoStop}%
\bibitem [{\citenamefont {Evrard}\ \emph
  {et~al.}(2021{\natexlab{a}})\citenamefont {Evrard}, \citenamefont {Qu},
  \citenamefont {Dalibard},\ and\ \citenamefont {Gerbier}}]{Evrard_2021b}%
  \BibitemOpen
  \bibfield  {author} {\bibinfo {author} {\bibfnamefont {B.}~\bibnamefont
  {Evrard}}, \bibinfo {author} {\bibfnamefont {A.}~\bibnamefont {Qu}}, \bibinfo
  {author} {\bibfnamefont {J.}~\bibnamefont {Dalibard}},\ and\ \bibinfo
  {author} {\bibfnamefont {F.}~\bibnamefont {Gerbier}},\ }\bibfield  {title}
  {\bibinfo {title} {Coherent seeding of the dynamics of a spinor
  {Bose-Einstein} condensate: From quantum to classical behavior},\ }\href
  {https://doi.org/10.1103/PhysRevA.103.L031302} {\bibfield  {journal}
  {\bibinfo  {journal} {Phys. Rev. A}\ }\textbf {\bibinfo {volume} {103}},\
  \bibinfo {pages} {L031302} (\bibinfo {year}
  {2021}{\natexlab{a}})}\BibitemShut {NoStop}%
\bibitem [{\citenamefont {Nolan}\ \emph {et~al.}(2016)\citenamefont {Nolan},
  \citenamefont {Sabbatini}, \citenamefont {Bromley}, \citenamefont {Davis},\
  and\ \citenamefont {Haine}}]{Nolan_2016}%
  \BibitemOpen
  \bibfield  {author} {\bibinfo {author} {\bibfnamefont {S.~P.}\ \bibnamefont
  {Nolan}}, \bibinfo {author} {\bibfnamefont {J.}~\bibnamefont {Sabbatini}},
  \bibinfo {author} {\bibfnamefont {M.~W.~J.}\ \bibnamefont {Bromley}},
  \bibinfo {author} {\bibfnamefont {M.~J.}\ \bibnamefont {Davis}},\ and\
  \bibinfo {author} {\bibfnamefont {S.~A.}\ \bibnamefont {Haine}},\ }\bibfield
  {title} {\bibinfo {title} {Quantum enhanced measurement of rotations with a
  spin-1 {B}ose-{E}instein condensate in a ring trap},\ }\href
  {https://doi.org/10.1103/PhysRevA.93.023616} {\bibfield  {journal} {\bibinfo
  {journal} {Phys. Rev. A}\ }\textbf {\bibinfo {volume} {93}},\ \bibinfo
  {pages} {023616} (\bibinfo {year} {2016})}\BibitemShut {NoStop}%
\bibitem [{\citenamefont {Kasevich}\ and\ \citenamefont
  {Chu}(1992)}]{kasevich_1992}%
  \BibitemOpen
  \bibfield  {author} {\bibinfo {author} {\bibfnamefont {M.}~\bibnamefont
  {Kasevich}}\ and\ \bibinfo {author} {\bibfnamefont {S.}~\bibnamefont {Chu}},\
  }\bibfield  {title} {\bibinfo {title} {Measurement of the gravitational
  acceleration of an atom with a light-pulse atom interferometer},\ }\href
  {https://doi.org/10.1007/BF00325375} {\bibfield  {journal} {\bibinfo
  {journal} {Applied Physics B}\ }\textbf {\bibinfo {volume} {54}},\ \bibinfo
  {pages} {321} (\bibinfo {year} {1992})}\BibitemShut {NoStop}%
\bibitem [{Note1()}]{Note1}%
  \BibitemOpen
  \bibinfo {note} {This also encompasses the measurement of e.g., coherences
  between the different $m_F$ levels by combining population measurements with
  linear operations that couple the internal states.}\BibitemShut {Stop}%
\bibitem [{\citenamefont {Pezz\`e}\ \emph {et~al.}(2015)\citenamefont
  {Pezz\`e}, \citenamefont {Hyllus},\ and\ \citenamefont
  {Smerzi}}]{Pezze_2015}%
  \BibitemOpen
  \bibfield  {author} {\bibinfo {author} {\bibfnamefont {L.}~\bibnamefont
  {Pezz\`e}}, \bibinfo {author} {\bibfnamefont {P.}~\bibnamefont {Hyllus}},\
  and\ \bibinfo {author} {\bibfnamefont {A.}~\bibnamefont {Smerzi}},\
  }\bibfield  {title} {\bibinfo {title} {Phase-sensitivity bounds for two-mode
  interferometers},\ }\href {https://doi.org/10.1103/PhysRevA.91.032103}
  {\bibfield  {journal} {\bibinfo  {journal} {Phys. Rev. A}\ }\textbf {\bibinfo
  {volume} {91}},\ \bibinfo {pages} {032103} (\bibinfo {year}
  {2015})}\BibitemShut {NoStop}%
\bibitem [{\citenamefont {Caves}\ \emph {et~al.}(1991)\citenamefont {Caves},
  \citenamefont {Zhu}, \citenamefont {Milburn},\ and\ \citenamefont
  {Schleich}}]{Caves_SqueezedCoherent_1991}%
  \BibitemOpen
  \bibfield  {author} {\bibinfo {author} {\bibfnamefont {C.~M.}\ \bibnamefont
  {Caves}}, \bibinfo {author} {\bibfnamefont {C.}~\bibnamefont {Zhu}}, \bibinfo
  {author} {\bibfnamefont {G.~J.}\ \bibnamefont {Milburn}},\ and\ \bibinfo
  {author} {\bibfnamefont {W.}~\bibnamefont {Schleich}},\ }\bibfield  {title}
  {\bibinfo {title} {Photon statistics of two-mode squeezed states and
  interference in four-dimensional phase space},\ }\href
  {https://doi.org/10.1103/PhysRevA.43.3854} {\bibfield  {journal} {\bibinfo
  {journal} {Phys. Rev. A}\ }\textbf {\bibinfo {volume} {43}},\ \bibinfo
  {pages} {3854} (\bibinfo {year} {1991})}\BibitemShut {NoStop}%
\bibitem [{\citenamefont {Dowling}\ \emph {et~al.}(1994)\citenamefont
  {Dowling}, \citenamefont {Agarwal},\ and\ \citenamefont
  {Schleich}}]{Dowling_1994}%
  \BibitemOpen
  \bibfield  {author} {\bibinfo {author} {\bibfnamefont {J.~P.}\ \bibnamefont
  {Dowling}}, \bibinfo {author} {\bibfnamefont {G.~S.}\ \bibnamefont
  {Agarwal}},\ and\ \bibinfo {author} {\bibfnamefont {W.~P.}\ \bibnamefont
  {Schleich}},\ }\bibfield  {title} {\bibinfo {title} {Wigner distribution of a
  general angular-momentum state: Applications to a collection of two-level
  atoms},\ }\href {https://doi.org/10.1103/PhysRevA.49.4101} {\bibfield
  {journal} {\bibinfo  {journal} {Phys. Rev. A}\ }\textbf {\bibinfo {volume}
  {49}},\ \bibinfo {pages} {4101} (\bibinfo {year} {1994})}\BibitemShut
  {NoStop}%
\bibitem [{\citenamefont {Koczor}\ \emph {et~al.}(2020)\citenamefont {Koczor},
  \citenamefont {Zeier},\ and\ \citenamefont {Glaser}}]{Koczor_2020}%
  \BibitemOpen
  \bibfield  {author} {\bibinfo {author} {\bibfnamefont {B.}~\bibnamefont
  {Koczor}}, \bibinfo {author} {\bibfnamefont {R.}~\bibnamefont {Zeier}},\ and\
  \bibinfo {author} {\bibfnamefont {S.~J.}\ \bibnamefont {Glaser}},\ }\bibfield
   {title} {\bibinfo {title} {Fast computation of spherical phase-space
  functions of quantum many-body states},\ }\href
  {https://doi.org/10.1103/PhysRevA.102.062421} {\bibfield  {journal} {\bibinfo
   {journal} {Phys. Rev. A}\ }\textbf {\bibinfo {volume} {102}},\ \bibinfo
  {pages} {062421} (\bibinfo {year} {2020})}\BibitemShut {NoStop}%
\bibitem [{\citenamefont {Holland}\ and\ \citenamefont
  {Burnett}(1993)}]{Holland_1993}%
  \BibitemOpen
  \bibfield  {author} {\bibinfo {author} {\bibfnamefont {M.~J.}\ \bibnamefont
  {Holland}}\ and\ \bibinfo {author} {\bibfnamefont {K.}~\bibnamefont
  {Burnett}},\ }\bibfield  {title} {\bibinfo {title} {Interferometric detection
  of optical phase shifts at the heisenberg limit},\ }\href
  {https://doi.org/10.1103/PhysRevLett.71.1355} {\bibfield  {journal} {\bibinfo
   {journal} {Phys. Rev. Lett.}\ }\textbf {\bibinfo {volume} {71}},\ \bibinfo
  {pages} {1355} (\bibinfo {year} {1993})}\BibitemShut {NoStop}%
\bibitem [{\citenamefont {Kitagawa}\ and\ \citenamefont
  {Ueda}(1993)}]{Kitagawa1993}%
  \BibitemOpen
  \bibfield  {author} {\bibinfo {author} {\bibfnamefont {M.}~\bibnamefont
  {Kitagawa}}\ and\ \bibinfo {author} {\bibfnamefont {M.}~\bibnamefont
  {Ueda}},\ }\bibfield  {title} {\bibinfo {title} {{Squeezed spin states}},\
  }\href {https://doi.org/http://dx.doi.org/10.1103/PhysRevA.47.5138}
  {\bibfield  {journal} {\bibinfo  {journal} {Phys. Rev. A}\ }\textbf {\bibinfo
  {volume} {47}},\ \bibinfo {pages} {5138} (\bibinfo {year}
  {1993})}\BibitemShut {NoStop}%
\bibitem [{\citenamefont {L{\"{u}}cke}\ \emph {et~al.}(2014)\citenamefont
  {L{\"{u}}cke}, \citenamefont {Peise}, \citenamefont {Vitagliano},
  \citenamefont {Arlt}, \citenamefont {Santos}, \citenamefont {T{\'{o}}th},\
  and\ \citenamefont {Klempt}}]{Lucke2014}%
  \BibitemOpen
  \bibfield  {author} {\bibinfo {author} {\bibfnamefont {B.}~\bibnamefont
  {L{\"{u}}cke}}, \bibinfo {author} {\bibfnamefont {J.}~\bibnamefont {Peise}},
  \bibinfo {author} {\bibfnamefont {G.}~\bibnamefont {Vitagliano}}, \bibinfo
  {author} {\bibfnamefont {J.}~\bibnamefont {Arlt}}, \bibinfo {author}
  {\bibfnamefont {L.}~\bibnamefont {Santos}}, \bibinfo {author} {\bibfnamefont
  {G.}~\bibnamefont {T{\'{o}}th}},\ and\ \bibinfo {author} {\bibfnamefont
  {C.}~\bibnamefont {Klempt}},\ }\bibfield  {title} {\bibinfo {title}
  {{Detecting Multiparticle Entanglement of Dicke States}},\ }\href
  {https://doi.org/10.1103/PhysRevLett.112.155304} {\bibfield  {journal}
  {\bibinfo  {journal} {Phys. Rev. Lett.}\ }\textbf {\bibinfo {volume} {112}},\
  \bibinfo {pages} {155304} (\bibinfo {year} {2014})}\BibitemShut {NoStop}%
\bibitem [{\citenamefont {Wineland}\ \emph {et~al.}(1994)\citenamefont
  {Wineland}, \citenamefont {Bollinger}, \citenamefont {Itano},\ and\
  \citenamefont {Heinzen}}]{Wineland1994a}%
  \BibitemOpen
  \bibfield  {author} {\bibinfo {author} {\bibfnamefont {D.}~\bibnamefont
  {Wineland}}, \bibinfo {author} {\bibfnamefont {J.}~\bibnamefont {Bollinger}},
  \bibinfo {author} {\bibfnamefont {W.}~\bibnamefont {Itano}},\ and\ \bibinfo
  {author} {\bibfnamefont {D.}~\bibnamefont {Heinzen}},\ }\bibfield  {title}
  {\bibinfo {title} {{Squeezed atomic states and projection noise in
  spectroscopy}},\ }\href {https://doi.org/10.1103/PhysRevA.50.67} {\bibfield
  {journal} {\bibinfo  {journal} {Phys. Rev. A}\ }\textbf {\bibinfo {volume}
  {50}},\ \bibinfo {pages} {67} (\bibinfo {year} {1994})}\BibitemShut {NoStop}%
\bibitem [{\citenamefont {Zhang}\ \emph {et~al.}(2013)\citenamefont {Zhang},
  \citenamefont {Li},\ and\ \citenamefont {Jin}}]{Zhang_2013}%
  \BibitemOpen
  \bibfield  {author} {\bibinfo {author} {\bibfnamefont {Y.-M.}\ \bibnamefont
  {Zhang}}, \bibinfo {author} {\bibfnamefont {X.-W.}\ \bibnamefont {Li}},\ and\
  \bibinfo {author} {\bibfnamefont {G.-R.}\ \bibnamefont {Jin}},\ }\bibfield
  {title} {\bibinfo {title} {Quantum interferometer with two-mode squeezed
  vacuum: $\hat{S}_z^2$ measurement},\ }\href
  {https://doi.org/10.1088/1674-1056/22/11/114206} {\bibfield  {journal}
  {\bibinfo  {journal} {Chinese Physics B}\ }\textbf {\bibinfo {volume} {22}},\
  \bibinfo {pages} {114206} (\bibinfo {year} {2013})}\BibitemShut {NoStop}%
\bibitem [{\citenamefont {Rowe}\ \emph {et~al.}(2001)\citenamefont {Rowe},
  \citenamefont {de~Guise},\ and\ \citenamefont {Sanders}}]{Rowe_2001}%
  \BibitemOpen
  \bibfield  {author} {\bibinfo {author} {\bibfnamefont {D.~J.}\ \bibnamefont
  {Rowe}}, \bibinfo {author} {\bibfnamefont {H.}~\bibnamefont {de~Guise}},\
  and\ \bibinfo {author} {\bibfnamefont {B.~C.}\ \bibnamefont {Sanders}},\
  }\bibfield  {title} {\bibinfo {title} {Asymptotic limits of {SU}(2) and
  {SU}(3) wigner functions},\ }\href {https://doi.org/10.1063/1.1358305}
  {\bibfield  {journal} {\bibinfo  {journal} {Journal of Mathematical Physics}\
  }\textbf {\bibinfo {volume} {42}},\ \bibinfo {pages} {2315} (\bibinfo {year}
  {2001})}\BibitemShut {NoStop}%
\bibitem [{Note2()}]{Note2}%
  \BibitemOpen
  \bibinfo {note} {In fact, seeded states are technically always superior
  according to this metric for $n_s < N_+$. However, this improvement is
  vanishingly small beyond $\sigma \approx 35$.}\BibitemShut {Stop}%
\bibitem [{\citenamefont {Tal-Ezer}\ and\ \citenamefont
  {Kosloff}(1984)}]{Kosloff_JCP1984}%
  \BibitemOpen
  \bibfield  {author} {\bibinfo {author} {\bibfnamefont {H.}~\bibnamefont
  {Tal-Ezer}}\ and\ \bibinfo {author} {\bibfnamefont {R.}~\bibnamefont
  {Kosloff}},\ }\bibfield  {title} {\bibinfo {title} {An accurate and efficient
  scheme for propagating the time dependent {S}chr\"odinger equation},\ }\href
  {https://doi.org/10.1063/1.448136} {\bibfield  {journal} {\bibinfo  {journal}
  {The Journal of Chemical Physics}\ }\textbf {\bibinfo {volume} {81}},\
  \bibinfo {pages} {3967} (\bibinfo {year} {1984})}\BibitemShut {NoStop}%
\bibitem [{\citenamefont {Evrard}\ \emph
  {et~al.}(2021{\natexlab{b}})\citenamefont {Evrard}, \citenamefont {Qu},
  \citenamefont {Dalibard},\ and\ \citenamefont {Gerbier}}]{Evrard_2021a}%
  \BibitemOpen
  \bibfield  {author} {\bibinfo {author} {\bibfnamefont {B.}~\bibnamefont
  {Evrard}}, \bibinfo {author} {\bibfnamefont {A.}~\bibnamefont {Qu}}, \bibinfo
  {author} {\bibfnamefont {J.}~\bibnamefont {Dalibard}},\ and\ \bibinfo
  {author} {\bibfnamefont {F.}~\bibnamefont {Gerbier}},\ }\bibfield  {title}
  {\bibinfo {title} {From many-body oscillations to thermalization in an
  isolated spinor gas},\ }\href
  {https://doi.org/10.1103/PhysRevLett.126.063401} {\bibfield  {journal}
  {\bibinfo  {journal} {Phys. Rev. Lett.}\ }\textbf {\bibinfo {volume} {126}},\
  \bibinfo {pages} {063401} (\bibinfo {year} {2021}{\natexlab{b}})}\BibitemShut
  {NoStop}%
\bibitem [{\citenamefont {Lyu}\ \emph {et~al.}(2020)\citenamefont {Lyu},
  \citenamefont {Lv},\ and\ \citenamefont {Zhou}}]{Zhou_2020}%
  \BibitemOpen
  \bibfield  {author} {\bibinfo {author} {\bibfnamefont {C.}~\bibnamefont
  {Lyu}}, \bibinfo {author} {\bibfnamefont {C.}~\bibnamefont {Lv}},\ and\
  \bibinfo {author} {\bibfnamefont {Q.}~\bibnamefont {Zhou}},\ }\bibfield
  {title} {\bibinfo {title} {Geometrizing quantum dynamics of a {Bose-Einstein}
  condensate},\ }\href {https://doi.org/10.1103/PhysRevLett.125.253401}
  {\bibfield  {journal} {\bibinfo  {journal} {Phys. Rev. Lett.}\ }\textbf
  {\bibinfo {volume} {125}},\ \bibinfo {pages} {253401} (\bibinfo {year}
  {2020})}\BibitemShut {NoStop}%
\bibitem [{\citenamefont {Szigeti}\ \emph {et~al.}(2017)\citenamefont
  {Szigeti}, \citenamefont {Lewis-Swan},\ and\ \citenamefont
  {Haine}}]{Szigeti_2017}%
  \BibitemOpen
  \bibfield  {author} {\bibinfo {author} {\bibfnamefont {S.~S.}\ \bibnamefont
  {Szigeti}}, \bibinfo {author} {\bibfnamefont {R.~J.}\ \bibnamefont
  {Lewis-Swan}},\ and\ \bibinfo {author} {\bibfnamefont {S.~A.}\ \bibnamefont
  {Haine}},\ }\bibfield  {title} {\bibinfo {title} {Pumped-up su(1,1)
  interferometry},\ }\href {https://doi.org/10.1103/PhysRevLett.118.150401}
  {\bibfield  {journal} {\bibinfo  {journal} {Phys. Rev. Lett.}\ }\textbf
  {\bibinfo {volume} {118}},\ \bibinfo {pages} {150401} (\bibinfo {year}
  {2017})}\BibitemShut {NoStop}%
\bibitem [{\citenamefont {Niezgoda}\ \emph {et~al.}(2019)\citenamefont
  {Niezgoda}, \citenamefont {Kajtoch}, \citenamefont {Dzieka{\'n}ska},\ and\
  \citenamefont {Witkowska}}]{niezgoda2019optimal}%
  \BibitemOpen
  \bibfield  {author} {\bibinfo {author} {\bibfnamefont {A.}~\bibnamefont
  {Niezgoda}}, \bibinfo {author} {\bibfnamefont {D.}~\bibnamefont {Kajtoch}},
  \bibinfo {author} {\bibfnamefont {J.}~\bibnamefont {Dzieka{\'n}ska}},\ and\
  \bibinfo {author} {\bibfnamefont {E.}~\bibnamefont {Witkowska}},\ }\bibfield
  {title} {\bibinfo {title} {Optimal quantum interferometry robust to detection
  noise using spin-1 atomic condensates},\ }\href
  {https://doi.org/10.1088/1367-2630/ab4099} {\bibfield  {journal} {\bibinfo
  {journal} {New Journal of Physics}\ }\textbf {\bibinfo {volume} {21}},\
  \bibinfo {pages} {093037} (\bibinfo {year} {2019})}\BibitemShut {NoStop}%
\bibitem [{\citenamefont {Guo}\ \emph {et~al.}(2021)\citenamefont {Guo},
  \citenamefont {Chen}, \citenamefont {Liu}, \citenamefont {Xue}, \citenamefont
  {Chen}, \citenamefont {Cao}, \citenamefont {Mao}, \citenamefont {Tey},\ and\
  \citenamefont {You}}]{Guo_2021}%
  \BibitemOpen
  \bibfield  {author} {\bibinfo {author} {\bibfnamefont {S.-F.}\ \bibnamefont
  {Guo}}, \bibinfo {author} {\bibfnamefont {F.}~\bibnamefont {Chen}}, \bibinfo
  {author} {\bibfnamefont {Q.}~\bibnamefont {Liu}}, \bibinfo {author}
  {\bibfnamefont {M.}~\bibnamefont {Xue}}, \bibinfo {author} {\bibfnamefont
  {J.-J.}\ \bibnamefont {Chen}}, \bibinfo {author} {\bibfnamefont {J.-H.}\
  \bibnamefont {Cao}}, \bibinfo {author} {\bibfnamefont {T.-W.}\ \bibnamefont
  {Mao}}, \bibinfo {author} {\bibfnamefont {M.~K.}\ \bibnamefont {Tey}},\ and\
  \bibinfo {author} {\bibfnamefont {L.}~\bibnamefont {You}},\ }\bibfield
  {title} {\bibinfo {title} {Faster state preparation across quantum phase
  transition assisted by reinforcement learning},\ }\href
  {https://doi.org/10.1103/PhysRevLett.126.060401} {\bibfield  {journal}
  {\bibinfo  {journal} {Phys. Rev. Lett.}\ }\textbf {\bibinfo {volume} {126}},\
  \bibinfo {pages} {060401} (\bibinfo {year} {2021})}\BibitemShut {NoStop}%
\bibitem [{\citenamefont {Yuen}(1976)}]{Yuen_1976}%
  \BibitemOpen
  \bibfield  {author} {\bibinfo {author} {\bibfnamefont {H.~P.}\ \bibnamefont
  {Yuen}},\ }\bibfield  {title} {\bibinfo {title} {Two-photon coherent states
  of the radiation field},\ }\href {https://doi.org/10.1103/PhysRevA.13.2226}
  {\bibfield  {journal} {\bibinfo  {journal} {Phys. Rev. A}\ }\textbf {\bibinfo
  {volume} {13}},\ \bibinfo {pages} {2226} (\bibinfo {year}
  {1976})}\BibitemShut {NoStop}%
\bibitem [{\citenamefont {Feng}\ \emph {et~al.}(2015)\citenamefont {Feng},
  \citenamefont {Wang}, \citenamefont {Yang},\ and\ \citenamefont
  {Jin}}]{WignerD_2015}%
  \BibitemOpen
  \bibfield  {author} {\bibinfo {author} {\bibfnamefont {X.~M.}\ \bibnamefont
  {Feng}}, \bibinfo {author} {\bibfnamefont {P.}~\bibnamefont {Wang}}, \bibinfo
  {author} {\bibfnamefont {W.}~\bibnamefont {Yang}},\ and\ \bibinfo {author}
  {\bibfnamefont {G.~R.}\ \bibnamefont {Jin}},\ }\bibfield  {title} {\bibinfo
  {title} {High-precision evaluation of {W}igner's $d$ matrix by exact
  diagonalization},\ }\href {https://doi.org/10.1103/PhysRevE.92.043307}
  {\bibfield  {journal} {\bibinfo  {journal} {Phys. Rev. E}\ }\textbf {\bibinfo
  {volume} {92}},\ \bibinfo {pages} {043307} (\bibinfo {year}
  {2015})}\BibitemShut {NoStop}%
\bibitem [{\citenamefont {Tajima}(2015)}]{WignerD2_2015}%
  \BibitemOpen
  \bibfield  {author} {\bibinfo {author} {\bibfnamefont {N.}~\bibnamefont
  {Tajima}},\ }\bibfield  {title} {\bibinfo {title} {Analytical formula for
  numerical evaluations of the {Wigner} rotation matrices at high spins},\
  }\href {https://doi.org/10.1103/PhysRevC.91.014320} {\bibfield  {journal}
  {\bibinfo  {journal} {Phys. Rev. C}\ }\textbf {\bibinfo {volume} {91}},\
  \bibinfo {pages} {014320} (\bibinfo {year} {2015})}\BibitemShut {NoStop}%
\bibitem [{\citenamefont {Bohnet}\ \emph {et~al.}(2016)\citenamefont {Bohnet},
  \citenamefont {Sawyer}, \citenamefont {Britton}, \citenamefont {Wall},
  \citenamefont {Rey}, \citenamefont {Foss-Feig},\ and\ \citenamefont
  {Bollinger}}]{Bohnet2016}%
  \BibitemOpen
  \bibfield  {author} {\bibinfo {author} {\bibfnamefont {J.~G.}\ \bibnamefont
  {Bohnet}}, \bibinfo {author} {\bibfnamefont {B.~C.}\ \bibnamefont {Sawyer}},
  \bibinfo {author} {\bibfnamefont {J.~W.}\ \bibnamefont {Britton}}, \bibinfo
  {author} {\bibfnamefont {M.~L.}\ \bibnamefont {Wall}}, \bibinfo {author}
  {\bibfnamefont {A.~M.}\ \bibnamefont {Rey}}, \bibinfo {author} {\bibfnamefont
  {M.}~\bibnamefont {Foss-Feig}},\ and\ \bibinfo {author} {\bibfnamefont
  {J.~J.}\ \bibnamefont {Bollinger}},\ }\bibfield  {title} {\bibinfo {title}
  {Quantum spin dynamics and entanglement generation with hundreds of trapped
  ions},\ }\href {https://doi.org/10.1126/science.aad9958} {\bibfield
  {journal} {\bibinfo  {journal} {Science}\ }\textbf {\bibinfo {volume}
  {352}},\ \bibinfo {pages} {1297} (\bibinfo {year} {2016})}\BibitemShut
  {NoStop}%
\end{thebibliography}%

\appendix

\section{Analytic treatment in the undepleted pump regime \label{app:UPA}}
The results of Secs.~\ref{sec:Analytics}-\ref{sec:DetectionNoise} in the main text follow from an analytic solution of the dynamics generated by, 
\begin{equation}
    \hat{H}_{\mathrm{UP}} = gN \left( \hat{a}_1\hat{a}_{-1} + \hat{a}^{\dagger}_1\hat{a}^{\dagger}_{-1} \right) , 
\end{equation}
which is quadratic in bosonic creation/annihilation operators and thus exactly solvable. In the following subsections we present the relevant solutions of the dynamics in both the Schr\"{o}dinger and Heisenberg pictures, and present details of key expressions presented in the main text. 

\subsection{Entangled input state}
Evolution under $\hat{H}_{\mathrm{UP}}$ is equivalent to bosonic two-mode squeezing that is studied in quantum optics. As a consequence, it is straightforward to adopt known results from the quantum optics literature to analytically solve the dynamics of the time-evolved state $\vert \psi^{s,d}_{t} \rangle_{\mathrm{UP}} = e^{-i\hat{H}_{\mathrm{UP}}t}\vert \psi^{s,d}_0 \rangle$ \cite{Caves_SqueezedCoherent_1991}. The combination of initial seeding and subsequent squeezing evolution means that the generated state within the undepleted pump regime can be identified as the well understood two-mode \emph{squeezed coherent} state (TMSC) \cite{Yuen_1976}. Specifically, $\vert \psi^{s,d}_{t} \rangle_{\mathrm{UP}} = \hat{S}(r(t),\pi/4)\hat{D}(\alpha_1,\alpha_{-1})\vert 0, 0 \rangle$ where $\hat{S}(r,\phi) = e^{r(\hat{a}_1\hat{a}_{-1}e^{-2i\phi}+\hat{a}^{\dagger}_1\hat{a}^{\dagger}_{-1}e^{2i\phi})}$ is the two-mode squeezing operator and $r(t) = gNt$, $\hat{D}(\alpha,\beta) = e^{\alpha\hat{a}^{\dagger}_1 - \alpha^*\hat{a}_1}e^{\beta\hat{a}^{\dagger}_{-1} - \beta^*\hat{a}_1}$ is the two-mode coherent displacement operator with $\alpha_{\pm 1}$ the initial coherent amplitude of the $m_F = \pm 1$ modes,  and $\vert 0, 0 \rangle$ is the bosonic vacuum state for the $m_F = \pm1$ modes. 

For a single initial seed, the time-evolved state in the Fock basis can be written as $\vert \psi^s_t \rangle_{\mathrm{UP}} = \sum_{n,m=0}^{\infty} c_{n,m}(t) \vert n, m\rangle$ with expansion coefficients \cite{Caves_SqueezedCoherent_1991}, 
\begin{multline}
    c_{n,m}(t) = e^{-n_s/2} e^{i(m+n)\pi/4} \sqrt{\frac{m!}{n!}}   \\
    \times \mathrm{sech}(gNt) \left[ \sqrt{n_s}e^{i(\theta_s - \pi/4)} \mathrm{sech}(gNt) \right]^{n-m} \\ \times \left[-\mathrm{tanh}(gNt) \right]^m L_{m}^{n-m}(0) , 
\end{multline}
for $m \geq n$ and zero otherwise. The state can be equivalently written in the collective spin basis, e.g., Eq.~(\ref{eqn:psit_SU2}) of the main text, by making the correspondence $J = (n+m)/2$ and $m_z = (m-n)/2$. We use the latter to plot the Wigner function \cite{Dowling_1994} of the generated state on a collective Bloch sphere, as in Fig.~\ref{fig:fig2}. 

\subsection{Expectation values}
Expectation values for the population dynamics, QFI and metrological sensitivity are instead most easily computed by treating the dynamics in the Heisenberg picture. The two-mode squeezing generated by the Hamiltonian $\hat{H}_{\mathrm{UP}}$ has the solution \cite{lewis2013epr}, 
\begin{equation}
    \hat{a}_{\pm1}(\tau) = \mathrm{cosh}(\tau)\hat{a}_{\pm 1} (0) - i\mathrm{sinh}(\tau)\hat{a}^{\dagger}_{\mp 1}(0) \label{eqn:UP_Heisenberg}
\end{equation}
where $\tau = gNt$. It is straightforward to use Eq.~(\ref{eqn:UP_Heisenberg}) in combination with the given initial states to compute all relevant correlation functions of the system after the period of spin-changing collisions, such as Eqs.~(\ref{eqn:Np})-(\ref{eqn:NpFluct}) in the main text. These expressions will naturally involve the interaction time $\tau$, but this can be replaced with the number of scattered atoms $\bar{n}$ by using, 
\begin{equation}
    \tau = \frac{1}{2}\mathrm{log}\left[ \frac{1 + n_s + \bar{n} + \sqrt{\bar{n}(2 + 2n_s + \bar{n})}}{1 + n_s} \right] ,
\end{equation}
which follows from the definition $N_+ = \langle \hat{n}_1(\tau) + \hat{n}_{-1}(\tau) \rangle = \bar{n} + n_s$.

\subsection{MZ sensitivity}
Relevant correlations at the output of the MZ interferometer can also be obtained by apply a series of linear transformations on Eq.~(\ref{eqn:UP_Heisenberg}) corresponding to the beam-splitter and mirror elements. Specifically, denoting $\hat{b}_{\pm1}$ as the bosonic annihilation operator of the $m_F = \pm1$ modes after the full MZ sequence we have the relation, 
\begin{equation}
    \hat{b}_{\pm1} = \cos\left(\frac{\varphi}{2}\right)\hat{a}_{\pm1}(\tau) - i\sin\left(\frac{\varphi}{2}\right)\hat{a}_{\mp1}(\tau) . \label{eqn:MZ_transformation}
\end{equation}

With Eq.~(\ref{eqn:MZ_transformation}) we then obtain the following required correlation functions. For a single initial seed, \begin{widetext}
\begin{equation}
    \begin{gathered}
        \langle \hat{J}_z(\varphi) \rangle_s =  -\frac{n_s}{2}\cos (\varphi ) + \frac{n_s\cos(2\theta_s)  \sqrt{\bar{n} \left(\bar{n}+2 n_s+2\right)}}{2(n_s+1)}\sin (\varphi ) , \label{eqn:JzJz2Jz4} \\
        \langle \hat{J}^2_z(\varphi)\rangle_s = \frac{\bar{n} \sin ^2(\varphi ) \left(\bar{n}+2 n_s+2\right) \left(n_s^2 \cos (4\theta_s )+(n_s+4) n_s+2\right)}{8(n_s+1)^2} - \frac{n_s \cos (2 \theta_s ) \sin (2 \varphi )
        \sqrt{\bar{n} \left(\bar{n}+2 n_s+2\right)}}{4} \\ + \frac{n_s (n_s \cos (2 \varphi )+n_s+2)}{8} , \\
        \langle \hat{J}^4_z(\varphi)\rangle_s = -\frac{3}{32} - \frac{5}{4}\langle \hat{J}_z^2(\varphi) \rangle_s + \frac{1}{32}\sum_{j=0}^{4} \mathcal{C}^s_j \cos^{4-j}(\varphi)\sin^j(\varphi)
    \end{gathered}
\end{equation}
where 
\begin{equation}
    \begin{gathered}
        \mathcal{C}^s_0 =  \left(2 n_s^4+12 n_s^3+24 n_s^2+12 n_s+3\right) , \\
        \mathcal{C}^s_1 = -\frac{n_s (n_s+3) \left(2 n_s^2+6 n_s+3\right) \sqrt{\bar{n} \left(\bar{n}+2
        n_s+2\right)}}{ (n_s+1)} , \\
        \mathcal{C}^s_2 = \frac{\left(2 n_s^4+14 n_s^3+25 n_s^2+12 n_s+3\right) \bar{n} \left(\bar{n}+2
        n_s+2\right)+(2 n_s+1) (n_s+1)^4}{ (n_s+1)^2} , \\
        \mathcal{C}^s_3 = -\frac{n_s (n_s+3) \sqrt{\bar{n} \left(\bar{n}+2 n_s+2\right)} \left\{2 [n_s
        (n_s+6)+3] \bar{n} \left(\bar{n}+2 n_s+2\right)+3 (2 n_s+1) (n_s+1)^2\right\}}{
        (n_s+1)^3} , \\
        \mathcal{C}^s_4 =  \frac{2 \bar{n} \left(\bar{n}+2 n_s+2\right) \left\{[n_s (n_s+6)+3]
   \bar{n}+3 (n_s+1)^2\right\} \left\{[n_s (n_s+6)+3] \bar{n}+(n_s+1) [n_s (2
   n_s+9)+3]\right\}}{(n_s+1)^4}+6 n_s (n_s+2)+3 .
    \end{gathered}
\end{equation}
For an initial state with dual seeds, 
\begin{equation}
    \begin{gathered}
        \langle \hat{J}_z(\varphi) \rangle_d = -\frac{n_s(1+n_s + \bar{n})\mathrm{sin}(2\theta_s)}{2(1+n_s)} \mathrm{sin}(\varphi) , \label{eqn:JzJz2Jz4_d} \\
        \langle \hat{J}^2_z(\varphi)\rangle_d =  \frac{n_s(2+n_s)}{8} + \frac{\bar{n} \sin ^2(\varphi ) \left(\bar{n}+2 n_s+2\right)}{4} - \frac{n_s^2}{8}\cos(2\varphi) , \\
        \langle \hat{J}^4_z(\varphi)\rangle_d = -\frac{3}{32} - \frac{5}{4}\langle \hat{J}_z^2(\varphi) \rangle_d + \frac{1}{32}\sum_{j=0}^{4} \mathcal{C}^d_j \cos^{4-j}(\varphi)\sin^j(\varphi)
    \end{gathered}
\end{equation}
where
\begin{equation}
    \begin{gathered}
        \mathcal{C}^d_0 =  3 \left(2 n_s^2+4 n_s+1\right) , \\
        \mathcal{C}^d_1 = 0 , \\
        \mathcal{C}^d_2 = \frac{\left\{n_s [n_s (2 n_s+7)+12]+3\right\} \bar{n} \left(\bar{n}+2 n_s+2\right)+(2 n_s+1)
   (n_s+1)^4}{(n_s+1)^2} , \\
        \mathcal{C}^d_3 = 0 , \\
        \mathcal{C}^d_4 = \frac{2 \bar{n} \left(\bar{n}+2 n_s+2\right)}{(1+n_s)^4} \left\{ 2 \left(n_s^2+6 n_s+3\right)^2 (n_s+1) \bar{n}+\left(n_s^2+6 n_s+3\right)^2
   \bar{n}^2+\left(2 n_s^4+18 n_s^3+51 n_s^2+36 n_s+9\right) (n_s+1)^2 \right\} \\ + 2 n_s (6 + 12 n_s + 6 n_s^2 + n_s^3) + 3 .
    \end{gathered}
\end{equation}
\end{widetext}
In all expressions we have again replaced the natural dependence on $\tau$ with $\bar{n}$. 

The sensitivity of the MZ interferometer with measurement signals $\hat{J}_z$ or $\hat{J}_z^2$ can be constructed using the results of Eqs.~(\ref{eqn:JzJz2Jz4}) and (\ref{eqn:JzJz2Jz4_d}). In the main text we presented the results for the former under the simplification that $\theta_s = 0,\pi/4$, but in full generality we obtain:
\begin{widetext}
\begin{equation}
    (\Delta\varphi)^2_{\hat{J}_z,s} = \frac{n_s (n_s+1) \left[n_s+1 -\cos (2 \theta_s ) \sin (2 \varphi ) \sqrt{\bar{n} \left(\bar{n}+2
    n_s+2\right)}\right]+(2 n_s+1) \bar{n}^2 \sin ^2(\varphi )+2 (n_s+1) (2
    n_s+1) \bar{n} \sin ^2(\varphi )}{n_s^2 \left[\cos (2 \theta_s ) \cos (\varphi ) \sqrt{\bar{n}
    \left(\bar{n}+2 n_s+2\right)}+(n_s+1) \sin (\varphi )\right]^2} , \label{eqn:SensJz_theta_s}
\end{equation}
and 
\begin{multline}
    (\Delta\varphi)^2_{\hat{J}_z,d} = \frac{\csc ^2(2\theta_s ) \sec ^2(\varphi )}{n_s^2\left(\bar{n}+n_s+1\right)^2} \Bigg\{ (2 n_s+1) \bar{n} \sin ^2(\varphi ) \left(\bar{n}+2 n_s+2\right) \\
    + n_s (n_s+1) \left[n_s+1-\cos (2 \theta ) \sin (2 \varphi ) \sqrt{\bar{n} \left(\bar{n}+2 n_s+2\right)}\right] \Bigg\} . \label{eqn:SensJz_theta_d}
\end{multline}
The expressions for the sensitivity obtained with $\hat{J}_z^2$ are much more involved and not useful to reproduce.
\end{widetext}

We use Eqs.~(\ref{eqn:SensJz_theta_s}) and (\ref{eqn:SensJz_theta_d}) to analytically compute the optimal sensitivity as a function of $\varphi$, $n_s$ and $\theta_s$. The results of this are quoted in Sec.~\ref{sec:OptimalSensitivity}, but we briefly discuss our procedure here. 

First, it is straightforward to determine that $\theta_s = 0$ and $\theta_s = \pi/4$ are optimal seed phases for the single and dual seed initial states, respectively, as they maximize the amplitude of the signal $\langle \hat{J}_z(\varphi) \rangle$. The optimal working point $\varphi_{\mathrm{opt}}$ is then obtained from Eq.~(\ref{eqn:SensJz_theta_d}) by minimizing the numerator. We find $\varphi_{\mathrm{opt}} = 0$ and thus $(\Delta\varphi)^2_{\hat{J}_z,d}\vert_{\varphi_{\mathrm{opt}}} = (1+n_s)^2/[n_s(1+n_s+\bar{n})]$ as per Eq.~(\ref{eqn:JzSensOpt_d}) of the main text. Moreover, this sensitivity is minimized for $n^{\mathrm{opt}}_s \simeq 1$ in the limit of large $\bar{n} \gg n_s, 1$. 

On the other hand, optimizing Eq.~(\ref{eqn:SensJz_theta_s}) is more involved. First, we empirically identify that in the limit of $n_s \gtrsim 1$ (i.e., when the sensitivity is meaningfully useful) the sensitivity is minimized in the neighbourhood of $\varphi \approx 0$. This motivates an expansion of $(\Delta\varphi)^2_{\hat{J}_z,s}$ as a Maclaurin series in $\varphi$, e.g., $(\Delta\varphi)^2_{\hat{J}_z,s} \approx a_0 + a_1\varphi + a_2\varphi^2 + \mathcal{O}(\varphi^3)$ where $a_0,a_1,a_2,...$ can be computed from derivatives of Eq.~(\ref{eqn:SensJz_theta_s}). Retaining only terms to quadratic order, it is straightforward to obtain the minimum of the sensitivity as $(\Delta\varphi)^2_{\hat{J}_z,d}\vert_{\varphi_{\mathrm{opt}}} = a_0 - a_1^2/4a_2 = (1+n_s)^3/[\bar{n}^2n_s(1+2n_s)]$ at $\varphi_{\mathrm{opt}} = a_1/(2_a2) = [n_s(1+n_s)]/[\bar{n}(1+2n_s)]$. Subsequent minimization with respect to $n_s$ gives $n^{\mathrm{opt}}_s \simeq (1+\sqrt{3})/2$ in the limit of large $\bar{n} \gg n_s, 1$. For completeness, we follow a similar recipe to obtain the results (\ref{eqn:JzSens_thetasfluct}) and (\ref{eqn:JzSens_nsfluct}) where $\theta_s$ or $n_s$ are allowed to randomly fluctuate due to imperfect state preparation.

The results for $(\Delta\varphi)^2_{\hat{J}_z}$ given above can be readily extended to include detection noise $\sigma$, characterized by the relations (\ref{eqn:sigma_relations}) of the main text. As the ideal ($\sigma = 0$) working point coincides with the maximum magnitude of the slope $\partial_{\varphi}\langle \hat{J}_z \rangle$ we can assume that to a good approximation $\varphi_{\mathrm{opt}}$ is unchanged for modest detection noise.
Under this assumption, the optimal sensitivity for $\sigma \neq 0$ can be estimated as
\begin{multline}
        (\Delta\varphi)^2_{\hat{J}_z,s,\sigma}\vert_{\varphi_{\mathrm{opt}}} \simeq \Bigg[ (\Delta\varphi)^2_{\hat{J}_z,s,\sigma=0} \\ + \frac{\sigma^2}{\vert \partial_{\varphi}\langle \hat{J}_z \rangle_{\sigma=0} \vert^2} \Bigg] \Bigg\vert_{\varphi=\varphi_{\mathrm{opt}}} .
\end{multline}
Substitution of relevant expressions into this formula leads directly to Eqs.~(\ref{eqn:SigmaSensJz}) and (\ref{eqn:SigmaSensJz_d}) of the main text. 

Finally, the DNR for $(\Delta\varphi)^2_{\hat{J}_z}$ can also be computed in the absence of detection noise [Eq.~(\ref{eqn:JzDNR}) of the main text]. For an initial state with two seeds this can be accomplished exactly by finding  values of $\varphi$ that solve $(\Delta\varphi)^2_{\hat{J}_z,d} = 1/\sqrt{N_+}$ using Eq.~(\ref{eqn:SensJz_theta_d}) in the limit of large $\bar{n}\gg 1, n_s$. For a single seed we again use the Maclaurin series expansion of $(\Delta\varphi)^2_{\hat{J}_z,s}$ and solve the same equality to obtain an identical result.

\section{Efficient numerical calculation of dynamics and classical Fisher information \label{app:CFI}}

\subsection{Quantum Dynamics}
Figures \ref{fig:fig5} and \ref{fig:fig6} of the main text present results based on the full quantum dynamics within the single-mode approximation. To be concrete, we use an efficient Chebyshev expansion approach~\cite{Kosloff_JCP1984} to numerically solve the dynamics of the system in the Schr\"{o}dinger picture. Calculations are implemented using OpenMP on a HPC node and a simulation involving $N \sim 10^4$ particles typically takes $\sim 1-5$hrs using $\sim 20$ CPU cores.

Our numerical calculations assume the initial $m_F = 0$ BEC is prepared in a Fock state of $N$ atoms before the $n_s$ seed atoms are coherently transferred into the $m_F = +1$ state. This approach becomes formally equivalent to the initial coherent seeds we consider in our analytic calculations in the limit of $N \to \infty$ \cite{Jianwen_2019}, but is primarily useful to reduce computational overhead by working in a regime of fixed total particle number $\hat{N} = \hat{n}_0 + \hat{n}_1 + \hat{n}_{-1}$.  

Our calculations use the full Hamiltonian
\begin{multline}
    \hat{H}_{\mathrm{sim}} = g\left( \hat{a}_0\hat{a}_0 \hat{a}^{\dagger}_1\hat{a}^{\dagger}_{-1} + h.c. \right)+
g \hat{n}_0 \left( \hat{n}_1 + \hat{n}_{-1} \right) \\
- q(\hat{n}_1 + \hat{n}_{-1})  +
\frac{g}{2}\left(\hat{n}_1 - \hat{n}_{-1}\right)^2- g n_s \left(\hat{n}_1 - \hat{n}_{-1} \right), \label{spin_collision}
\end{multline}
which includes all terms of $\hat{H}$ [Eq.~(\ref{eqn:H})] in the main text except for the irrelevant linear Zeeman shift. The last term $- g n_s \left(\hat{n}_1 - \hat{n}_{-1} \right)$ of $\hat{H}_{\mathrm{sim}}$ is added to cancel off the mean-field contribution of the spin-precession generated by $\frac{g}{2}\left(\hat{n}_1 - \hat{n}_{-1}\right)^2$ (discussed in more detail in Appendix~\ref{app:Hamiltonian}). This mimics the experimental calibration of the relationship between initial seeding phase and the beam-splitter phase.

\subsection{CFI}
The CFI can be evaluated according to Eq.~(\ref{eqn:CFI}) of the main text. However, in practice we find it is simpler to compute the CFI in the equivalent form
\begin{equation}
    F_C(\varphi) = \sum_{J,m_z} \frac{1}{P_{J,m_z}(\varphi)} \left[ \frac{\partial P_{J,m_z}(\varphi)}{\partial \varphi} \right]^2, 
\end{equation}
where $P_{J,m_z}(\varphi) \equiv \vert c^{\varphi}_{J,m_z}$ is the joint distribution function for the total spin and mean projection along $z$ obtained in terms of the expansion coefficients of the collective spin state after the phase-shift is imprinted, e.g., $\vert \psi^{\varphi}_{\mathrm{BS}} \rangle = \hat{U}_{\mathrm{BS}} \hat{U}_{\varphi} \vert \psi^s_{\tau} \rangle$. 

We obtain $\vert \psi^{\varphi}_{\mathrm{BS}} \rangle$ and the associated $c^{\varphi}_{J,m_z}$ using a Chebyshev expansion approach, i.e., by treating the rotation operator $\exp\left(-i\hat{\bf{{S}}}\cdot{\bf{n}}\varphi\right)$ as a time propagation operator where the angle $\varphi$ plays the role of time and $\hat{\bf{{S}}}\cdot{\bf{n}}$ is the effective Hamiltonian. This approach enables us to avoid the compatutationally expensive evaluation of Wigner-D matrices in a very high angular momentum space \cite{WignerD_2015, WignerD2_2015}.

To account for detection noise we compute the convolved distribution $|\tilde{P}_{J, m_z}|^2$ using,
\begin{equation}
\label{convolution}
|\tilde{P}_{J, m_z}|^2 =\frac{1}{2\pi\sigma^2} \sum_{J^{\prime}, m_z^{\prime}} |P_{J^{\prime}, m_z^{\prime}}|^2e^{-\frac{(J-J^{\prime})^2+(m_z-m_z^{\prime})^2}{2\sigma^2}}.
\end{equation}
In our numerical calculation we truncate the double sum so that it only includes the relevant region $|J^{\prime}-J|, |m_z^{\prime}-m_z|\in [0, 10\sigma]$. A brute force calculation of the convolution is not appropriate as the evaluation of the sum scales as $O(100\sigma^2\mathcal{N})$ where $\mathcal{N}$ is the size of the Hilbert space, which becomes slow for large $\sigma$ or $\mathcal{N}$. 
Instead, we decouple the expression (\ref{convolution}) into a pair of sums, 
\begin{eqnarray}
\label{convolution1}
|\mathcal{P}_{J, m_z^{\prime}}|^2 = \frac{1}{\sqrt{2\pi\sigma^2}}\sum_{J^{\prime}} |P_{J^{\prime}, m_z^{\prime}}|^2e^{-\frac{(J-J^{\prime})^2}{2\sigma^2}},\\\nonumber
|\tilde{P}_{J, m_z}|^2 = \frac{1}{\sqrt{2\pi\sigma^2}}\sum_{m_z^{\prime}} |\mathcal{P}_{J, m_z^{\prime}}|^2e^{-\frac{(m_z-m_z^{\prime})^2}{2\sigma^2}},
\end{eqnarray}
which only scales as $2\times O(10\sigma \mathcal{N})$.

\section{Role of terms $\propto (\hat{n}_1 - \hat{n}_{-1})^2$ in the dynamics \label{app:Hamiltonian}}
In the analytic treatment of Secs.~\ref{sec:Analytics}-\ref{sec:DetectionNoise} it was assumed that elastic collisions $\propto (\hat{n}_1 - \hat{n}_{-1})^2$ in $\hat{H}$ could be ignored. This was justified by noting that: i) the term commutes with the Hamiltonian and can be treated independently, and ii) it should be negligibly small compared to other contributions to the dynamics given that we typically assume $n_s \ll N$. The former point means that we can understand the influence of the additional term as simply a subsequent transformation of the state we generate, e.g., $\vert \psi_t \rangle \to e^{-igt(\hat{n}_1 - \hat{n}_{-1})^2} \vert \psi_t \rangle$, and input to the MZ interferometer. We use this to better justify point ii) in the following. 

The transformation, $e^{-igt(\hat{n}_1 - \hat{n}_{-1})^2}$, is equivalent a one-axis twisting (OAT) term in the collective spin picture -- $(\hat{n}_1 - \hat{n}_{-1})^2 \propto \hat{J}_z^2$ -- and this provides a useful way to quantify its impact on the physics we predict. In particular, the OAT evolution can be broken into two independent effects: 1) a mean-field rotation $\propto g t \langle \hat{J}_z \rangle \hat{J}_z \sim g n_s t \hat{J}_z$ of the state $\vert \psi^s_t \rangle$ about the Bloch sphere that deterministically shifts the seed phase, similar to the linear Zeeman shift, $\theta_s \to \theta_s + g n_s t$ (this is absent for states with dual seeding), and 2) a nonlinear shearing of the fluctuations of the state at a rate $\sim g\sqrt{n_s}$ (corresponding to the size of the quantum fluctuations in $J_z$ of the squeezed states we prepare). 

Both effects must be considered with respect to the relevant time-scale of the pair production process, e.g., $t \sim \mathcal{O}(1/gN)$. For example, the precession generates a drift of the optimal seed phase on the order of $gn_s t \sim n_s/N$. In principle, this drift is negligibly small for most cases we consider ($n_s \ll N$). However, it is also easily removed by appropriate calibration of the ensuing beam-splitter operations/spin rotations (identically to the linear Zeeman shift). For this reason we artificially remove this contribution in Fig.~\ref{fig:fig5} of the main text, to better match a real experiment (see prior discussion in Appendix~\ref{app:CFI}). On the other hand, the shearing contribution leads to an irreversible reduction in contrast, e.g., decrease in effective $\vert \langle \hat{J}_+ \rangle \vert$ of the squeezed state, that degrades the performance of the interferometer \cite{Kitagawa1993,Bohnet2016}. Fortunately, the much slower rate of this effect means that it can be easily neglected on the relevant time-scale of pair production, e.g., $g\sqrt{n_s}t \sim \sqrt{n_s}/N_0 \ll 1$ for all cases we consider in Sec.~\ref{sec:RealCalc}. We have checked this claim by constructing Fig.~\ref{fig:fig5}(b) with and without the $(\hat{n}_1 - \hat{n}_{-1})^2$ contribution and observed that the relevant results change negligibly for $n_s \lesssim 0.1N$.

\end{document}